\documentclass[journal]{IEEEtran}

\ifCLASSINFOpdf
\else
\fi
\usepackage[utf8]{inputenc}
\usepackage{amsmath,epsfig,comment}
\usepackage{amsthm}
\usepackage{tcolorbox}
\usepackage{amssymb}
\usepackage{mathrsfs}
\usepackage{amsmath}
\usepackage{tikz}
\usetikzlibrary{arrows,decorations.markings,automata}
\usepackage{hyperref}
\hypersetup{
	colorlinks=true,
	linkcolor=black,
	filecolor=magenta,
	urlcolor=cyan,
	citecolor=blue
}

\usepackage{cases} 
\usepackage{amssymb,amsmath,cite}
\usepackage{epsfig}
\usepackage{color}
\usepackage{bm}
\usepackage{graphicx}
\usepackage{subfigure}
\usepackage{algorithm}
\usepackage{algorithmic}
\usepackage{multirow}
\usepackage{array}
\usepackage{soul}
\usepackage{extarrows}

\usepackage{float}
\usepackage{booktabs}
\usepackage{multirow}
\usepackage{lipsum}

\newcommand{\diagg}{\operatorname{diag}}

\newcommand\Lc{\ensuremath{{\mathcal{L}}}}

\newcommand\tr{\ensuremath{{\rm Tr}}}

\DeclareMathAlphabet\mathbfcal{OMS}{cmsy}{b}{n}

\newcommand{\mat}[1]{\mathbf{#1}}

\def\A{\mathbf{A}}
\def\X{\mathcal{X}}

\def\d{\mathbf{d}}

\def\R{\mathbb{R}}
\def\C{\mathbb{C}}
\def\H{\mathbf{H}}

\def\y{\mathbf{y}}

\def\x{\mathbf{x}}
\def\h{\mathbf{h}}

\def\v{\mathbf{v}}

\def\T{\mathsf{T}}

\makeatletter
\newcommand{\pushright}[1]{\ifmeasuring@#1\else\omit\hfill$\displaystyle#1$\fi\ignorespaces}
\newcommand{\pushleft}[1]{\ifmeasuring@#1\else\omit$\displaystyle#1$\hfill\fi\ignorespaces}
\makeatother

\newcommand{\bb}{\mathbf{b}}
\newcommand{\bc}{\mathbf{c}}
\newcommand{\be}{\mathbf{e}}
\newcommand{\bh}{\mathbf{h}}

\newcommand{\bU}{\mathbf{U}}
\newcommand{\bW}{\mathbf{W}}
\newcommand{\bu}{\mathbf{u}}
\newcommand{\bv}{\mathbf{v}}

\newcommand{\bV}{\mathbf{V}}
\newcommand{\bT}{\mathbf{T}}

\newcommand{\bw}{\mathbf{w}}

\newcommand{\bx}{\mathbf{x}}
\newcommand{\bX}{\mathbf{X}}
\newcommand{\bd}{\mathbf{d}}
\newcommand{\bp}{\mathbf{p}}
\newcommand{\bq}{\mathbf{q}}
\newcommand{\br}{\mathbf{r}}
\newcommand{\bs}{\mathbf{s}}
\newcommand{\bt}{\mathbf{t}}
\newcommand{\by}{\mathbf{y}}
\newcommand{\bz}{\mathbf{z}}
\newcommand{\bA}{\mathbf{A}}
\newcommand{\bB}{\mathbf{B}}

\newcommand{\bH}{\mathbf{H}}

\newcommand{\bQ}{\mathbf{Q}}

\newcommand{\bgam}{\boldsymbol{\gamma}}
\newcommand{\bka}{\boldsymbol{\kappa}}
\newcommand{\cB}{\mathcal{B}}

\def\st{\mathrm{s.t.}}
\def\tr{\mathrm{tr}}
\def\cCN{\mathcal{CN}}
\def\d{\ldots}
\renewcommand{\v}[1]{\boldsymbol{\mathbf{#1}}}
\newcommand{\mv}[1]{\boldsymbol{\mathbf{#1}}}
\newcommand{\m}[1]{\boldsymbol{\mathbf{#1}}}
\def\hermitian{\dagger}
\def\transpose{\mathsf{T}}
\def\cK{\mathcal{K}}

\def\cM{\mathcal{M}}
\def\fk{~k \in \cK}
\def\fm{~m \in \cM}

\newtheoremstyle{noparens}%
{}{}%
{\itshape}{}%
{\bfseries}{.}%
{ }%
{\thmname{#1}\thmnumber{ #2}\mdseries\thmnote{ #3}}

\theoremstyle{noparens}

\begin{document}
%
\title{A Survey of Recent Advances in Optimization Methods for Wireless Communications}
%
%
%

\author{
	\IEEEauthorblockN{Ya-Feng~Liu, \IEEEmembership{Senior Member, IEEE}, Tsung-Hui Chang, \IEEEmembership{Fellow, IEEE}, Mingyi Hong, \IEEEmembership{Senior Member, IEEE}, \\ 
	Zheyu Wu, \IEEEmembership{Graduate Student Member, IEEE}, Anthony Man-Cho So, \IEEEmembership{Fellow, IEEE}, \\ Eduard A. Jorswieck, \IEEEmembership{Fellow, IEEE},  and Wei Yu, \IEEEmembership{Fellow, IEEE}}
	\thanks{Manuscript submitted to \emph{IEEE Journal on Selected Areas in Communications} on January 16, 2024, revised April 14, 2024 and May 29, 2024, accepted May 30, 2024. 
	The work of Ya-Feng Liu and Zheyu Wu was supported by the National Natural Science Foundation of China (NSFC) under Grant 12371314, Grant 12022116, Grant 12021001, and Grant 12288201. The work of Tsung-Hui Chang was supported by the Shenzhen Science and Technology Program under Grant RCJC20210609104448114 and ZDSYS20230626091302006, the NSFC under Grant 62071409, and by the Guangdong Provincial Key Laboratory of Big Data Computing. The work of Mingyi Hong was supported by the NSF under Grant EPCN-2311007 and Grant CNS-2003033. The work of Eduard A. Jorswieck was supported by the Federal Ministry of Education and Research of Germany in the program of ``Souveraen. Digital. Vernetzt.'' Joint project 6G-RIC, project identification number: 16KISK031.  The work of Wei Yu was supported by the Natural Sciences and Engineering Research Council (NSERC) via a Discovery grant and via the Canada Research Chairs program.	
		\emph{(Corresponding author: Wei Yu)}
		}

	\thanks{Ya-Feng Liu and Zheyu Wu are with the State Key Laboratory of Scientific and Engineering Computing, Institute of Computational Mathematics and Scientific/Engineering Computing, Academy of Mathematics and Systems Science, Chinese Academy of Sciences, Beijing 100190, China (e-mails: \{yafliu, wuzy\}@lsec.cc.ac.cn).}
	\thanks{Tsung-Hui Chang is with the School of Science and Engineering, The Chinese University of Hong Kong, Shenzhen, Shenzhen, China, and Shenzhen Research Institute of Big Data (e-mail: tsunghui.chang@ieee.org).}
	\thanks{Mingyi Hong is with the Department of Electrical and Computer Engineering, University of Minnesota, Minneapolis, MN, 55455, USA (e-mail: mhong@umn.edu).}
	\thanks{Anthony Man-Cho So is with the Department of Systems Engineering and Engineering Management, The Chinese University of Hong Kong, HKSAR, China (e-mail: manchoso@se.cuhk.edu.hk).}
	\thanks{Eduard A. Jorswieck is with the Institute for Communication Technology, Technische Universitat Braunschweig, Germany (email: e.jorswieck@tu-braunschweig.de).}
	\thanks{Wei Yu is with The Edward S. Rogers Sr. Department of Electrical and Computer Engineering, University of Toronto, Toronto, Ontario M5S 3G4, Canada (e-mail: weiyu@ece.utoronto.ca).}
}

\maketitle
\begin{abstract}

Mathematical optimization is now widely regarded as an indispensable modeling and solution tool for the design of wireless communications systems. 
While optimization has played a significant role in the revolutionary progress in wireless communication and networking technologies 
from 1G to 5G and onto the future 6G, the innovations in wireless technologies have also substantially transformed the nature of the underlying mathematical optimization problems upon which the system designs are based and have sparked significant innovations in the development of methodologies to understand, to analyze, and to solve those problems. 
In this paper, we provide a comprehensive survey of recent advances in mathematical optimization theory and algorithms for wireless communication system design. We begin by illustrating common features of mathematical optimization problems arising in wireless communication system design. We discuss various scenarios and use cases and their associated mathematical structures from an optimization perspective. We then provide an overview of recently developed optimization techniques in areas ranging from nonconvex optimization, global optimization, and integer programming, to distributed optimization and learning-based optimization. The key to successful solution of mathematical optimization problems is in carefully choosing or developing suitable algorithms (or neural network architectures) that can exploit the underlying problem structure. We conclude the paper by identifying several open research challenges and outlining future research directions.

\end{abstract}



%
\IEEEpeerreviewmaketitle

\section{Introduction}
\subsection{Evolution of Wireless Cellular Communication Systems: From 1G to 6G}
Wireless communication technology has progressed dramatically in the last several decades. Wireless communication systems have impacted our society in profound ways and have become an integral part of our daily lives. 
The development of wireless communication technology is continuously driven by the requirements imposed by newly emerging use cases---such as aggregate/peak data rate, latency, cost and energy consumption, spectrum and energy efficiency, connectivity density, and many others. These ever more stringent key performance indicators (KPIs) have propelled innovations in both the physical and networking layer technologies from 1G to current 5G \cite{5G} in the past several decades, and will continue to do so into the era of future 6G wireless systems \cite{saad2019vision,dang2020should,you2021towards,jiang2021road,ITU-R2023}. 
These innovations include but are not limited to: advanced massive multiple-input multiple-output (MIMO) technologies \cite{bjornson2017massive,ZhangBMNYL20a,bjornson2023twenty} such as coordinated/cooperative beamforming \cite{gershman2010convex,elbir2023twenty}, 
hybrid beamforming \cite{zhang_molisch_hybrid} and symbol-level precoding\cite{CItutorial}, new waveforms \cite{liu2022evolution} ranging from time-division multiple access, code-division multiple access, orthogonal frequency-division multiple access to nonorthogonal multiple access, 
novel access protocols and paradigms \cite{Liu2018b,ChenX2021} such as grant-free multiple access, new networking architectures \cite{simeone2016cloud,ngo2017cell} such as cloud radio access network (C-RAN) and cell-free (massive) MIMO, as well as advanced signal processing algorithms such as 
efficient compressed sensing techniques. 

\begin{figure}[!t]
	\centering
	\includegraphics[width=0.35\textwidth]{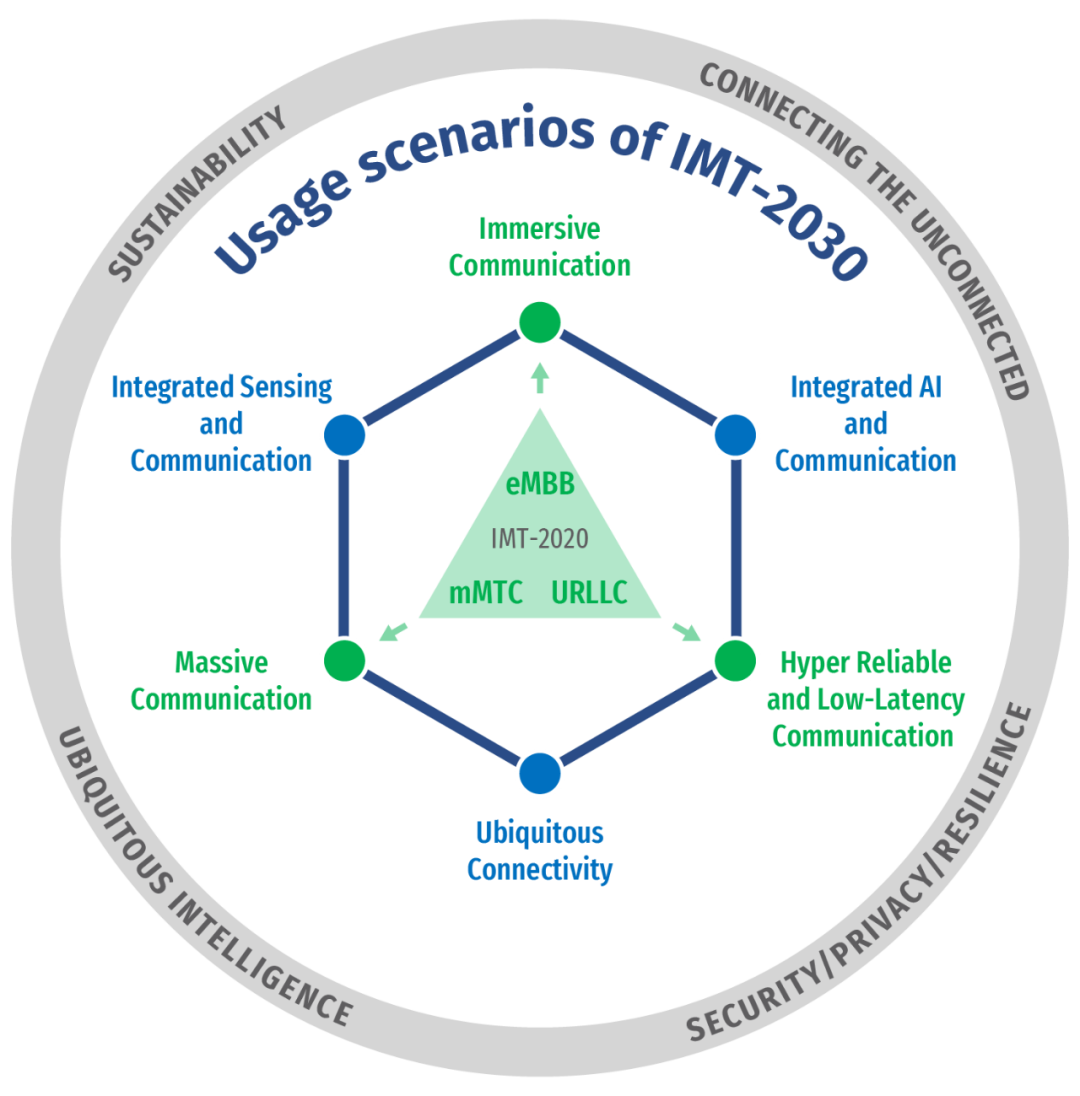}
	\caption{Six usage scenarios and overarching goals of IMT-2030\cite{ITU-R2023}. }
	\label{fig:6Gusecases}
\end{figure}

The 5G cellular system is currently being standardized and deployed
worldwide. It can provide services for enhanced mobile broadband (eMBB),
massive machine-type communication (mMTC), as well as ultra-reliable and
low-latency communication (URLLC) for both the conventional human-type and new
Internet-of-Things (IoT) users. The 6G, as the successor of 5G and to be
commercialized around 2030, is now in the research spotlight.  
A recent milestone in the development of 6G is the
Recommendation for IMT-2030 \cite{ITU-R2023}. It is drafted by the International
Telecommunication Union 
and provides guidelines on the framework and overall objectives of
6G. In particular, it defines six major usage scenarios of 6G, as shown in
Fig.~\ref{fig:6Gusecases}.  The additional use cases of 
integrated sensing and communications (ISAC), integrated AI, and ubiquitous connectivity,
at the intersections of eMBB, URLLC, and mMTC, are expected to 
be the drivers of wireless technology developments in the coming decade. 


{
	\begin{figure*}[!t]
		\centering
		\includegraphics[width=0.6\textwidth]{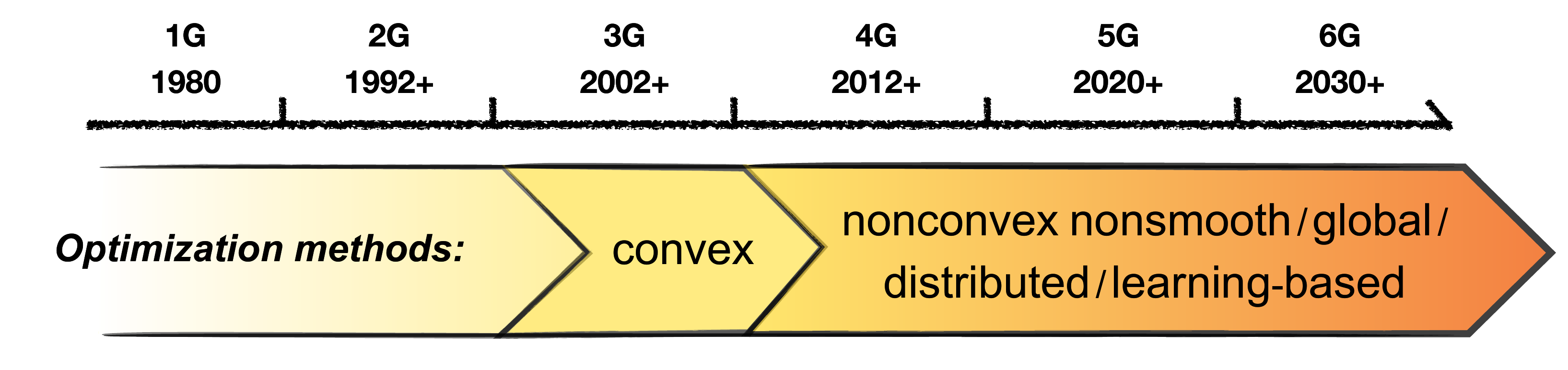}
		\caption{The evolution of wireless communication systems and the development of optimization methods are closely interwoven with each other. In particular, convex optimization techniques influenced the design of 3G communication system, whereas mathematical optimization problems arising from 4G--6G communication system design call for and have driven the development of new and advanced optimization theory, algorithms, and techniques such as nonconvex nonsmooth optimization, global and integer optimization, parallel and distributed optimization, and learning-based optimization.}
		\label{fig:opt_commun}
\end{figure*}}

\subsection{Central Role of Mathematical Optimization in Wireless Communication System Design}

Mathematical optimization is at the core of all of the above mentioned wireless communication technologies. It is widely recognized as a powerful and indispensable modeling and solution tool in the systematic design of wireless communication systems. 
Indeed, many problems arising from wireless communication system design can be formulated as mathematical optimization problems and efficiently solved by leveraging suitable optimization algorithms and techniques. Mathematical programming is in fact now a common language for wireless communication researchers. 
Fig.~\ref{fig:opt_commun} illustrates the progression of major (of course, not all) optimization methods that play central roles in and make strong impacts on different generations of wireless communication systems. The boundaries between these generations are of course porous. 

Convex optimization has played a vital role in 3G wireless research and has been by far the most extensively adopted paradigm for tackling wireless communication applications; see \cite{luo2003applications,chiang2005geometric,luo2006introduction,palomar2010convex} and the references therein. In some sense, once the problem is formulated and recognized as a convex optimization problem, efficient solutions are often in sight, 
as convex optimization problems, even complex ones such as second-order cone programs (SOCPs) and semidefinite programs (SDPs), possess favorable theoretical and computational properties and can be tackled by efficient and mature solvers such as SeDuMi, SDPT3, and SDPNAL+. Indeed, many problems of practical interest in 3G wireless communication system design have been shown to admit convex formulations/reformulations or good convex approximations/relaxations \cite{luo2010}.

Compared to 3G, technological advancements in previous 4G, current 5G, and future 6G wireless communication systems have substantially changed the structures and nature of mathematical optimization problems behind the system design and posed serious challenges in understanding, analyzing, and solving the corresponding optimization problems. For instance, most of the problems become ``non-'' problems, i.e., they are nonconvex, nonsmooth, non-Lipschitz, nonseparable, and nondeterministic, 
and the design variables of the problems range from continuous to integer or even mixed. 
These new features of mathematical optimization problems urgently call for and indeed have driven the development of many new and advanced optimization theory, algorithms, and techniques such as nonconvex nonsmooth optimization, fractional programming (FP), global and integer optimization, distributed optimization, sparse optimization, and learning-based optimization. These form the subject of this paper.

\subsection{Goals of the Paper}
The goals of this paper are as follows:
\begin{itemize}
	\item \emph{Provide an overview of recent advances in mathematical optimization theory and algorithms:} The first goal of this paper is to provide a survey of recent advances in mathematical optimization theory and algorithms for wireless communication system design. In particular, this paper surveys recent advances in nonconvex nonsmooth optimization, global optimization, distributed optimization, and learning-based optimization. The focus is on their theoretical properties as well as successful applications of mathematical optimization techniques in the design of wireless communication systems.
	\item \emph{Guide the choice and development of suitable algorithms for solving structured optimization problems:} The second goal of this paper is to give some guidance on how to choose or to develop suitable algorithms for solving mathematical optimization problems based on their special structures and features. To achieve this goal, the current paper analyzes and highlights the structures and features of the underlying optimization problems and clarify how the associated algorithms utilize these unique problem structures and features. 
	\item \emph{Promote the cross-fertilization of ideas in mathematical optimization and wireless communications:} The final goal of this paper is to promote the cross-fertilization of research agendas in mathematical optimization and wireless communications. On the one hand, advanced optimization tools and techniques enable innovations in understanding, analyzing, and solving optimization problems from wireless communications; on the other hand, novel applications arising from wireless communications have driven and will continue to drive the development of new optimization theory and algorithms. As can be seen from Fig.~\ref{fig:opt_commun}, the evolution of wireless communication systems and the development of  optimization methods are closely interwoven. The cross-fertilization of ideas in mathematical optimization and wireless communications have led and will continue to lead to fruitful outcomes. 
\end{itemize}

\subsection{Structure of the Paper}

The structure of this paper is as follows. We first list some mathematical optimization problems arising from wireless communication system design and discuss their special structures and challenges from the mathematical optimization perspective in Section \ref{sec:problem}. Then we review recent advances in structured nonconvex optimization in Section \ref{sec:nonconvex}, which covers FP, sparse optimization, proximal gradient (PG) algorithms, penalty methods, and duality-based algorithms. Next, we review recent advances in global optimization and distributed optimization in Sections \ref{sec:global} and \ref{sec:distributed}, respectively; we review learning-based optimization with and without the channel state information (CSI) in Sections \ref{sec:learning} and \ref{sec:no:csi}, respectively.  In Section \ref{sec:open}, we give some open research questions and future research directions. Finally, we conclude the paper in Section \ref{sec:conclusion}. 

This survey differs from many others in the wireless communications literature that typically target specific
technologies (e.g., ISAC \cite{liu2022integrated}, reconfigurable intelligent surfaces (RIS) \cite{di2020smart}, non-orthogonal multiple-access (NOMA) \cite{liu2022evolution}, massive connectivity \cite{Liu2018b}, massive random access \cite{ChenX2021}, etc.). Perhaps the most relevant survey papers to this paper are
\cite{luo2003applications,chiang2005geometric,luo2006introduction,palomar2010convex},
which provide the state-of-the-art in convex optimization for communications and signal
processing at the time. However, it has been more than a decade since those works are published,
while significant innovations have taken place in wireless technology (from 3G to 5G and
beyond) as well as in mathematical optimization theory and algorithms. 
The topics of this paper include nonconvex nonsmooth optimization, global optimization, distributed optimization, and learning-based optimization, all of which have not been covered in \cite{luo2003applications,chiang2005geometric,luo2006introduction,palomar2010convex}.    

\emph{Notation.} 
We adopt the following standard notation in this paper. Lower and upper case letters in bold are used for vectors and
matrices, respectively. For any given matrix $\m{A}$, $\m{A}^\hermitian,$ $\m{A}^\transpose,$ and $\m{A}^{-1}$ denote the conjugate transpose, the transpose, and the inverse (if invertible) of $\m{A}$, respectively; 
$\m{A}^{(m, n)}$ denotes the entry on the $m$-th row and the $n$-th column of $\m{A};$ and $\m{A}^{(m_1:m_2, n_1:n_2)}$ denotes a submatrix of $\m{A}$ by taking the rows from $m_1$ to $m_2$ and columns from $n_1$ to $n_2,$ respectively. For any given complex matrix $\m{A}$, we use $\mathrm{Re}(\m{A})$ and $\mathrm{Im}(\m{A})$ to denote its real and imaginary parts, respectively. All of the above usages also apply to vectors and scalars. 
$\lVert \m{x} \rVert_2$ denotes the $\ell_2$-norm of the vector $\m{x}$. In some cases, the index $2$ is omitted. $\m{A} \bullet \m{B} = \tr(\m{A} \m{B})$ is the trace matrix product. 
We use $\cCN(\m{\mu}, \m{Q})$ to denote the complex Gaussian distribution with mean $\m{\mu}$ and covariance $\m{Q}$. Finally, we use $\m{I}$ to denote the identity matrix of an appropriate size, $\m{0}$ to denote the all-zero matrix of an appropriate size, and $\textrm{i}$ to denote the imaginary unit (which satisfies $\textrm{i}^2=-1$).

{
	\begin{figure*}[!t]
		\centering 		
		\includegraphics[width=0.9\textwidth]{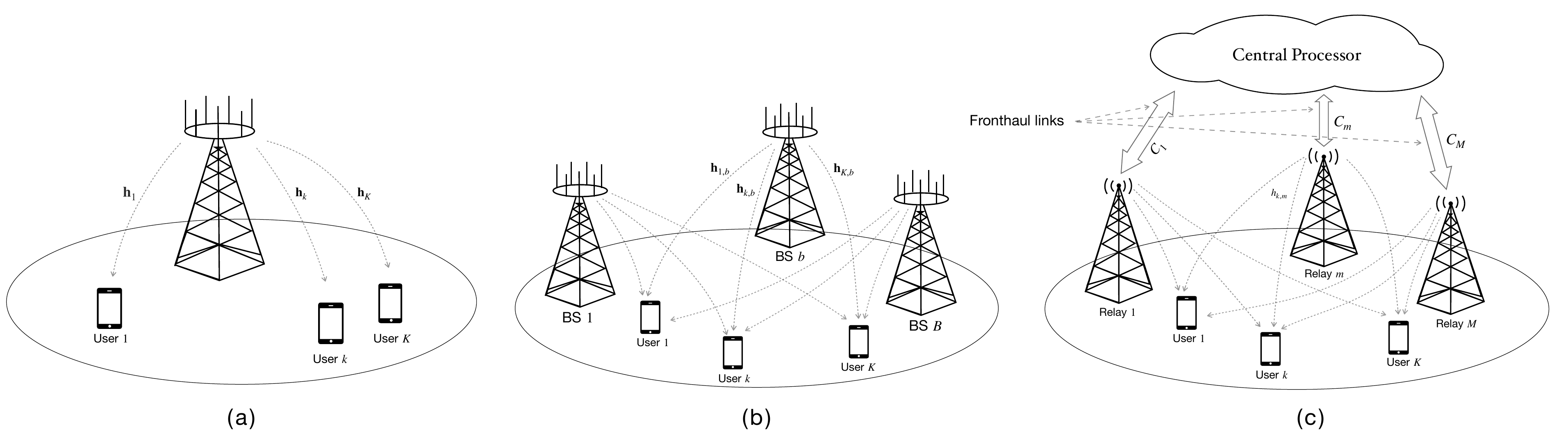}
		\caption{Beamforming in wireless communication systems: (a) Downlink multi-user MIMO system considered in problems \eqref{problem:sumrate}, \eqref{problem:QoS}, 
		and \eqref{problem:jabf}; (b) Cooperative cellular system considered in problem \eqref{problem:clustering}; and (c) Cooperative cellular system with finite-capacity fronthaul links 
			considered in problem \eqref{problem:qoscompression}.}
		\label{fig:commun_system}
\end{figure*}}

\section{Optimization Problems in Wireless Communications: Structures and Challenges}\label{sec:problem}

In this section, we first list some of the mathematical optimization
problems arising from wireless communication system design in various 
use cases in Section \ref{subsection:problem}. Some of these problems
are classic in communication system design but unique challenges 
appear due to the new communication scenarios in 5G
or 6G; some of these problems are new. 
We then summarize the challenging features of these problems from the
optimization perspective in Sections \ref{subsection:challenge} and
\ref{subsection:feature}. Recognizing the specific structures of the optimization 
problems is the first step towards their efficient solution. 

\subsection{Optimization Problems in Wireless Communications}\label{subsection:problem}

Optimization problems can be classified according to the nature of the optimization
variables and the analytic properties of the objective and constraint functions, e.g., 
linear vs. nonlinear, unconstrained vs. constrained, smooth vs. nonsmooth, 
convex vs. nonconvex, stochastic vs. deterministic, integer vs. continuous, etc. 
Below we give important examples of optimization problems commmonly encountered in wireless communication system design according to such classification. 

\subsubsection{Optimization Problems with Continuous Variables}






Beamforming refers to a signal processing technique which combines elements in an antenna array to shape and focus an electromagnetic wave toward certain desired directions/locations and eliminate interferences to the others \cite{elbir2023twenty}. 
Recent advances in beamforming techniques in wireless communications lead to many interesting structured signal processing and optimization problems \cite{gershman2010convex,elbir2023twenty}. 
Beamformer design, which is often coupled with power control, is an example of continuous optimization problems.

\paragraph{Downlink Beamforming} Consider the downlink multi-user MIMO system in Fig.~\ref{fig:commun_system}(a), where the transmitter is equipped with $M$ antennas,\footnote{The ``transmitter'' and ``antenna'' here are abstract concepts, which can be a multi-antenna base station (BS) in the single-cell case, or a virtually cooperative transmitter consisting of many relay-like BSs in C-RAN \cite{simeone2016cloud} and cell-free MIMO \cite{ngo2017cell} cases.} and sends the data to $K$ individual users/receivers each equipped with a single antenna. Let $\cK = \left\{ 1, 2, \ldots, K \right\}$ denote the set of the receivers. The transmitter can direct a beam to each receiver in such a way that its own signal is enhanced and the interference to the other receivers is depressed. Let $\bh_k \in \mathbb{C}^M$
denote the channel vector between the transmitter and the $k$-th receiver, and let $\bv_k\in \mathbb{C}^M$ denote the beamforming vector (also called the precoding vector) used for receiver $k$ by the transmitter. Assume that $s_k\sim \cCN(0, 1)$ is the signal information for user $k$. The transmitted signal is given by $\bx=\sum_{j\in\cK}\bv_js_j$, and the received signal
at the $k$-th receiver is given by 
\begin{equation}
y_k=\bh_k^\dagger \bx+z_k=\bh_k^\dagger\left(\sum_{j\in\cK}\bv_j s_j \right)+z_k,~k\in\cK,
\end{equation}
where $z_k$ is the additive white Gaussian noise (AWGN) with
variance $\sigma_k^2$.
Then, the signal-to-interference-and-noise-ratio (SINR) of the $k$-th receiver can be written as
\begin{equation}\label{eqn:SINR}\displaystyle\textrm{SINR}_k\left(\left\{\bv_k\right\}\right)=\frac{|\bh_k^\dagger \bv_k|^2}{\sum_{j\neq k}|\bh_k^\dagger \bv_j|^2+ \sigma_k^2},~k\in\cK.
\end{equation}

There are two well-studied formulations of the downlink beamforming design problem, which arise from different perspectives. 
From the perspective of the system operator, the downlink beamforming design problem is often formulated as a system utility maximization problem under a total power constraint.
For example, adopting the sum rate of all users as the system utility, the downlink beamforming design problem can be formulated as
\begin{equation}\label{problem:sumrate}
\begin{aligned}
\max_{\left\{\bv_k\right\}}~& \sum_{k\in\cK}\log\left(1+\textrm{SINR}_k(\left\{\bv_k\right\})\right)\\
\text{\normalfont s.t. }~& \sum_{k\in\cK} \|\bv_k\|^2\leq P,
\end{aligned}
\end{equation} where $P$ is the transmitter's power budget.

From a different perspective, the downlink beamforming design problem can also be formulated as a total power minimization problem under the users' quality-of-service (QoS) constraints as follows: 
\begin{equation}\label{problem:QoS}
\begin{aligned}
\min_{\left\{\bv_k\right\}}~& \sum_{k\in\cK} \|\bv_k\|^2 \\
\text{\normalfont s.t. }~& \textrm{SINR}_k(\left\{\bv_k\right\}) \geq \gamma_k,~k\in\cK,
\end{aligned}
\end{equation} where $\gamma_k$ is the SINR target of user $k$. It is worth noting that the optimization formulation (\ref{problem:QoS}) is considerably more tractable from a computational point of view than (\ref{problem:sumrate}), because the former can often be converted into a convex form \cite{luo2006introduction}, while no convex reformulation is known for the latter.

\paragraph{Hybrid Beamforming} 
Massive MIMO, which deploys hundreds or even thousands of antennas at the BS,  
is a key technology for significantly improving the spectrum and energy efficiency of wireless communication systems \cite{5G,massivemimo2,massivemimo3}. 
However, scaling the numbers of radio-frequency (RF) chains and analog-to-digital converters (ADCs)/digital-to-analog converters (DACs) with the number of antennas would result in high hardware complexity and high power consumption. For this reason, instead of using the classical fully-digital beamforming technique, massive MIMO systems are often implemented in a hybrid analog-digital beamforming architecture \cite{zhang_molisch_hybrid,ayach_hybrid,AD2,AD3} in which a large-antenna array is driven by only a limited number of RF chains, call it $N_{\mathrm{RF}}$. In this case, the transmit signal, instead of being the form $\bx=\sum_{j\in\cK}\bv_js_j$, now has the following structure:
\begin{equation}
\bx = \bV_\mathrm{RF} \sum_{j\in\cK}\bv_{j} s_j,
\label{eq:hybrid_beamforming}
\end{equation}
where $\bV_\mathrm{RF}$ is an $M \times N_{\mathrm{RF}}$ analog beamforming matrix, typically implemented using phase shifters, i.e., its entries are complex numbers with unit magnitude, while 
 $\left\{\bv_{j}\right\}$ are digital beamformers of dimension $N_{\mathrm{RF}}$. The joint design of analog beamformer $\bV_{\mathrm{RF}}$ and digital beamformers $\left\{\bv_{j}\right\}$ poses a unique challenge in optimization.

\paragraph{RIS Beamforming} 

 An RIS is a metasurface, consisting of many small reconfigurable passive low-cost reflecting elements that can easily introduce a controlled individual phase shift to the impinging electromagnetic wave \cite{di2020smart}. These RIS elements can jointly provide passive beamforming that can effectively enhance the propagation condition over wireless channels 
by directing electromagnetic radiation toward the intended direction. 
Structurally, the optimization of RIS phase shifts to maximize the signal-to-noise-ratio (SNR) has a similar form as the optimization of hybrid beamformers. 

The overall downlink channel model for an RIS-assisted communication scenario has the following form. 
 Let $\mathbf h_k^{\rm d}\in\mathbb{C}^{M}$ denote the direct channel from the BS to {user $k$}, and $\mathbf h_k^{\rm r}\in\mathbb{C}^{N_{\mathrm{RIS}}}$ denote the channel from the RIS to user $k$, and $\mathbf G\in\mathbb{C}^{N_{\mathrm{RIS}} \times M}$ denote the channel from the BS to the RIS. 
Let the RIS reflection coefficients be denoted by $\mathbf \Omega =[e^{\textrm{i}\omega_1},e^{\textrm{i}\omega_2},\ldots,e^{\textrm{i}\omega_{N_{\mathrm{RIS}}}}]^\transpose$, where $\omega_i\in(-\pi,\pi]$ is the phase shift of the $i$-th element. Then, the received signal at user $k$ can be represented as
\begin{align}
    y_k &= (\mathbf h_k^{\rm d}+ \mathbf G^\transpose\diagg(\mathbf \Omega)\mathbf h_k^{\rm r})^\dagger
	\sum_{j\in \cK} \mathbf v_j s_j + n_k. 
\label{eq:RIS_beamforming}
\end{align} 
It would be of interest to jointly optimize the unit-modulus RIS phase shifters matrix $\mathbf{\Omega}$ and the beamformers $\left\{\bv_j\right\}$.

\paragraph{Joint BS Clustering and Beamformer Design} Beamforming can also be performed across multiple BSs. Consider a cooperative cellular network consisting of a (large) set of densely deployed BSs (e.g., macro/micro/pico BSs), denoted by $\mathcal{B}=\{1,2,\dots, B\}$, that provide services to a set of users, denoted by $\mathcal{K}=\{1,2,\dots, K\},$ as depicted in Fig.~\ref{fig:commun_system}(b). Assume that each BS is equipped with $M$ antennas and each user is equipped with a single antenna. Let $\h_{k,b}\in\C^M$ be the channel between BS $b$ and user $k$, and let $\h_{k}=[\h_{k,1}^\mathsf{T},\h_{k,2}^\mathsf{T},\dots,\h_{k,B}^\mathsf{T}]^\mathsf{T}\in\C^{MB}$ be the channel between all the BSs and user $k$. In addition, let $\bv_{k,b}\in\C^M$ be the beamforming vector of BS $b$ for user $k$, and let $\bv_k=[\bv_{k,1}^\mathsf{T},\bv_{k,2}^\mathsf{T},\dots,\bv_{k,B}^\mathsf{T}]^\mathsf{T}\in\C^{MB}$. If all the $B$ BSs are allowed to share data and fully cooperate with each other, then they can be treated as a virtual BS with $MB$ antennas. In this case, the network reduces to the downlink multi-user MIMO channel in Fig.~\ref{fig:commun_system}(a). 
In practice, full cooperation among all the BSs is impractical, as it would result in a large signaling overhead. A popular strategy to reduce the overhead of the above network is user-centric BS clustering \cite{zhang2009networked,ng2010linear}, i.e., each user is served by only a small number of BSs. 

With the above setup, the SINR of user $k$ can be expressed as \eqref{eqn:SINR}. If we wish to pursue a sparse solution in which each user is served by a small cluster of BSs, we can consider an optimization problem similar to \eqref{problem:sumrate}, but with an additional mixed $\ell_2/\ell_1$ regularization term to induce a group-sparse structure in each $\bv_k$. 
Specifically, the problem is formulated in \cite{Hong2013} as
\begin{equation}\label{problem:clustering}
\begin{aligned}
\max_{\{\bv_k\}}~&\sum_{k\in\cK}\left(\log(1+\text{SINR}_k\left(\{\bv_k\}\right))-\rho\sum_{b\in\cB}\|\bv_{k,b}\|_2\right)\\
\text{s.t. }~&\sum_{k\in\cK}\|\bv_{k,b}\|_2^2\leq P_b,~b\in\cB,
\end{aligned}
\end{equation} where $P_b$ is the power budget of BS $b$ and $\rho$ is the parameter that controls the group sparsity of the vectors $\left\{\bv_{k,b}\right\},$ i.e., the coordination overhead between different BSs. In particular, if $\bv_{k,b}=\bm{0},$ then BS $b$ does not cooperate in serving user $k.$ We want to point out that the regularizer $\sum_{b\in\cB}\|\bv_{k,b}\|_2$ in problem \eqref{problem:clustering} is nonsmooth.

\paragraph{Joint Downlink Beamforming and Fronthaul Compression} Consider now a more practical cooperative cellular network (e.g., C-RAN) consisting of one central processor (CP) and $M$ single-antenna relay-like BSs (called relays for short in the rest of the paper), 
which cooperatively serve $K$ single-antenna users, as shown in Fig.~\ref{fig:commun_system}(c). In such a network, the users and the relays are connected by noisy wireless channels, and the relays and the CP are connected by noiseless fronthaul links of finite capacities. Let $\cM = \left\{ 1, 2, \ldots, M \right\}$ denote the set of relays (i.e., antennas). 

The compression model from the CP to the relays plays a central role in formulating the joint beamforming and compression problem. 
The ideal beamformed signal at the CP is $\sum_{k\in\cK} \v{v}_k s_k,$ where $\v{v}_k = [v_{k,1}, v_{k,2}, \d , v_{k,M}]^\transpose\in \mathbb{C}^M$ is the beamforming vector for user $k$. However, the transmitted signal from the CP to the relays needs to be first compressed (through quantization) due to the limited capacities of the fronthaul links. Let the compression error be modeled as $\v{e} = [e_1, e_2, \d, e_M]^\transpose \sim \cCN(\m{0}, \m{Q}),$ where $e_m$ denotes the quantization noise for compressing the signal to relay $m$, and $\m{Q}$ is the covariance matrix of the quantization noise. Then, the transmitted signal (by treating all relays as a virtual transmitter) is 
\begin{equation}
\v{x} = \sum_{j\in\cK} \v{v}_j s_j + \v{e} 
\label{equ:model_R}
\end{equation} 
and the received signal at user $k$ is 
\begin{equation}
y_k = \v{h}_k^\hermitian \left( \sum_{j\in\cK} \v{v}_j s_j \right) + \v{h}_k^\hermitian \v{e} + z_k,~k\in\cK. 
\label{equ:DBM}
\end{equation} 
In this case, 
the SINR of user $k$ is
\begin{equation*}
\textrm{SINR}_k\left(\left\{\bv_k\right\},\m{Q} \right)= \frac{|\v{h}_k^\hermitian \v{v}_k|^2}{\sum_{j\neq k} |\v{h}_k^\hermitian \v{v}_j|^2 + \v{h}_k^\hermitian \m{Q} \v{h}_k + \sigma_k^2},~k\in\cK, 
\end{equation*} and the compression rate of relay $m\in\cM$ can be expressed below, if we adopt the information-theoretically optimal multivariate compression strategy (with the compression order from relay $ M $ to relay $ 1 $) \cite{park2013JointPrecodingMultivariate}: 
\begin{equation*}
\begin{aligned}
C_m\left(\left\{\bv_k\right\},\m{Q}\right)= \log_2 \left( \frac{\sum_{k\in\cK} |v_{k,m}|^2 + \m{Q}^{(m,m)}}{ \m{Q}^{(m:M, m:M)}/\m{Q}^{(m+1:M, m+1:M)} } \right),
\end{aligned} 
\end{equation*} 
where $\m{Q}^{(m:M, m:M)}/\m{Q}^{(m+1:M, m+1:M)}\triangleq\m{Q}^{(m,m)} - \m{Q}^{(m,m+1:M)} (\m{Q}^{(m+1:M, m+1:M)})^{-1} \m{Q}^{(m+1:M, m)}.$
The QoS-constrained joint beamforming and compression design problem can then be formulated as
\begin{equation}\label{problem:qoscompression}
\begin{aligned}
\min_{\{\v{v}_k\}, \m{Q}\succeq \m{0}}~& \sum_{k\in\cK} \|\v{v}_k\|^2 + \tr(\m Q)\\
\st~~~~~ & \textrm{SINR}_k\left(\left\{\bv_k\right\},\m{Q} \right) \geq \gamma_k, \fk{},\\
& C_m\left(\left\{\bv_k\right\},\m{Q}\right) \leq {C}_m, \fm{}, 
\end{aligned}
\end{equation} where $C_m$ is the fronthaul capacity of relay $m.$


\subsubsection{Optimization Problems with Integer Variables}

When optimization is performed at the level of constellation symbols, which comes from 
a discrete set, it gives rise to a discrete optimization problem. In the following, we describe two 
optimization problems with integer/discrete variables in the context of massive MIMO.

\paragraph{MIMO Detection} MIMO detection is an example of a discrete optimization problem in digital communications. Although MIMO detection has been extensively studied for more than two decades, it has gained renewed interest in the context of massive MIMO \cite{So2017,RN113}. 
Consider a (massive) MIMO channel model in the uplink:
\begin{equation}\label{rHv}\by=\bH\bs+\v{z},\end{equation} 
where $\by \in\mathbb{C}^{M}$ is the vector of received signals, $\bH\in\mathbb{C}^{M\times K}$ is 
an $M\times K$ complex channel matrix (for $K$ inputs and $M$ outputs with $M\geq K$), $\bs\in\mathcal{S}^{K}$ is the vector of transmitted
symbols by the user terminals, and $\v{z}\in\mathbb{C}^{M}$ is an AWGN with zero mean. 
We consider the cases where $\mathcal{S}$ is the $(4u^2)$-QAM constellation 
\begin{equation}\label{QAM}\mathcal{Q}_u=\left\{z\in \mathbb{C} \mid \mathrm{Re}(z),\mathrm{Im}(z)=\pm1,\pm3,\ldots,\pm{\left(2u-1\right)}\right\}
\end{equation} or the $L$-PSK constellation $\mathcal{S}_L$
\begin{equation}\label{MPSK}
\mathcal{S}_L = \left\{\exp({2\pi\textrm{i}(\ell-1)}/{L})\mid \ell=1,2,\ldots,L\right\}.\end{equation}
The MIMO detection problem is to recover the vector of transmitted symbols $\bs$ from the vector of received signals $\by$ based on the knowledge of the channel matrix $\bH$.
The standard mathematical formulation of the MIMO detection problem is 
\begin{equation}\label{problem:mimodec}\begin{aligned}
\displaystyle \min_{\bx \in \mathbb{C}^{K}} ~& \left\|\bH\bx-\by\right\|_2^2 \\
\mbox{s.t.~} ~& {\bx}\in \mathcal{S}^K.
\end{aligned}\end{equation} 
One of the new challenges of solving the above problem in the massive MIMO context is the large problem size, which prevents the use of many algorithms (e.g., semidefinite programming relaxation (SDR)-based algorithms \cite{ma2004semidefinite,jalden2006detection}) that may be efficient when the problem size is small to medium. 

\paragraph{Symbol-Level Precoding} In the downlink, when the CSI is available at the transmitter, it is also possible to formulate an optimization problem of designing the transmitted signal, so that the desired received signal aligns with the constellation point. Consider a downlink scenario and let $\bs \in \mathbb{C}^K$ be a set of desired constellation points (corresponding to multiple users), the BS may try to construct the downlink transmit signal $\bx$ so that after going through the channel $\bH,$ the received signal would align with the desired $\bs$ as closely as possible. This technique is called symbol-level precoding\cite{CItutorial}, because a different $\bx$ is designed for each symbol $\bs$---in contrast to the beamforming technique where the same beamformer is used for the entire channel coherence time. This formulation is appealing in the massive MIMO context. When the BS is equipped with many antennas, it is possible to restrict the transmit signal to be discrete, e.g., $\mathcal{X} = \{\pm 1\pm\textrm{i}\}$, which simplifies implementation, and still provides excellent performance. 
In this case, the problem formulation becomes \cite{SQUID}
\begin{equation}\label{problem:symbol_level_precoding}
\begin{aligned}
\displaystyle \min_{\bx \in \mathbb{C}^{M}} ~& \left\|\bH\bx-\bs\right\|_2^2 \\
\mbox{s.t.~} ~& {\bx}\in \mathcal{X}^M,
\end{aligned}\end{equation} 
which is a discrete optimization problem. We mention here that it is possible to consider also the joint optimization of constellation range \cite{sohrabi_liu_yu_one-bit} in this problem formulation.

\subsubsection{Optimization Problems with Mixed Variables}
In power control and beamforming design problems, when admission control, user scheduling, and/or BS-user association are involved, the corresponding optimization problems would have mixed variables, i.e., both continuous and integer (in particular, binary on-and-off) variables. Let us consider two examples.

\paragraph{Joint Admission Control and Multicast Beamforming}
Consider the same downlink multi-user MIMO system in Fig.~\ref{fig:commun_system}(a) as in problems \eqref{problem:sumrate} and \eqref{problem:QoS}. Different from problems \eqref{problem:sumrate} and \eqref{problem:QoS}, here we assume that the intended information from the transmitter to all of the $K$ users is the same, i.e., the transmitter multicasts the common information to $K$ users simultaneously \cite{Sidiropoulos2006}. Let $\bw$ be the multicast beamforming vector used by the transmitter. In this case, the SNR of the $k$-th user is given by $$\textrm{SNR}_k(\bw)=\frac{|\bh_k^\dagger \bw|^2}{ \sigma_k^2}.$$ If user $k$ is admitted to be served by the transmitter, then its QoS constraint $\textrm{SNR}_k(\bw)\geq \gamma_k$ should be satisfied, where $\gamma_k$ is the given SNR threshold of user $k.$ When the transmitter cannot simultaneously support all users (because, e.g., the number of users is too large), admission control \cite{zander1992performance,zander1992distributed} is needed to select a subset of users to serve at their SNR targets. 

One possible problem formulation for admission control is to maximize the number of admitted users under the power budget constraint \cite{MitliagkasSS11,liu13tspjpac}, and another possible formulation is to select a subset of users with a given cardinality to minimize the total transmit power \cite{xu2014semidefinite,ni2018mixed}. 
In particular, given $1\leq \hat K\leq K,$ the joint admission control and multicast beamforming (JABF) design problem of selecting a subset of $\hat K$ users with the minimum total transmit power is formulated in \cite{ni2018mixed} as
\begin{equation}\label{problem:jabf}\begin{aligned}
\displaystyle \min_{\bw,\,\bm{\beta}} ~& \|\bw\|^2 \\
\mbox{s.t.~} ~& \displaystyle {|\tilde{\bh}_k^{\dagger}\bw|^2} \geq \beta_k,~k\in\cK,\\
~&~\sum_{k\in\cK} \beta_k \geq \hat K,~\beta_k\in\left\{0,1\right\},~k\in\cK,
\end{aligned}\end{equation}
where 
$\bm{\beta}=[\beta_1,\beta_2,\ldots,\beta_K]^\T$ is a binary vector with $\beta_k$ modeling whether user $k$ is selected, $\bw$ is a continuous multicast beamforming vector, and $\tilde{\bh}_k$ is redefined as $\bh_k/(\sigma_k\sqrt{\gamma_k})$ for ease of notation.
This is a mixed continuous/discrete optimization problem.

\paragraph{Joint Uplink Scheduling and Power Control} Consider the uplink of a wireless cellular network, where a single-antenna BS is associated with several single-antenna users in each cell, and the users are scheduled for uplink transmission within each cell. Let $\cB$ denote the set of cells/BSs in the network, $\cK_i$ denote the set of users who are associated with BS $i$,  $\kappa_i\in\cK_i$ denote the user to be scheduled for transmission at cell $i$, and $p_k$ denote the transmit power of the scheduled user $k$. Assume that $s_{\kappa_i}\sim\mathcal{CN}(0,1)$ is the transmitted signal  of user $\kappa_i$.  Then, the received signal at BS $i$ is
	$$y_i=\sum_{j\in\cB}{p_{\kappa_j}h_{i,\kappa_j}s_{\kappa_j}}+z_i,$$
	where $h_{i,\kappa_j}$ is the uplink channel coefficient from user $\kappa_j$ to BS $i$ and $z_i\sim\mathcal{CN}(0,\sigma_i^2)$ is the AWGN. Given a set of weights $\left\{w_k\right\}$ that reflects the user priorities and adopting the weighted sum rate as the system performance metric, the joint uplink scheduling and power control problem is formulated in \cite{FP2} as
	\begin{equation}\label{problem:scheduling}
	\begin{aligned}
	\max_{\bka,\, \bp}~&\sum_{i\in\cB}w_{\kappa_i}\log\left(1+\frac{|h_{i,\kappa_i}|^2p_{\kappa_i}}{\sum_{j\neq i}|h_{i,\kappa_j}|^2p_{\kappa_j}+\sigma_i^2}\right)\\
	\text{s.t.~}~&0\leq p_k\leq P_k,~k\in\cup_{i\in\cB}~\cK_i,\\
	&\kappa_i\in\cK_i\cup\{\emptyset\},~i\in\cB,
	\end{aligned}
	\end{equation}
	where $\kappa_i=\emptyset$ means that no user in cell $i$ is scheduled for transmission. The discrete scheduling variables $\{\kappa_i\}$ and the continuous power control variables $\{p_k\}$ are coupled in problem \eqref{problem:scheduling}, making it a mixed continuous/discrete optimization problem. 
	
	\subsubsection{Remarks}
	In the above, we have only listed a few of wireless communication scenarios that give rise to interesting optimization formulations. As device technology, network architecture, deployment use cases, and new application scenarios for 5G and 6G continue to evolve, novel optimization formulations will continue to emerge. For example, the incorporation of RIS in the wireless environment not only makes the optimization problem high-dimensional but also poses challenges in channel modeling and estimation, which motivate learning-based optimization without explicit channel estimation.	As another example, mMTC service with sporadic device activities gives rise to sparse optimization problems. For eMBB services, large-scale C-RAN or cell-free networks give rise to novel formulations of the joint optimization of fronthaul compression and data transmission. Moreover, the incorporation of artificial intelligence (AI) into wireless networks (e.g., federated learning) gives rise to interesting distributed optimization problem settings. Finally, deep learning may provide an alternative path to the traditional optimization paradigm. These novel problem settings and their associated solution techniques will be the main focus of the rest of this paper.

\subsection{Challenges from the Optimization Perspective}\label{subsection:challenge}
The innovations in the wireless communication system architecture (from 3G to 5G and beyond) in the last two decades have substantially changed the structures and the nature of optimization problems arising from system design. This makes these problems more challenging to analyze and to solve. We briefly summarize these challenges below.
\begin{itemize}
	\item \emph{Large dimensionality and high nonlinearity.} The dimension of optimization problems becomes much larger, and the objective and constraint functions become highly nonlinear. The larger dimensionality comes from the larger number of system parameters and variables, such as the large number of antennas deployed at the BS in the massive MIMO system and the large number of users \cite{bjornson2017massive}, the large number of devices in the massive machine-type communication network with a potentially sparsity structure\cite{Liu2018b}, the large number of subcarriers, and the large number of passive reflection elements in the RIS-aided communication system \cite{wu2019towards,di2020smart}.~
	Nonlinearity may be caused by the coupling of design variables and complicated expressions of the objective function and constraints with respect to the variables. For instance, reflective beamforming vectors and transmit beamforming vectors are multiplicatively coupled and further composed with fractional and logarithmic functions in RIS-aided communication systems \cite{wu2019towards,di2020smart}; the performance metric of target estimation (e.g., the Cram\'er-Rao bound of radar sensing \cite{liu2022CramerRaoBoundOptimization}) is usually a highly nonlinear function of the design variables in ISAC systems\cite{liu2022integrated}.
	
	\item \emph{Lack of favorable properties.} Many of the aforementioned optimization problems become ``non-'' problems, i.e., they are either nonconvex, nonsmooth, non-Lipschitz, nonseparable, or nondeterministic. There are two reasons for the frequent occurrences of these new types of optimization problems. First, some nonconvex nonsmooth non-Lipschitz regularization terms are often needed in optimization problems to promote certain desired structures in their solutions (e.g., sparsity, low-rankness, and fairness) \cite{zhang2009networked,ng2010linear,Hong2013}, especially in cooperative communication networks (see, e.g., problem \eqref{problem:clustering}). Second, nonconvex and nonsmooth terms are helpful in transforming certain structured optimization problems with discrete variables into ``easy'' (globally/locally) equivalent problems with continuous variables, which facilitates algorithm design. 
		The lack of favorable properties (see, e.g., \cite{li2020understanding} for a discussion) necessitates a judicious treatment of both the theoretical and algorithmic aspects of optimization.
	
	\item \emph{Mixed-integer variables.} In various system design scenarios, both continuous and discrete variables can appear in the associated optimization problems. Some examples include 
		the admission control and user scheduling problems \eqref{problem:jabf} and \eqref{problem:scheduling}. The integer variables often make the optimization problems significantly more difficult to solve than their continuous counterparts. For problems with only integer variables, the ``brute force'' enumeration (of all feasible points) is guaranteed to find a global solution. However, this approach is not feasible for solving large-scale problems, as its complexity grows exponentially with the number of variables. Therefore, special attention and advanced optimization theory and techniques are needed to tackle large-scale problems with (mixed) integer variables.
\end{itemize}
\subsection{Structural Properties of Optimization Problems}\label{subsection:feature}

The unique difficulties and structural features of optimization problems arising from wireless system design have driven the development of many new and advanced optimization theory and algorithms. 
The basic features include analytic properties of the functions in its objective and constraints (e.g., convexity, smoothness, and monotonicity), type of its design variables (e.g., continuous, integer, or both), and the degree of coupling of its design variables (e.g., the variables are fully separable or coupled in a structured manner). More advanced features include (but are not limited to) hidden convexity\footnote{A nonconvex optimization problem is said to have hidden convexity if it admits an equivalent convex reformulation \cite{xia2020survey}.} (of a seemingly nonconvex problem) and zero duality gap, computational complexity status, easy projection property (onto its feasible set), tight global bounds (of its objective function), and ``simple'' structured conditions that its solution(s) should satisfy. 

Recognizing the special structures of optimization problems is of paramount importance, as it allows us to select suitable tools for analyzing them and algorithms for tackling them. In the remaining part of this subsection, we use some problems listed in Section \ref{subsection:problem} to elucidate the above discussion.

Consider the (massive) MIMO detection problem \eqref{problem:mimodec}. Although it has integer (discrete) variables, different variables are fully decoupled in the constraint and thus the feasible set enjoys an easy-projection property. Moreover, the objective function of problem \eqref{problem:mimodec} is quadratic and hence has a Lipschitz-continuous gradient. These features suggest that problem \eqref{problem:mimodec} is amenable to the gradient projection (GP) algorithm. 
We introduce GP and the more general PG algorithms in Section \ref{subsection:proximal}. 

As another example, consider the RIS and hybrid beamforming problems (\ref{eq:hybrid_beamforming}) and (\ref{eq:RIS_beamforming}). To account for the phase-only constraint (which is quadratic), it is possible to use SDR \cite{Wu_Zhang_IRS,hailiang_twc_2021}, but its complexity is not scalable. Alternatively, GP can be used.
A straightforward GP optimization for the analog beamformer would involve taking a gradient step and then projecting the result onto the unit-modulus domain by retaining only its phases. 
However, a better approach {is} to recognize that the unit-modulus constraints form a Riemannian manifold \cite{yu_shen_alternating}, so instead of taking a Euclidean gradient followed by projection, a faster algorithm can be devised by first projecting the gradient vector onto the tangent space of the complex circle manifold. To further speed up convergence, the conjugate gradient version of this idea may be used. This gives the so-called Riemannian conjugate gradient method; see \cite{yu_shen_alternating} and a specific application of this algorithm in \cite{jiang_nulling}.
This is an example of how taking advantage of the problem structure can enable faster convergence of the algorithm.


Hidden convexity is an important feature to recognize (if it exists). 
Consider the beamforming design problems \eqref{problem:QoS} and \eqref{problem:qoscompression}. 
While the objective functions are convex (and very simple) and design variables are continuous, the constraints are complicated and the number of constraints (related to the numbers of users and relays in the considered system) is large. By examining the constraints in problems \eqref{problem:QoS} and \eqref{problem:qoscompression} more carefully, both of them turn out to admit convex reformulations (assuming that the constraints are feasible), and we can show that the duality gap between the primal and dual problems is zero. These features suggest that duality-based algorithms are suitable for solving problems \eqref{problem:QoS} and \eqref{problem:qoscompression}.  We introduce duality-based algorithms in Section \ref{subsection:duality}.

Before leaving this subsection, let us comment on the computational complexity of optimization problems that arise in wireless communication system design. Determining the complexity class of an optimization problem (e.g., (strongly) NP-hard or polynomial-time solvable) provides valuable information about what lines of approaches are more promising. Once a problem is shown to be ``hard'', the search for an efficient exact algorithm should often be accorded lower priority. Instead, less ambitious goals, such as looking for algorithms that can solve various special cases of the general problem efficiently; looking for algorithms that, though not guaranteed to have a polynomial-time complexity, run quickly most of the time; or relaxing the problem and looking for an algorithm that can find an approximate solution efficiently, should be considered. Compared with convexity and nonconvexity, which can provide useful intuition on the easiness/hardness of an optimization problem, computational complexity theory is a more robust and reliable tool for characterizing the tractability/intractability of an optimization problem. Back to the optimization problems discussed in this section, although problems \eqref{problem:QoS} and \eqref{problem:qoscompression} and some special cases of other problems admit simple closed-form solutions or are polynomial-time solvable \cite{Hanly99powercontrol,hanly95,Foschini93,zander1992performance,zander1992distributed,liu2013maxsimo,fan2022EfficientlyGloballySolving,wiesel2006LinearPrecodingConic,
	schubert2005IterativeMultiuserUplink, 	
	yu2007TransmitterOptimizationMultiantenna,  
	dahrouj2010CoordinatedBeamformingMulticell}, a small variant of these problems (e.g., (\ref{problem:sumrate})) can be (strongly) NP-hard \cite{luo_complexity,liu2017dynamic,liu2014complexity,liu11tspbeamforming,liu2013max,Sidiropoulos2006,computational_complexity,luo2007approximation,journal_zheyu,NET-036,hong12survey}, which means that there does not exist a (pseudo-) polynomial-time algorithm that can solve the corresponding problem to global optimality unless $\mathcal{P}=\mathcal{NP}.$     
Understanding this complexity analysis is essential for algorithm design.

\section{Structured Nonconvex Optimization} \label{sec:nonconvex}
Although the optimization problems presented in the previous section are nonconvex in general, in this section, we discuss how their special structures can be exploited to design tailored algorithms that can find high-quality locally optimal or suboptimal solutions of those problems in an efficient manner.
We should point out that these algorithms do not involve global optimization techniques and are generally not guaranteed to find a globally optimal solution. We survey advanced global optimization algorithms and techniques in Section \ref{sec:global}.

This section is organized as follows. We first review two useful transformations for tackling FPs in Section \ref{subsection:fractional}. These transformations can then be used to efficiently solve the sum-rate maximization problem \eqref{problem:sumrate} and the corresponding scheduling problem \eqref{problem:scheduling}. Second, we review sparse optimization theory and techniques in Section \ref{subsec:sparse}. Sparse optimization is useful for solving and analyzing optimization problems whose solutions admit a sparse structure, e.g., sparse channel estimation and sparse device activity detection problems. Third, we review the PG algorithm in Section \ref{subsection:proximal}, which is suitable for solving (not necessarily convex) optimization problems with a ``simple'' nonsmooth term in their objectives or a ``simple'' constraint. These include the MIMO detection problem \eqref{problem:mimodec} and the BS clustering and beamforming design problem \eqref{problem:clustering}. Fourth, we review the penalty method in Section \ref{subsection:penalty}. Such a method is generally suitable for tackling optimization problems in which the constraint can be decomposed into a simple convex constraint plus a simple penalty function. We demonstrate how the penalty method can be used to tackle 
the MIMO detection problem \eqref{problem:mimodec}. Finally, we review the (Lagrangian) duality-based algorithm in Section \ref{subsection:duality}, which can be used to solve (hidden) convex problems with many complicated constraints, such as the QoS-constrained joint beamforming and compression problem \eqref{problem:qoscompression}. 

\subsection{Fractional Programming} \label{subsection:fractional}
FP refers to a specific class of optimization problems that involve ratio terms. It plays a vital role in the design and optimization of wireless communication systems due to the ubiquitous fractional structure of various performance metrics related to communication links. Notably, the SINR (e.g., in \eqref{eqn:SINR}), which is naturally defined by a fractional function, is an essential quantity for the performance evaluation of wireless communication systems. In addition, energy efficiency (EE), defined as the ratio between the amount of transmitted data and consumed energy, is an important performance metric in the design of wireless communication systems \cite{Zappone,jiaqitcom2024}.

{Early works on FP mainly focus on single-ratio problems, particularly concave-convex single-ratio maximization problems, where the objective function contains a single ratio term with a nonnegative concave numerator and a positive convex denominator. To deal with single-ratio FP problems, two classic techniques are the Charnes-Copper transform and the Dinkelbach's transform. Both methods ensure the convergence to the global optimum of concave-convex single-ratio FP problems and have been extensively applied to solve EE maximization problems for wireless communication systems \cite{Zappone}.
	Though working well for single-ratio FP, 
	the aforementioned techniques cannot be easily generalized to multiple-ratio cases, which are more prevalent in system-level communication network design (as the overall system performance typically involves multiple ratio terms). A prominent recent advance in FP is \cite{FP1,FP2}, where new transforms for solving multiple-ratio FP problems are developed. In this subsection, we review the transforms and methods proposed in \cite{FP1,FP2} and their applications in solving two important problems arising from wireless communication system design. 
}
\subsubsection{Two FP Transforms}
We now review the two FP transforms proposed in \cite{FP1,FP2}.  

\paragraph{Quadratic Transform} The first transform is designed for  the sum-of-ratio FP problem
	\begin{equation}\label{prob:multiratio}
	\begin{aligned}
	\max_{\x\in\mathcal{X}}~\sum_{i=1}^I\frac{A_i(\x)}{B_i(\x)},
	\end{aligned}
	\end{equation}
	where $A_i(\cdot)\geq 0$ and $B_i(\cdot)>0$ on $\mathcal{X}$ for all $i\in\{1,2,\dots,I\}$. {The quadratic transform} \cite{FP1} of the multi-ratio FP problem \eqref{prob:multiratio} is defined as 
	\begin{equation}\label{quadtransform}
	\begin{aligned}
	\max_{\x\in\mathcal{X},\y\in\mathbb{R}}~\sum_{i=1}^I \left(2y_i\sqrt{A_i(\x)}-y_i^2B_i(\x)\right).
	\end{aligned}
	\end{equation}
	It has been shown in \cite{FP1} that the problem \eqref{quadtransform} is equivalent to the sum-of-ratio FP problem \eqref{prob:multiratio}, which can be easily seen by substituting the optimal solution \begin{equation}\label{quady}y_i^\ast=\frac{\sqrt{A_i(\x)}}{B_i(\x)},~i=1,2,\ldots,I\end{equation} into the objective function of problem \eqref{quadtransform}. The quadratic transform decouples the numerator and the denominator of each ratio term and is of particular interest when the transformed problem \eqref{quadtransform} is convex in $\x$ for a given $\y$ (e.g., each $A_i(\cdot)$ is concave and so is $\sqrt{A_i(\cdot)}$, each $B_i(\cdot)$ is convex, and $\mathcal{X}$ is convex), in which case alternating optimization (AO) over $\bx$ and $\by$ can be efficiently performed and is guaranteed to converge to a stationary point of problems  \eqref{prob:multiratio} and \eqref{quadtransform}. It is worth noting that the quadratic transform can be extended to tackle more general sum-of-functions-of-ratio problems (where the functions are required to be nondecreasing) \cite[Corollary 2]{FP1} and to the matrix case \cite[Theorem 1]{FP:MM}. 

\paragraph{Lagrangian Dual Transform} {This second transform is tailored for the sum-rate maximization problem. 
	Specifically, consider the general sum-of-logarithm maximization problem
	\begin{equation}\label{prob:sumlog}
	\begin{aligned}
	\max_{\x\in\mathcal{X}}~\sum_{i=1}^I\log\left(1+\frac{A_i(\bx)}{B_i(\bx)}\right),
	\end{aligned}
	\end{equation}
	where $A_i(\cdot)\geq 0,$  $B_i(\cdot)>0$ on $\mathcal{X}$, and ${A_i(\cdot)}/{B_i(\cdot)}$ can be physically interpreted as an SINR term (which includes the SINRs in problems \eqref{problem:sumrate} and \eqref{problem:scheduling} as special cases). The {Lagrangian dual transform} of problem \eqref{prob:sumlog}  is defined as \cite{FP2}
	\begin{equation}\label{lagtransform}
	\begin{aligned}
	\max_{\x\in\mathcal{X},\bgam}~\sum_{i=1}^I\left(\log\left(1+\gamma_i\right)-\gamma_i\right)+\sum_{i=1}^I\frac{(1+\gamma_i)A_i(\x)}{A_i(\x)+B_i(\x)},
	\end{aligned}
	\end{equation}where $\bgam=[\gamma_1,\gamma_2,\ldots,\gamma_I]^{\T}\in\mathbb{R}^I.$ The Lagrangian dual transform in \eqref{lagtransform} is equivalent to problem \eqref{prob:sumlog}, which can be seen by substituting the optimal solution \begin{equation}\label{laggamma}\gamma_i^\ast=\frac{{A_i(\x)}}{B_i(\x)},~i=1,2,\ldots,I\end{equation} into the objective function of problem \eqref{lagtransform}. Compared with problem \eqref{prob:sumlog}, its Lagrangian dual transform \eqref{lagtransform} has the advantage of moving the SINRs outside of the logarithmic functions, which allows for a subsequent quadratic transform.   }

\subsubsection{Application Examples}  Now, let us apply the quadratic transform and the Lagrangian dual transform to solve two important problems in wireless communications.
	
\paragraph{Downlink Beamforming for Sum-Rate Maximization} Consider first the sum-rate downlink beamforming design problem \eqref{problem:sumrate}. {An efficient FP-based approach for solving the sum-rate maximization problem \eqref{problem:sumrate} is to first reformulate the sum-of-logarithm form into a sum-of-ratio form using the Lagrangian dual transform and then apply the quadratic transform to the latter.  More specifically, by applying the Lagrangian dual transform to problem \eqref{problem:sumrate}, we obtain 
	\begin{align}
	\max_{\left\{\bv_k\right\},\bgam}~& \sum_{k\in\cK}\left(\log(1+\gamma_k)-\gamma_k\right)+\sum_{k\in\cK}\frac{(1+\gamma_k)|\bh_k^\dagger\bv_k|^2}{\sum_{j\in\cK}|\bh_k^\dagger\bv_j|^2+\sigma_k^2} \nonumber \\
	\text{\normalfont s.t. ~}~& \sum_{k\in\cK} \|\bv_k\|^2\leq P.
	\label{problem:sumrate2}
	\end{align}
	When $\{\bv_k\}$ is fixed, the optimal $\bgam$ of problem \eqref{problem:sumrate2} has a closed-form solution, which takes the form in \eqref{laggamma}.
%
	To update $\{\bv_k\}$ for a fixed $\bgam$, the quadratic transform can be applied to the sum-of-ratio term in \eqref{problem:sumrate2}.  In particular, by treating $(1+\gamma_k)|\bh_k^\dagger\bv_k|^2$ as the numerator and $\sum_{j\in\cK}|\bh_k^\dagger\bv_j|^2+\sigma_k^2$ as the denominator and applying the quadratic transform, we obtain the problem
	\begin{align}
	\max_{\{\bv_k\},\by} \quad & \sum_{k\in\cK}\Bigg(2\sqrt{1+\gamma_k}\mathrm{Re}(y_k^\dagger \bh_k^\dagger \bv_k)\Bigg. \nonumber \\ & \qquad \Bigg.-|y_k|^2\left(\sum_{j\in \cK}|\bh_k^\dagger\bv_j|^2+\sigma^2_k\right)\Bigg) \nonumber \\
	\text{\normalfont s.t. ~} \quad & \sum_{k\in\cK} \|\bv_k\|^2\leq P,
		\label{problem:sumrate3}
	\end{align}
	where a constant term depending on $\bgam$ is omitted. With $\{\bv_k\}$ fixed, the above problem has a closed-form solution in $\by$, which takes the form in \eqref{quady} with $A_k=(1+\gamma_k)|\bh_k^\dagger\bv_k|^2$ and $B_k=\sum_{j\in\cK}|\bh_k^\dagger\bv_j|^2+\sigma_k^2.$ 
	When $\by$ is fixed, the above problem has the following solution in $\{\bv_k\}$:
	\begin{equation*}\label{sumrate:v}
\bv_k=y_k\sqrt{1+\gamma_k}\left(\sum_{j\in\cK}|y_j|^2\bh_j\bh_j^\dagger+\lambda\mathbf{I}\right)^{-1}\bh_k,~k\in\cK,
	\end{equation*}
	where $\lambda\geq 0$ is the optimal Lagrange multiplier associated with the total power constraint that can be efficiently determined by a bisection search. By updating $\bgam$, $\by$, and $\{\bv_k\}$ in an alternating fashion as described above, we obtain an efficient FP algorithm, which is guaranteed to converge to a stationary point of the sum-rate maximization problem \eqref{problem:sumrate2} \cite[Appendix A]{FP2}.} {We wish to remark here that, in addition to the downlink MIMO channel, the above FP techniques can be applied to solve sum-rate maximization problems in much more general channels (e.g., the MIMO interfering broadcast channel).}

{It is interesting to note that the above FP algorithm is equivalent to the well-known weighted minimum-mean-square-error (WMMSE) algorithm \cite{WMMSE1,WMMSE2}. In fact, the WMMSE algorithm, which is originally derived based on a signal minimum mean-square-error analysis,  can also be derived by applying the quadratic and Lagrangian dual transforms in a similar way as the above FP algorithm. The only difference is that, when applying the quadratic transform to problem \eqref{problem:sumrate2}, the WMMSE algorithm treats $|\bh_k^\dagger \bv_k|^2$ as the numerator and $(1+\gamma_k)$ as a scaling factor in front of the fractional term, which leads to a problem different from \eqref{problem:sumrate3} and hence different update rules for $\by$ and $\bv_k$; see the details in \cite[Section VI]{FP2}. While the two algorithms are fundamentally equivalent for the sum-rate maximization problem \eqref{problem:sumrate}, their different treatments of the term $(1 + \gamma_k)|\bh_k^\dagger\bv_k|^2$ lead to different algorithms for solving sum-rate maximization problems in more complicated scenarios. The above FP algorithm is thus preferable from a practical perspective, because it often yields a problem amenable to distributed optimization, as demonstrated below.}

\paragraph{Joint Uplink Scheduling and Power Control for Sum-Rate Maximization} {We now briefly review the application of FP techniques to solving the joint uplink scheduling and power control problem \eqref{problem:scheduling}. The main idea is to reformulate the problem appropriately with the aid of the two FP transforms, so that the resulting problem is amenable to AO and is in a distributed form that allows for per-cell scheduling and power update. To be specific, by first applying the Lagrangian dual transform to problem \eqref{problem:scheduling}, we get
\begin{align}
	\max_{\bka,\bp,\bgam} \quad & \sum_{i\in\cB}w_{\kappa_i}\left(\log(1+\gamma_i)-\gamma_i\right) \nonumber \\
& \qquad +\sum_{i\in\cB}\frac{w_{\kappa_i}(1+\gamma_i)|h_{i,\kappa_i}|^2p_{\kappa_i}}{\sum_{j\in\cB}|h_{i,\kappa_j}|^2p_{\kappa_j}+\sigma_i^2} \nonumber \\
	\text{s.t. } \quad &0\leq p_k\leq P,~k\in\cup_{i\in\cB}~\cK_i, \nonumber \\
	&\kappa_i\in\cK_i\cup\{\emptyset\},~i\in\cB.
	\label{problem:scheduling2}
\end{align}
	When $(\bka,\bp)$ are fixed, the optimal $\bgam$ of problem \eqref{problem:scheduling2} has a closed-form solution that takes the form in \eqref{laggamma}.
	To optimize $(\bka,\bp)$ in \eqref{problem:scheduling2} with a fixed $\gamma$, we further apply the quadratic transform to the sum-of-ratio term  in \eqref{problem:scheduling2}, which, after some simple manipulations, yields the following equivalent problem:
	\begin{align}
	\max_{\bka,\bp,\by} \quad & \sum_{i\in\cB}\Bigg(w_{\kappa_i}(\log(1+\gamma_i)-\gamma_i)\nonumber \\ & \quad +2y_i\sqrt{w_{\kappa_i}(1+\gamma_i)|h_{i,\kappa_i}|^2p_{\kappa_i}} \nonumber \\ & \qquad -y_i^2\sigma_i^2-\sum_{j\in\cB}y_j^2|h_{j,\kappa_i}|^2p_{\kappa_i}\Bigg) \nonumber \\
	\text{s.t. } \quad & 0\leq p_k\leq P,~k\in\cup_{i\in\cB}~\cK_i, \nonumber \\
	&\kappa_i\in\cK_i\cup\{\emptyset\},~i\in\cB.
\label{problem:scheduling3}
	\end{align}
	A favorable structure of problem \eqref{problem:scheduling3} is that the scheduling and power variables of each cell, i.e., $(\kappa_i, p_{\kappa_i})$, are decoupled when $\by$ is fixed, thus allowing the scheduling and power optimization to be performed independently within each cell. More details on the solution of the scheduling and power control subproblem \eqref{problem:scheduling3} and the AO algorithm for solving the joint uplink scheduling and power control problem \eqref{problem:scheduling} can be found in \cite{FP2}.
}

\subsubsection{Remarks}
{We conclude this subsection with further remarks on the quadratic and Lagrangian dual transforms from both optimization and application perspectives.  
	From the optimization perspective, the principle behind the two transforms is to lift complicated low-dimensional problems to equivalent high-dimensional ones where optimization is easier to do by appropriately introducing some auxiliary variables. The key is to ensure that the lifted high-dimensional problem is easy to solve with respect to each variable block (e.g., being convex or admitting closed-form solutions) so that AO techniques such as the block coordinate descent (BCD) algorithm can be applied. AO algorithms can efficiently find a stationary point of the lifted problem, which is also a stationary point of the original problem. It is also interesting to note that the BCD algorithm for solving the quadratic problem \eqref{quadtransform} lies in the minorization-maximization (MM) framework for solving the original problem \eqref{prob:multiratio} \cite{FP:MM}}. 

From the application perspective, the quadratic and Lagrangian dual transforms are crucial tools for solving problems with fractional structures that arise from wireless communication system design. For instance, the Lagrangian dual transform significantly simplifies the structure of the sum-rate maximization problem by moving the SINRs outside the nonlinear logarithmic functions. This is particularly advantageous when complicated variables are involved in the SINR expressions, such as the discrete scheduling variables in \eqref{problem:scheduling}, the multiplicatively coupled variables in RIS-aided systems \cite{FPRIS2}, and hybrid beamforming \cite{FPHB}. Moreover, when appropriately utilized and implemented, the two transforms can enable the problem to be reformulated into a form that allows for distributed optimization (e.g., within each cell), which is favorable for the design and optimization of wireless cellular networks.

\subsection{Sparse Optimization}\label{subsec:sparse}
{Sparse optimization refers to a class of problems whose solution exhibits an inherent sparse structure. Here, sparsity means that only a small fraction of the entries in the solution vector is nonzero. Driven by the emergence and success of compressed sensing (CS) \cite{CS}---a signal acquisition paradigm designed to recover a sparse signal from a small set of incomplete measurements---sparse optimization has received significant attention over the past few decades. In this subsection, we first briefly review the theory and models associated with CS and sparse optimization. Then, we introduce two successful applications of CS and sparse optimization to wireless communication system design, which are localized statistical channel modeling \cite{shutao2023_10299600} and device activity detection in mMTC \cite{bockelmann2016massive}.}
\subsubsection{Compressed Sensing and Recovery Conditions} 
In various real-world applications, signals are (approximately) sparse or have a sparse representation under a certain basis. By exploiting the inherent sparsity of the true signal, CS enables the reconstruction of the original signal from only a small number of observations (e.g., from an underdetermined linear system), thereby significantly reducing the burden of sample acquisition, data storage, and computation.  Mathematically, the reconstruction process of the sparse signal  can be formulated as the following optimization problem:
\begin{equation}\label{CSproblem}
\begin{aligned}
\min_{\x}~& \|\x\|_0\\
\text{s.t. }~&\bA\x=\y.
\end{aligned}
\end{equation}
Here, $\x\in\R^n$ is the sparse signal to be recovered, $\|\x\|_0$ denotes the $\ell_0$-norm that counts the numbers of nonzero entries in $\x$, $\y=\bA\x$ is an underdetermined system with the observation $\y\in\R^m$, $\bA\in\R^{m\times n}$ is the sensing matrix, and $m\ll n$. 
It is generally NP-hard to solve problem \eqref{CSproblem} \cite{natarajan1995sparse}. An alternative model that enables the design of computationally efficient recovery algorithms is given by
 \begin{equation}\label{CSproblem2}
\begin{aligned}
\min_{\x}~& \|\x\|_1\\
\text{s.t. }~&\bA\x=\y,
\end{aligned}
\end{equation}
where the nonconvex term $\|\cdot\|_0$ in \eqref{CSproblem} is replaced by the convex term $\|\cdot\|_1$ in \eqref{CSproblem2}. 
In practice, the measurements may include some noise, i.e.,  $\y=\bA\x+\mathbf{z}$ with $\|\mathbf{z}\|_2\leq \epsilon$. In this case, the reconstruction problem can be formulated as
 \begin{equation}\label{CSproblem3}
\begin{aligned}
\min_{\x}~& \|\x\|_1\\
\text{s.t. }~&\|\bA\x-\y\|_2\leq\epsilon.
\end{aligned}
\end{equation}
The above constrained problem can be further recast into the unconstrained problem
 \begin{equation}\label{CSproblem4}
\begin{aligned}
\min_{\x}~&\|\bA\x-\y\|_2^2+\lambda \|\x\|_1,
\end{aligned}
\end{equation} where $\lambda>0$ is a parameter that trades off the data fidelity term $\|\bA\x-\y\|_2^2$ and the sparsity term $\|\x\|_1$.
Problems \eqref{CSproblem3} and \eqref{CSproblem4} are connected in that for any $\epsilon$, there exists a $\lambda$ such that the two problems have the same solutions \cite{CSbook}. 

Two fundamental questions for the above reconstruction problems are (i) under what conditions can the formulations \eqref{CSproblem} and \eqref{CSproblem2} precisely recover any $k$-sparse signal $\x$ (i.e., $\|\x\|_0\leq k$) and (ii) how large is the recovery error when there is some noise in the measurements. These questions have been extensively studied,  yielding numerous remarkable and insightful results. One of the most well-known recovery conditions is the restricted isometry property (RIP) \cite{Candes2005} of the sensing matrix $\bA$. Specifically, a matrix $\bA$ is said to satisfy the RIP of order $k$ if there exists a constant $\delta_k\in(0,1)$ such that 
$$(1-\delta_k)\|\x\|_2^2\leq \|\bA\x\|_2^2\leq (1+\delta_k)\|\x\|_2^2$$ holds for all $k$-sparse vector $\x$.   Intuitively, RIP can be viewed as a characteristic that preserves the geometry (i.e., the distance) between sparse vectors. A smaller $\delta_k$ implies better preservation capability, making $\bA$ a more effective sensing matrix. 
 An intriguing recovery result characterized by RIP is that when $\bA$ satisfies RIP of order $2k$  with $\delta_{2k}<1$, any $k$-sparse signal can be exactly recovered by the $\ell_0$ minimization problem \eqref{CSproblem}. Furthermore, if $\delta_{2k}<\sqrt{2}-1$, then the solution of the $\ell_1$ minimization problem \eqref{CSproblem2} is the same as that of  \eqref{CSproblem}, i.e., the  $k$-sparse signal can also be recovered from \eqref{CSproblem2}. When the measurements are corrupted with noise,  the recovery error is in the order of  $\mathcal{O}(\epsilon)$ for problem \eqref{CSproblem3}  \cite{Candes2008}.
 The upper bound on the above RIP constant can be further improved; see, e.g., \cite{mo2011new}. It is worth mentioning that the RIP can be satisfied with high probability for a wide class of random matrices, including the i.i.d. Gaussian/Bernoulli matrices \cite{Gaussian} and the partial Fourier matrix \cite{Fourier}, when the number of measurements satisfies $$m>\mathcal{O}( k\log (n/k) ).$$ Finally, we remark that there are also many other important conditions based on which recovery results are established. Interested readers are referred to a comprehensive review of these conditions as well as efficient sparse signal recovery algorithms for solving the previous models in \cite{CSbook}.

\subsubsection{Application Examples} Sparse optimization and CS have found broad applications in wireless communication systems \cite{choi2017compressed}. In this subsection, we showcase the important role of sparse optimization and CS approaches in formulating and analyzing two optimization problems---localized statistical channel modeling \cite{shutao2023_10299600} and device activity detection in mMTC \cite{bockelmann2016massive}. 

\paragraph{Localized Statistical Channel Modeling} 
In network optimization\cite{10155734}, it is desirable to have a channel model that captures the specific multi-path topography and statistical properties of the targeted communication environment. The so-called localized statistical channel modeling (LSCM) \cite{shutao2023_10299600} aims to leverage the beam-wise reference signal received power (RSRP) measurements to estimate the angular power spectrum (APS) of the channel between the BS and the user.  

Consider a scenario in which the BS is equipped with a uniform rectangular array possessing $N_T = N_1 \times N_2$ antennas, and the user has only a single antenna.  The downlink channel between the $(x,y)$-th antenna and the user is denoted as $h_{x,y}(t),$ where $x = 0,1, \ldots, N_1 -1$ and $y = 0,1, \ldots, N_2 - 1$.  
In 5G networks, synchronization signals and CSI reference beam signals are regularly transmitted to the user.  The measured  RSRP of the $m$-th beam at the $t$-th time slot is given by \cite{shutao2023_10299600}
	\begin{equation} \label{eq: rsrp}
		\mathrm{rsrp}_{m}(t)=P\left|\sum_{x = 0}^{N_1 - 1}\sum_{y = 0}^{N_2 - 1} h_{x, y} (t) W_{x, y}^{(m)}\right|^{2},
	\end{equation}
where $  W_{x, y}^{(m)} = e^{j\phi_{x,y}^{(m)}}$ is the $(x, y)$-th entry of the precoding matrix $\boldsymbol{W}^{(m)} \in \mathbb{C}^{N_x \times N_y}$ for the $m\text{-th}$  beam, $\phi_{x,y}^{(m)}$ is the weight of  DFT matrix, and $P$ represents the transmit power.   Suppose that there are $M$ directional beams in total.  The expected beam-wise RSRP measurements ${\bf rsrp} \in \mathbb{R}^{M \times 1}$ is
	\begin{align} \nonumber
		{\bf rsrp}  = \left[{\rm RSRP}_{1}, {\rm RSRP}_{2}, \ldots, {\rm RSRP}_{m}, \ldots, {\rm RSRP}_{M}  \right]^{\rm T},
	\end{align}
	where  ${\rm{RSRP}}_{m} \triangleq \mathbb{E}\left[\mathrm{rsrp}_{m}(t) \right]$.
	 As demonstrated in \cite{shutao2023_10299600},  the beam-wise average RSRP measurements and the channel APS have the linear relationship
	\begin{align}
		 {\bf rsrp} = {\bf A}{\bf x},
	\end{align}
where ${\bf A} \in \mathbb{R}^{M \times N}$ is a sensing matrix depending on the beam waveform and antenna gains,  and ${\bf x} \in \mathbb{R}^{N}$ is the channel APS to be estimated.  Here,  the free space is discretized by $N $ equally spaced directions, and $N$ is usually a large number ($N \gg M$) for a high angular resolution. Due to the presence of a limited number of scatters around the BS,  there is a small angular spread in the angular domain, resulting in a sparse channel APS  $\bf x$.  To construct the localized statistical channel model,  we can formulate the sparse recovery problem  as
\begin{equation}\label{problem22}
	\begin{aligned}
		\ \min _{{\bf x}} &~ \|{\bf {Ax-rsrp}}\|_{2}^{2}\\
		\text {s.t.~} &~\|{\bf x}\|_{0} \leq K,~{\bf x} \geq \bm{0}, 
	\end{aligned}
\end{equation}
where $K$ is the maximum number of nonzero entries, representing the maximum number of channel paths; and the constraint ${\bf x} \geq \bm{0}$ is because the expectation of channel gain with respect to different angles is nonnegative. Efficient algorithms for solving problem \eqref{problem22} are proposed in  \cite{shutao2023_10299600}. 

The localized statistical channel model exhibits statistical indistinguishability from the true propagation environment, generating channels that are similar to real-world scenarios. The construction of the localized statistical channel model enables precise evaluation of the network performance and effectively facilitates the simulation for offline network optimization \cite{10155734, li2022real, zte_digital_twin_channel}.


\paragraph{Device Activity Detection in mMTC} 

Consider an uplink single-cell massive random access scenario \cite{ChenX2021} with $K\gg1 $ single-antenna devices potentially accessing a BS equipped with $M$ antennas, which corresponds to the uplink counterpart of the wireless system in Fig.~\ref{fig:commun_system}(a). A key feature of mMTC is that at any given time, only a small subset of users are active. To reduce the communication latency, grant-free random access schemes have been proposed in \cite{Liu2018b,Senel2018}, where the active devices directly transmit the data signals after transmitting their preassigned nonorthogonal signature sequences without first obtaining permissions from the BS. The BS identifies the active devices based on the received signature sequences. 
We now introduce two formulations of the {device activity} detection problem and their related detection theory.

We begin with the system model. For the purpose of
device identification, each device $k$ is preassigned a unique signature
sequence $\mathbf{s}_k=[s_{1k},s_{2k},\ldots, s_{Lk}]^\T\in \mathbb{C}^{L}$,
where $L$ is the sequence length. 
Let $a_{k}\in\{0,1\}$ denote the activity of device $k$, i.e., $a_k=1$ if the device is active and $a_k=0$ otherwise, and let $\mathbf{h}_{k}\in\C^M$ denote the (unknown) channel vector between device $k$ and the BS.
Then, the received signals $\mathbf{Y}\in \mathbb{C}^{L\times M}$ at the BS (in the pilot
phase) can be expressed as
\begin{align}\label{eq.sys}
\mathbf{Y}&=\sum_{k=1}^{K}a_{k}\mathbf{s}_{k}\mathbf{h}_{k}^\T +\mathbf{Z},
\end{align}
where $\mathbf{Z}\in \mathbb{C}^{L\times M}$ is the normalized effective i.i.d. Gaussian noise with variance $\sigma_{z}^2\mathbf{I}$. 

Define $\mathbf{S}=[\mathbf{s}_{1},\mathbf{s}_{2},\ldots,\mathbf{s}_{K}]\in\mathbb{C}^{L\times K}$ and $\mathbf{X}=[a_1\mathbf{h}_{1},a_2\mathbf{h}_{2},\ldots,a_K\mathbf{h}_{K}]^\T\in\mathbb{C}^{K\times M}.$ The received signals in \eqref{eq.sys} can then be rewritten as $\mathbf{Y}=\mathbf{S}\mathbf{X}+\mathbf{Z}.$ Since the user traffic is sporadic, i.e., only  $K_a\ll K$
 devices are active during each coherence interval, most rows of $\mathbf{X}$ will be zero. Based on this observation, the device activity detection problem can be formulated and analyzed using sparse optimization and CS approaches \cite{Liu2018massive,liu2018massive_partII}. For instance, the device activity detection problem can be formulated as
 \begin{equation}\label{CSactivity}
 \begin{aligned}
 \min_{\mathbf{X}}~&\|\mathbf{S}\mathbf{X}-\mathbf{Y}\|^2_F+\lambda \|\mathbf{X}\|_{2,1},
 \end{aligned}
 \end{equation} where $\|\mathbf{X}\|_{2,1}=\sum_{k=1}^K\sqrt{\sum_{m=1}^M X_{k,m}^2}$ is the $\ell_{2,1}$-norm, which is effective in promoting the group sparsity of $\mathbf{X}$.
The works \cite{Liu2018massive,liu2018massive_partII} propose to use the (vector) approximate message passing (AMP) algorithm to solve the device activity detection problem and analyze the detection performance by utilizing the state evolution analysis. It has been shown in \cite{Liu2018massive} that as the number of antennas $M$ goes to infinity, the missed detection and false alarm probabilities can always be made to go to zero by the AMP approach that exploits the sparsity in the user activity pattern. Problem \eqref{CSactivity} can also be solved by the PG algorithm (see Section \ref{subsection:proximal}), which comes with strong convergence guarantees~\cite{ZZS15,ZS17}. Recent progress on using sparse optimization and CS approaches to solve the device activity problem has been made in \cite{Chen2018sparse,zhu2023message,rajoriya2023joint,zhang2023joint,li2022dynamic}.  
 
Note that the sparse optimization approach described above recovers not only the user activities, but also an estimate of their channels. If we are only interested in the user activities and \emph{not} the channels, then an alternative approach is to formulate the device activity detection problem as a maximum likelihood estimation (MLE) problem of the user activities only \cite{liu2023grant}. Instead of treating $\mathbf{h}_{k}$ as a deterministic {unknown} variable as in the above CS approach, the MLE approach exploits the distribution information in $\mathbf{h}_{k},$ i.e., $\mathbf{h}_{k}=\sqrt{g_k} \tilde{\mathbf{h}}_{k},$ where $g_{k}\geq 0$ is the large-scale fading component, and $\tilde{\mathbf{h}}_{k}\in \mathbb{C}^{M}$ is the Rayleigh fading component following $\mathcal{CN}(\mathbf{0},\mathbf{I})$. In this case, the received signals in \eqref{eq.sys} can be rewritten as
$\mathbf{Y}=\mathbf{S}\boldsymbol{\Gamma}^{1/2}\tilde{\mathbf{H}}+\mathbf{Z},$ where $\boldsymbol{\Gamma}=\operatorname{diag}(\gamma_1,\gamma_2,\ldots,\gamma_K)\in\mathbb{R}^{K \times K}$ with $\gamma_k=a_kg_k$ being a diagonal matrix indicating both the device activity $a_k$ and the large-scale fading component $g_k$, and $\tilde{\mathbf{H}}=[\tilde{\mathbf{h}}_{1},\tilde{\mathbf{h}}_{2},\ldots,\tilde{\mathbf{h}}_{K}]^\T\in\mathbb{C}^{K \times M}$ is the normalized channel matrix. Note that the columns of $\mathbf{Y}$,
denoted by $\mathbf{y}_m\in\mathbb{C}^{L}$, $1\leq m\leq M$, are independent and each column $\mathbf{y}_m$
follows the complex Gaussian distribution 
$\mathbf{y}_m \sim \mathcal{CN}\left(\mathbf{0},\boldsymbol\Sigma\right)$
with covariance matrix $$\boldsymbol\Sigma=\mathbb{E}\left[\mathbf{y}_m\mathbf{y}_m^\dagger\right]=\mathbf{S}\boldsymbol{\Gamma}\mathbf{S}^\dagger+\sigma_z^2\mathbf{I}.$$Therefore, the problem of maximizing the likelihood $p(\mathbf{Y}\,|\,\boldsymbol{\Gamma})$ can be equivalently formulated as
\begin{equation}\label{eq.prob1}
	\begin{aligned}
	\underset{\boldsymbol\Gamma}{\operatorname{min}}    ~& \log\det\left(\boldsymbol\Sigma\right)+ \operatorname{tr}\left(\boldsymbol\Sigma^{-1}\widehat{\boldsymbol{\Sigma}}\right)\\
	\operatorname{s.t.} ~&  \boldsymbol{\Gamma} \geq \bm{0},
	\end{aligned}
\end{equation} where $\widehat{\boldsymbol{\Sigma}}=\mathbf{Y}\mathbf{Y}^\dagger/M$
is the sample covariance matrix of the received signals averaged over different antennas. The formulation \eqref{eq.prob1} leads to the so-called covariance-based approach in the literature because it depends on $\mathbf{Y}$ only through its covariance $\widehat{\boldsymbol{\Sigma}}$. It has been shown in \cite{ChenZ2020,fengler2021non} that when $\left\{\mathbf{s}_k\right\}$ is uniformly drawn from the sphere of radius $\sqrt{L}$ in an
i.i.d. fashion and the number of active devices satisfies $$K_a\leq c_1L^2/\log^2(eK/L^2),$$ then the MLE formulation in \eqref{eq.prob1} is able to successfully detect the active devices with {probability}  at least $1-\exp(-c_2L),$ where $c_1$ and $c_2$ are two constants whose
values do not depend on $K_a,$ $K,$ and $L.$ This result shows that if the number of antennas $M$ at the BS goes to infinity, then the number of active devices that can be detected by the covariance-based approach scales quadratically with the length of the devices' signature sequence $L$. Covariance-based approaches and analyses have also been extended to the joint activity and data detection case \cite{ChenZ2020}, the multi-cell scenario \cite{chen2021sparsedection}, the more practical ways of generating signature sequences \cite{wang2023covariance}, the asynchronous scenario \cite{wang2022covariance,li2023asynchronous}, the case where the BS is equipped with low-resolution ADCs \cite{wang2023device}, and the unsourced random access scenario \cite{fengler2021non}.

\subsubsection{Remarks} 
{We conclude this subsection with a summary highlighting the crucial role that sparse optimization and CS play in wireless communication system design and analysis. First, sparse optimization is helpful in formulating optimization problems in wireless communications to promote desirable sparse structures in the solution, e.g., sparsity in the localized statistical channel model problem \eqref{problem22} and group sparsity in the joint BS clustering and beamformer design problem \eqref{problem:clustering} and device activity detection problem \eqref{CSactivity}. Compared to traditional formulations without exploiting sparsity (in the appropriate domain), sparse optimization formulations can significantly reduce signaling and training overhead associated with channel estimation \cite{choi2017compressed,shutao2023_10299600,rao2014distributed, cheng2017channel, ma2018sparse}. Second, analytical tools derived from CS, including recovery conditions and the AMP-based high-dimensional analysis, are useful for understanding and analyzing the theoretical performance of certain optimization models/algorithms in wireless communications. As an example, these tools are employed in the device activity detection problem to theoretically characterize the detection performance of both CS and covariance-based approaches \cite{Liu2018massive,liu2018massive_partII,ChenZ2020,fengler2021non}. }
\subsection{Proximal Gradient Algorithms} \label{subsection:proximal} 
In this subsection, we first motivate the development of the PG algorithm. 
Then, we demonstrate how two optimization problems from wireless communication system design can be tackled by the PG algorithm.    
\subsubsection{PG Algorithm, Interpretation, and Convergence Property} Consider the problem
\begin{equation}\label{problem:PG}\min_{\x\in\R^n} f(\x)+g(\x),
\end{equation} where $f(\cdot)$ is a smooth function with Lipschitz continuous gradient and $g(\cdot)$ is a nonsmooth function. In \eqref{problem:PG}, none of the functions $f(\cdot),~g(\cdot),$ and $f(\cdot)+g(\cdot)$ is required to be convex. For $r\geq 1,$ the $r$-th iteration of the PG algorithm reads
\begin{equation}\label{PGupdate}
\x^{r+1}\in\text{prox}_{\alpha_r g}\left(\x^r-\alpha_r\nabla f(\x^r)\right),
\end{equation}
where $\alpha_r$ is the step size at the $r$-iteration and  $\text{prox}_{\alpha_rg}(\cdot)$ is the so-called proximal operator defined as 
 \begin{equation}\label{proximal}
 \text{prox}_{\alpha_rg}(\bv)\in\arg\min_{\y\in\R^n}\left\{g(\y)+\frac{1}{2\alpha_r}\|\y-\bv\|_2^2\right\}.
 \end{equation} 
 
By definition of the proximal operator in \eqref{proximal}, we can rewrite \eqref{PGupdate} into the following equivalent form:
 \begin{multline}\x^{r+1}\in\hspace{-0.05cm}\arg\min_{\bx \in \R^n} \left\{f(\x^{r})+\nabla {f(\x^{r})}^\mathsf{T}(\x-\x^{r})\right.\\\left.+\frac{1}{2\alpha_r}\|\x-\x^{r}\|^2_2
 +g(\x)\right\}.\label{interpret}\end{multline}
 Then, it is clear that the update in \eqref{PGupdate} can be interpreted as minimizing an approximation of the original objective function at each iteration. In particular, the right-hand side of \eqref{interpret} approximates the smooth term $f(\cdot)$ by its first-order Taylor's expansion at point $\bx^r$ plus a quadratic term and keeps the nonsmooth term $g(\cdot)$ unchanged.  In addition, when $\alpha_r\in (0, 1/L]$, where $L$ is the Lipschitz constant of $\nabla{f}$, the minimized function on the right-hand side of \eqref{interpret} is an upper bound of the objective function in \eqref{problem:PG}. As such, the PG algorithm falls under the MM framework \cite[Section 4.2]{proximal}.
 Obviously, the efficiency of the PG algorithm highly depends on that of computing the proximal operator in \eqref{proximal}. Fortunately, for many nonsmooth functions of practical interest, their proximal operators either admit a closed-form solution (e.g., $\|\cdot\|_{0.5},$ $\|\cdot\|_1,\|\cdot\|_2,\|\cdot\|_\infty$) or can be efficiently computed; see \cite[Page 177]{beckbook} for a summary of such examples. 
 
The PG algorithm enjoys nice theoretical convergence properties.  In the case where both $f$ and $g$ are closed proper convex functions, the PG algorithm with a fixed step size $\alpha_r=\alpha\in(0,2/L]$ is guaranteed to converge to the optimal solution of problem \eqref{problem:PG} \cite{Combettes2011}. For the nonconvex case, it is shown in \cite{Attoch2013} that the iterates generated by the PG algorithm converge to a critical point of problem \eqref{problem:PG} with $0<\underline{\alpha}<\alpha_r<\bar{\alpha}<1/L$, as long as  $f+g$ is proper, closed, and satisfies the Kurdyka-{\L}ojasiewicz property. These conditions are quite mild and are satisfied by a rich class of functions (e.g., semi-algebraic functions) \cite{Attouch2010}. Furthermore, there have been efforts in establishing the convergence of the inexact PG algorithm \cite{Attoch2013}. These results allow for an error in the calculation of the proximal operator at each iteration, thus offering flexibility in cases where the proximal operator lacks a closed-form solution and needs to be computed numerically.

It is worth noting an interesting special case where the nonsmooth term $g(\cdot)$ in problem \eqref{problem:PG} is the indicator function of a closed set $\mathcal{C}\subseteq\R^n.$ In this case, problem \eqref{problem:PG} reduces to the constrained problem 
$$\min_{\x\in\mathcal{C}} f(\x), $$ the proximal operator in \eqref{proximal} reduces to the familiar projection operator 
\begin{equation*}
\text{proj}_{\mathcal{C}}(\bv)=\arg\min_{\y\in\mathcal{C}}\left\{\frac{1}{2}\|\y-\bv\|_2^2\right\},
\end{equation*} and the PG algorithm reduces to the GP algorithm where \eqref{PGupdate} is replaced by
\begin{equation*}
\x^{r+1}\in\text{proj}_{\mathcal{C}}\left(\x^{r}-\alpha_r\nabla f(\x^{r})\right).
\end{equation*}

Due to its simple implementation, computational efficiency, and appealing theoretical properties, the PG algorithm is widely adopted for solving optimization problems that involve simple (nonconvex) nonsmooth terms (e.g., smooth problems with simple constraints). 

\subsubsection{Application Examples}
In this subsection, we apply the PG and GP algorithms to solve two fundamental problems in wireless communications, namely, massive MIMO detection \eqref{problem:mimodec} and joint BS clustering and beamformer design \eqref{problem:clustering}.

\paragraph{Massive MIMO Detection} We first review the application of the GP algorithm for solving the MIMO detection problem in \eqref{problem:mimodec}. As discussed in Section \ref{subsection:problem}, a new challenge for MIMO detection is the significant increase in problem size driven by the massive MIMO technology. In the context of massive MIMO, classic  MIMO detection algorithms/techniques that work well for small-to-median scale systems (e.g., SDR-based algorithms \cite{ma2004semidefinite,jalden2006detection}) become impractical, as their computational complexities grow quickly with the problem size. 

Motivated by the above, the following low-complexity GP algorithm 
$$\x^{r+1}=\text{Proj}_{\mathcal{S}^K}\left(\x^{r}-2\alpha_r\H^\dagger(\H\x^{r}-\by)\right),$$is proposed in \cite{So2017} to solve the massive MIMO detection problem, where $\alpha_r>0$ is the step size, $\text{Proj}_{\mathcal{S}^K}(\cdot)$ denotes the projection operator onto the set $\mathcal{S}^K$, and $\mathcal{S}$ is either the PSK constellation set in \eqref{MPSK}  or the QAM constellation set in \eqref{QAM}. The dominant computational cost at each iteration of the above GP algorithm lies in two matrix-vector multiplications and one projection onto $\mathcal{S}^K$. Since the set $\mathcal{S}^K$ is fully decoupled among different components, the projection onto $\mathcal{S}^K$ reduces to $K$ projections onto $\mathcal{S}.$ Moreover, the discrete set $\mathcal{S}$ is symmetric and highly structured. Hence, the projection $\text{Proj}_\mathcal{S}(\cdot)$ and consequently the projection $\text{Proj}_{\mathcal{S}^K}(\cdot)$ are easily computable. This makes the above GP algorithm extremely efficient and particularly suitable for solving large-scale MIMO detection problems arising from massive MIMO systems. 

In addition to its low per-iteration computational complexity, the above GP algorithm enjoys strong theoretical guarantees. It has been shown in \cite{So2017}  that under mild conditions (roughly speaking, when the noise variance is small and the ratio  $M/K$ is large), the iterates generated by the GP algorithm will converge to the true symbol vector $\mathbf{s}$ within a finite number of iterations. This result is somewhat surprising and is much stronger than the general convergence result for GP algorithms. First, the GP algorithm for solving nonconvex problems is generally not guaranteed to converge to an optimal solution but only to a critical point. Second, there is generally no theoretical guarantee that the maximum likelihood (ML) estimator (i.e., the optimal solution of problem \eqref{problem:mimodec}) is the true symbol vector $\bs$. This strong convergence result in \cite{So2017} is obtained by carefully exploiting the structure of problem \eqref{problem:mimodec}, particularly the special structure of the discrete set $\mathcal{S}$ and the statistical property of the channel matrix $\bH$; see the detailed proof in \cite[Theorem 1]{So2017}.

\paragraph{Joint BS Clustering and Beamformer Design} The PG algorithm plays an important role in solving the joint BS clustering and beamformer design problem \eqref{problem:clustering}.  
Observe that problem \eqref{problem:clustering} is challenging to solve, as the variables $\{\bv_{k,b}\}$ are coupled in both the objective function and the constraint, and the objective function has a nonsmooth term and is highly nonlinear. To tackle problem \eqref{problem:clustering}, it is useful to reformulate it into the following equivalent form using the technique similar to that in the FP and WMMSE approaches \cite{FP1,WMMSE1,WMMSE2}:
\begin{equation}\label{problem:clustering2}
\begin{aligned}
\min_{\{u_k\},\{w_k\},\{\bv_k\}}~&\sum_{k\in \cK}\left(w_ke_k-\log w_k+\rho\sum_{b\in\cB}\|\bv_{k,b}\|_2\right)\\
\text{s.t. }\qquad~&\sum_{k\in\cK}\|\bv_{k,b}\|_2^2\leq P_b,~b\in\cB,
\end{aligned}
\end{equation}
where $e_k$ is the mean squared error (MSE) for user $k$ given by 
$$e_k=\left|u_k\h_k^\dagger\bv_k-1\right|^2+\sum_{j\neq k}\left|u_k\h_k^\dagger\bv_j\right|^2+\sigma^2|u_k|^2.$$
A desirable property of the above reformulation \eqref{problem:clustering2} is that the problem is convex with respect to each of the variable blocks $\bv=\{\bv_k\}$, $\bu=\{u_k\}$, and $\bw=\{w_k\}$ (with the other two blocks being fixed), making it amenable to the BCD algorithm. In particular, when $\bu$ and $\bv$ are fixed, the problem in terms of $\bw$ admits the closed-form solution $w_k=e_k^{-1}$ for all~$k\in\cK;$ when $\bw$ and $\bv$ are fixed, the problem in terms of $\bu$ also has a closed-form solution. 
Below we consider the solution of the problem in terms of $\bv$ with fixed $\bu$ and $\bw$.

Since the constraint in \eqref{problem:clustering2} is separable in the beamforming vectors of different BSs, we can apply the BCD algorithm again to solve the $\bv$-subproblem by treating $\{\bv_{k,b}\}_{k\in\cK}$ as one block of variables. Specifically, the $b$-th subproblem takes the form
\begin{equation}\label{problem:clustering3}
\begin{aligned}
\min_{\{\bv_{k,b}\}_{k\in\cK}}~&\sum_{k\in\cK}\left(\bv_{k,b}^\dagger\bQ_{k,b}\bv_{k,b}-2\mathrm{Re}\left(\bd_{k,b}^\dagger\bv_{k,b}\right)+\rho\|\bv_{k,b}\|_2\right)\\
\text{s.t. }\quad~&\sum_{k\in\cK}\|\bv_{k,b}\|_2^2\leq P_b,
\end{aligned}
\end{equation}
where $\bQ_{k,b}\in\C^{M\times M}$ and $\bd_{k,b}\in\C^M$ are constants depending on the other blocks of the variables; see their explicit expressions in \cite{Hong2013}. The objective function in \eqref{problem:clustering3} is separable among different $k\in\cK$ and each of them is a simple quadratic function plus a convex nonsmooth $\ell_2$-norm. However, the presence of the quadratic constraint complicates the solution of the problem and makes the PG algorithm not efficient.\footnote{It might be possible to directly apply the PG algorithm to solve problem \eqref{problem:clustering3} by treating the sum of $\rho\sum_{k\in\cK}\|\bv_{k,b}\|_2$ and the indicator function of the feasible set of problem \eqref{problem:clustering3} as the nonsmooth term $g(\bx)$ in \eqref{problem:PG}. However, the proximal operator of this nonsmooth function is not easily computable, which makes the corresponding PG algorithm less efficient.} To overcome this difficulty, we consider the dual of problem \eqref{problem:clustering3} as follows: 
\begin{equation*}
\begin{aligned}
\max_{\lambda_b\geq 0}\min_{\{\bv_{k,b}\}}\sum_{k\in\cK}\left(\bv_{k,b}^\dagger\bQ_{k,b}\bv_{k,b}-2\mathrm{Re}\left(\bd_{k,b}^\dagger\bv_{k,b}\right)+\rho\|\bv_{k,b}\|_2\right)\\
+\lambda_b\left(\sum_{k\in\cK}\|\bv_{k,b}\|_2^2- P_b\right).~~
\end{aligned}
\end{equation*}
The above dual reformulation leads to an efficient algorithm for solving the subproblem in \eqref{problem:clustering3}. First, for a given $\lambda_b\geq 0$, the inner minimization problem over $\{\bv_{k,b}\}$ is separable among different $k\in\cK$, unconstrained, and convex. Hence, it can be efficiently solved to global optimality using the PG algorithm (as the proximal operator of the $\ell_2$-norm admits a closed-form solution). Second, the outer maximization problem over $\lambda_b\geq 0$ is a one-dimensional convex problem, whose solution can be quickly found via a simple bisection search. We remark here that even without the nonsmooth term in \eqref{problem:clustering3}, the solution of the corresponding $\bv$-subproblem (as in the FP and WMMSE approaches) also requires a bisection search; see the discussion below problem \eqref{problem:sumrate3} and \cite{FP1,WMMSE1,WMMSE2} for more details.

\subsubsection{Remarks}
We conclude this subsection with some remarks and conclusions drawn from the above two examples. 
First, in addition to the general theoretical convergence properties of the PG and GP algorithms\cite{Combettes2011,Attoch2013},  it is often possible to derive tailored results by carefully exploiting the special structure of the underlying problem. This is illustrated by the MIMO detection problem discussed earlier.
Second, many problems arising from wireless communication system design, though nonconvex and/or nonsmooth, have structured objective functions and/or constraints. Even though the PG and GP algorithms may not be directly applicable to tackling these problems, they 
can still play a vital role in solving these problems. By employing some approximations, equivalent reformulations, splitting techniques, or optimization frameworks like BCD or MM, these complicated problems often boil down to simple forms/subproblems that can be efficiently solved via the PG/GP algorithm; see more examples in \cite{Hong2013,Hong2016,Ibrahim2020}. 

\subsection{Penalty Methods}\label{subsection:penalty}

Penalty methods are an important class of methods for solving constrained optimization problems. The penalty methods look for the solution of the (complicated) constrained optimization problem by replacing it with a sequence of (relatively easy) unconstrained penalty subproblems. 
The objective function in the penalty subproblem is called the penalty function, which is formed by adding a penalty term to the objective function of the original constrained problem. The penalty term usually is a measure of the violation of the constraints of the original problem multiplied by a penalty parameter. Some important algorithms in this class include the quadratic penalty method and the augmented Lagrangian method.

Due to its simplicity, the penalty method has been widely studied and used to solve constrained optimization problems from various applications. A crucial concept associated with the penalty method is the \emph{exactness} of the penalty function. A penalty function is said to be exact if the unconstrained penalty problem with a sufficiently large penalty parameter would eventually share the same solution with the original constrained problem. The exactness of the penalty function plays a vital role in reducing and avoiding the ill-conditioning in the corresponding penalty method. Therefore, the choice of the penalty function in the corresponding penalty method is of fundamental importance to its numerical performance, and different choices of penalty terms would generally lead to different penalty methods. In this subsection, instead of reviewing the classic quadratic penalty methods, we review the recent penalty method developed in \cite{GEMM} for solving problems with integer/discrete variables arising from wireless communication system design.  

\subsubsection{Penalty Methods for a Class of Optimization Problems with Structured Constraints}
Let us first take the MIMO detection problem \eqref{problem:mimodec} as an example to illustrate how the penalty method can be applied to solve the optimization problem with binary variables. Consider the case in which the constellation is $L$-PSK given in \eqref{MPSK}. 
For notational simplicity,~
let $\bs=[s_1, s_2, \ldots, s_L]^\T\in \mathbb{C}^{L}$ be the vector of all constellation symbols, where $s_\ell =\exp({2\pi\textrm{i}(\ell-1)}/{L}).$~
We introduce the auxiliary variable $\bt=[\bt_1^{\T}, \bt_2^{\T}, \ldots, \bt_n^{\T}]^\T\in \mathbb{R}^{Ln},$ where $\bt_i=[t_{i,1},t_{i,2},\ldots, t_{i,L}]^{\T}\in \mathbb{R}^{L}.$ Then, for each $x_i^{\ast}$ of $\bx^{\ast},$ we have $x_i^{\ast} = \bt_i^{\T} \bs$ for some $\bt_i\in \mathbb{R}^{L}$ that has one entry equal to one and all other entries equal zero.
Then, problem \eqref{problem:mimodec} with $\mathcal{S}=\mathcal{S}_L$ can be equivalently rewritten as \cite{near_ml_decoding}
\begin{equation}\label{problem:t}
\begin{array}{rl}
\displaystyle \min_{\bt} & \bt^\T \bQ \bt+2 \bc^{\T}\bt \\
\mbox{s.t.} & \be^\T\bt_i = 1,~\bt_i \in \left\{0, 1\right\}^{L},~i=1,2,\ldots,n,\end{array}
\end{equation} where $\bQ\in\mathbb{R}^{Ln\times Ln}$ and $\bc\in\mathbb{R}^{Ln}$ are constants depending on the problem inputs $\bH, \br,$ and $\bs$.
The constraints in \eqref{problem:t} with respect to $\bt_i$ (for $i=1,2,\ldots,n$) enforce an assignment, where each agent (corresponding to each of the $n$ users in \eqref{problem:t}) can only choose one and only one item from a given set of items (corresponding to the constellation set $\mathcal{S}_L$). Using the same trick, problem \eqref{problem:mimodec} with $\mathcal{S}=\mathcal{Q}_u$ has a reformulation similar to \eqref{problem:t}.


We can apply the popular negative square penalty \cite{GEMM,xia2010efficient} to the objective function of problem \eqref{problem:t} and obtain the penalty problem
\begin{equation}\label{problem:tpenalty}
\begin{array}{rl}
\displaystyle \min_{\bt} & \bt^\T \bQ \bt+2 \bc^{\T}\bt-\lambda\sum_{i=1}^n\|\bt_i\|_2^2 \\
\mbox{s.t.} & \be^\T\bt_i = 1,~\bm{0}\leq \bt_i\leq \bm{1},~i=1,2,\ldots,n,\end{array}
\end{equation} where $\lambda\geq 0$ is the penalty parameter. The penalty problem in \eqref{problem:tpenalty} can be understood via the following relaxation-tightening procedure. First, problem \eqref{problem:t} is relaxed to obtain problem \eqref{problem:tpenalty} without the negative square penalty term (and the feasible set of \eqref{problem:tpenalty} is the convex hull of the feasible set of \eqref{problem:t}). Then, the negative square penalty term $-\lambda\sum_{i=1}^n\|\bt_i\|_2^2$ is added to the objective function of the relaxed problem in order to minimize/penalize the relaxation gap and tighten the relaxation. Note that any $\bt_i$ that is feasible for problem \eqref{problem:t} is the solution of the problem of minimizing $-\|\bt_i\|_2^2$ over the simplex constraint, which is the intuition why the tighten procedure works. 
%
%

The classic penalty methods usually eliminate a constraint by penalizing it in the objective function. However, the goal of penalty methods here is to transform (or relax) the hard constrained problems into easier constrained subproblems (whose feasible sets are usually convex relaxations of the original ones) and, at the same time, appropriately minimize/penalize the relaxation gap. It can be seen that if the penalty parameter $\lambda$ is greater than the largest eigenvalue of $\bQ$ in \eqref{problem:tpenalty}, then the objective function in \eqref{problem:tpenalty} is strictly concave in its variable. Consequently, the solution of problem \eqref{problem:tpenalty} is achieved on the boundary of the feasible set, which is also the feasible set of problem \eqref{problem:t}. This shows the equivalence of the two problems and the exactness of the penalty function in \eqref{problem:tpenalty}. 
It has been shown in \cite{zhao2021efficient} that in the case where $\mathcal{S}=\mathcal{S}_L,$ problem \eqref{problem:tpenalty} with the diagonal entries of $\bQ$ being set to zero always has a binary solution (even though $\lambda=0$). In this way, the ill-conditioning in the penalty method has been fully eliminated by judiciously exploiting the special structure of the PSK constellation. Compared to the GP algorithm in \cite{So2017} for solving the MIMO detection problem \eqref{problem:mimodec} with $\mathcal{S}=\mathcal{S}_L,$ the algorithm in \cite{zhao2021efficient} is more robust to the choice of the initial point and can generally achieve a better detection performance at the cost of a higher computational time (as it is based on the higher-dimensional problem reformulation \eqref{problem:t}). 

From the above example and discussion, we can conclude that the penalty method is suitable for solving an optimization problem whose constraint can be decomposed into a simple convex constraint and a simple penalty function, i.e., the solution set of minimizing the penalty function over the simple convex constraint is equal to the feasible set of the original problem. 
Therefore, in addition to the above MIMO detection problem whose feasible set can be equivalently characterized by the assignment constraint, 
the penalty method and related ideas can be used to solve problems in much more general setups \cite{GEMM,penaltyproof,xia2010efficient}. First, the constraint in the problem can be more general; e.g., each agent $i$ can choose at most $k\geq 1$ items from a given set (i.e., $\be^\T\bt_i \leq k,~\bt_i \in \left\{0, 1\right\}^{L}$), or different $\bt_i$ and $\bt_j$ need to satisfy some linear constraints like in the permutation matrix case \cite{penaltyproof}. Second, the objective function in the problem needs not be quadratic but can be any smooth function (with a bounded Hessian in the bounded feasible set) \cite{GEMM,xia2010efficient}. Finally, in addition to the negative square penalty, there are other kinds of penalty functions such as the $\ell_q$ penalty \cite{penaltyproof} $\|\bt_i\|_q^{q}\triangleq\sum_{\ell=1}^L t_{i,\ell}^q$ with $q\in(0,1).$

\subsubsection{Remarks}
We conclude this subsection with some remarks on the advantages of applying the penalty methods to solve optimization problems with integer variables.~
The exactness result of the penalty function in the corresponding penalty method serves as a necessary theoretical guarantee that one can focus on the \emph{smooth/continuous} model (e.g., problem \eqref{problem:tpenalty}) of the original \emph{discrete} problem (e.g., problem \eqref{problem:t}) for the purpose of algorithm design. This is important and beneficial for the following reasons. First, it gives more freedom to design algorithms, since smooth/continuous problems are generally easier to handle than discrete problems. Second and more importantly, solving the smooth/continuous problem is more likely to find a high-quality suboptimal solution of the original problem with integer variables because the former has a larger search space in which the homotopy (sometimes called warm-start) technique \cite{homotopy3,penaltyproof} can help bypass bad local solutions. For instance, problem \eqref{problem:tpenalty} can be efficiently solved by the GP algorithm. 

{The recent work \cite{journal_zheyu} proposes a negative $\ell_1$ penalty method for solving the one-bit precoding problem formulated in \cite{CImodel}, which is a special case of symbol-level precoding\cite{CItutorial}. The optimization problem has a nonsmooth objective function and discrete variables. The resulting penalty problem in \cite{journal_zheyu} can be efficiently solved by the single-loop AO algorithm \cite{journal_zheyu,xu2020unified,HiBSA}, where a projection subproblem onto the simplex needs to be solved at each iteration. 
The above negative $\ell_1$ penalty method can also be extended to solve problems with more general discrete constraints such as the quantized constant envelope (QCE) precoding problem \cite{WuISAC2024}. Recent progress on the analysis of diversity order and (asymptotic) symbol error probability on one-bit and QCE precoding can be found in \cite{10017355} and \cite{10268958}, respectively.}


\subsection{Duality-Based Algorithms}\label{subsection:duality}
{L}{agrangian} duality, a principle that (convex) optimization problems can be viewed from either the primal or dual perspective, is a powerful tool for revealing the intrinsic structures of optimization problems arising from wireless communications. 
The celebrated uplink-downlink duality \cite{
	rashid-farrokhi1998TransmitBeamformingPower,
	viswanath2003SumCapacityVector,vishwanath2003DualityAchievableRates
}~
in the power control and beamforming design for wireless communications can be interpreted via Lagrangian duality \cite{song2007NetworkDualityMultiuser, yu2006UplinkdownlinkDualityMinimax}. The uplink-downlink duality refers to the fact that the minimum total power required to achieve a set of SINR targets in the downlink channel is equal to that to achieve the same set of SINR targets in a virtual dual uplink channel, when the uplink and downlink channels are the conjugate transpose of each other. Usually, uplink problems, e.g., the transmit power minimization problems subject to QoS constraints, can be solved efficiently and globally (via the fixed-point iteration algorithm). The uplink-downlink duality thus allows downlink problems to be efficiently solved by solving the relatively easy uplink counterparts. 

The line of algorithms based on Lagrangian duality and uplink-downlink duality generally enjoys two key features. One is its high computational efficiency as the algorithm often only involves simple fixed-point iterations, and the other is its global optimality. 
Therefore, duality-based algorithms have been widely studied for solving power control and beamforming design problems in various communication networks; see \cite{
	wiesel2006LinearPrecodingConic,
	schubert2005IterativeMultiuserUplink, 	
	yu2007TransmitterOptimizationMultiantenna,  
	dahrouj2010CoordinatedBeamformingMulticell}
and the references therein. In this section, we demonstrate how uplink-downlink duality \cite{liu2021UplinkdownlinkDualityMultipleaccess} leads to a duality-based fixed-point iteration algorithm \cite{fan2022EfficientlyGloballySolving} for solving the QoS-constrained joint beamforming and compression problem 
\eqref{problem:qoscompression} in the cooperative cellular network.

\subsubsection{Uplink-Downlink Duality}
The main results in \cite{liu2021UplinkdownlinkDualityMultipleaccess} are several duality relationships between the achievable rate regions of the multiple-access relay channel and the broadcast relay channel, as shown in \cite[Fig.~2]{liu2021UplinkdownlinkDualityMultipleaccess}, under the same sum-power constraint and individual fronthaul constraints. 
A complete summary of the obtained duality relationships can be found in \cite[Table I]{liu2021UplinkdownlinkDualityMultipleaccess}. 
Below we state one of the main results in \cite{liu2021UplinkdownlinkDualityMultipleaccess}. Under the same sum-power constraint and individual fronthaul capacity constraints, the achievable rate region of the multiple-access relay channel implementing Wyner-Ziv compression across the relays and linear decoding at the CP and that of the broadcast relay channel implementing multivariate compression across the relays and linear encoding at the CP are identical. 
The duality result is proved by showing that given the same fixed beamformers $\left\{\bar{\mv{u}}_k\right\}$ and under the same set of rate targets $\left\{R_k\right\}$, the optimal values of the downlink problem \eqref{problem:qoscompression} and its uplink counterpart \eqref{problem:uplink} (at the top of the next page) are the same, where $${\mv{\Gamma} =\sum\limits_{k\in\cK} p_k^{\rm ul}{\mv{h}}_{k}{\mv{h}}_{k}^\dagger+\sigma^2\mv{I}+{\rm diag}(q_{1}^{{\rm ul}},q_{2}^{{\rm ul}},\ldots,q_{M}^{{\rm ul}}),}$$ $p_k^{\rm ul}$ denotes the transmit power of user $k,$ and $q_{m}^{{\rm ul}}$ denotes the variance of the compression noise at relay $m.$
\begin{figure*}[t]\normalsize
	\begin{equation}\label{problem:uplink}\begin{aligned} \displaystyle\mathop{{\min}}_{\{p_k^{\rm ul}\},\{q_m^{\rm ul}\}} &~ \displaystyle\sum_{k\in\cK} p_k^{\rm ul}\\
		\mathrm {s.t.} ~~~ & ~ \sigma^2+\sum\limits_{j\neq k}p_j^{\rm ul}|\bar{\mv{u}}_k^\dagger\mv{h}_j|^2+\sum\limits_{m\in\cM}q_m^{\rm ul}|\bar{u}_{k,m}|^2 -\frac{p_k^{\rm ul}|\bar{\mv{u}}_k^\dagger\mv{h}_k|^2}{2^{R_k}-1}\ {\leq}\ 0,~k\in \mathcal{K},\\ & ~ 2^{C_m} {q_m^{\rm ul}}\  {\geq}\ \mv{\Gamma}^{(m,m)} - \mv{\Gamma}^{(m,1:m-1)} \left(\mv{\Gamma}^{(1:m-1,1:m-1)}\right)^{-1} \mv{\Gamma}^{(1:m-1,m)},~m\in \mathcal{M}, \\ & ~ p_k^{\rm ul} \ge 0,~k\in \mathcal{K}, \\ & ~ q_m^{\rm ul} \ge 0,~m\in \mathcal{M}.\end{aligned} 
	\end{equation}
	\hrulefill
\end{figure*}

\subsubsection{Duality-Based Algorithms}
Now, we review the duality-based algorithm in \cite{fan2022EfficientlyGloballySolving} for solving the joint beamforming and compression problem 
\eqref{problem:qoscompression}.
There are two key steps in the algorithm proposed in \cite{fan2022EfficientlyGloballySolving}. 
In the first step, the seemingly nonconvex problem \eqref{problem:qoscompression} is shown to be equivalent to the convex SDP \cite{liu2021UplinkdownlinkDualityMultipleaccess,fan2022EfficientlyGloballySolving}
\begin{equation}\label{SDR} 
\begin{aligned}
\min_{\{\m V_k\}, \m{Q}} &\quad \sum_{k\in\cK} \tr(\m V_k) + \tr(\m Q)\\
\st~~ &\quad a_k(\left\{\m{V}_k\right\},\m{Q}) \geq 0,\fk{}, \\
&\quad \m{B}_m(\left\{\m{V}_k\right\},\m{Q}) \succeq \m{0},\fm{},\\
&\quad \m{V}_k \succeq \m{0} ,\fk{}, 
\end{aligned}
\end{equation}where 
\begin{align*}
& a_k(\left\{\m{V}_k\right\},\m{Q}) = - \left( \sum_{j\neq k} \tr(\m V_k \m H_k) + \tr(\m{Q}\m{H}_k) + \sigma_k^2 \right) \\ 
&\pushright{+\frac{1}{{\gamma}_k}\tr(\m{V}_k \m{H}_k),}\\ 
& \m{B}_m(\left\{\m{V}_k\right\},\m{Q}) = {\eta}_m \begin{bmatrix}
\m{0}& \m{0} \\
\m{0}& \m{Q}^{(m:M, m:M)}
\end{bmatrix} \\
& \pushright{-~\m{E}_m^\hermitian \left(\sum_{k\in\cK} \v{v}_k \v{v}_k^\hermitian+ \m{Q}\right) \m{E}_m}, \\
& 	\m{H}_k =\v{h}_k\v{h}_k^\dagger,~k\in\cK~\text{and}~{\eta}_m=2^{C_m},~m\in\cM.
\end{align*}
In the above, $\m{E}_m$ denotes the all-zero matrix except the $m$-th diagonal entry being one.
Combining the classic Karush-Kuhn-Tucker (KKT) conditions with the specific structure of problem \eqref{SDR}, we obtain a set of enhanced KKT conditions; see Eqs. (7)--(16) in \cite{fan2022EfficientlyGloballySolving} for details.
The second step is to separate the enhanced KKT conditions into two sets. We first solve equations involving the dual variables. Then, we solve equations involving the primal variables. It is interesting and somewhat surprising that each set can be solved elegantly via a fixed-point iteration algorithm \cite{fan2022EfficientlyGloballySolving}. 

\subsubsection{Remarks}
In order for the above (Lagrangian) duality-based algorithms to globally and efficiently solve the underlying problems, there are generally two technical challenges. The first is to reformulate the problem of interest into an equivalent convex form. 
This step is essential to ensure the global optimality of the algorithm but can be highly nontrivial. The SDR \cite{luo2010} turns out to be a rather useful tool here. The second is to judiciously explore the problem's solution structure and carefully exploit it in algorithm design. Making use of the solution structure is of paramount importance to the computational efficiency of the algorithm.  

Duality is just one way that the KKT conditions can reveal structural insight of an optimization problem.  For solving power allocation and beamforming problems in multi-user communication scenarios, it is always worthwhile to carefully examine the KKT conditions. This holds true not only for convex but also for nonconvex optimization problems. In some cases, the problem structure reflected in the optimality conditions allows us to reformulate the problem into a convex form and to develop an efficient algorithm to find the global optimum. In the domain of power control for multi-user networks, this approach has successfully led to optimal algorithms like the iterative water-filling algorithms for both the multiple-access channel \cite{IWF04} and the interference channel \cite{yu_ginis,yu_raymond}.  

While the sum-rate maximization problem for the interference channel is already known to be NP-hard \cite{luo_complexity,NET-036}, modern multiple-access technologies, such as NOMA, lead to additional challenges \cite{Clerckx21} that include discrete variables to determine the optimal decoding orders. In a typical multi-carrier downlink setup, NOMA can specialize to different variants, including single-carrier NOMA (SC-NOMA), FDMA-NOMA, and hybrid-NOMA \cite{Rezvani22a}. 
In the single-cell setup, the optimum decoding order for each subchannel can be determined easily, and the optimal power control problem turns out to be a convex optimization problem. 
However, one would expect, in the multi-cell setup \cite{Rezvani22}, that the optimization of the decoding order and power allocation will be an NP-hard mixed-integer nonlinear optimization problem \cite[Corollary 1]{Rezvani22}. Nevertheless, in specific cases, for fixed decoding orders, by exploiting the KKT conditions, the optimal power allocation can be computed in closed form \cite[Proposition 1]{Rezvani22}. Then, the optimal decoding order only depends on the total power consumption at the BSs and the search space is significantly reduced, which can potentially lead to 
centralized or distributed algorithms for solving the joint rate and power allocation problem for multi-cell NOMA-assisted downlink networks. 

\section{Problem-Specific Global Optimization}\label{sec:global}

Global optimization algorithms and techniques aim to find global solutions of (hard) optimization problems.
Global optimization distinguishes itself from local or heuristic optimization by its focus on finding a global solution, as opposed to finding a local or suboptimal solution. Global optimization usually is much more difficult and  requires more careful algorithmic design than local optimization.



In this section, we survey recent advances in problem-specific global optimization techniques that are closely related to wireless communication system design.~ 
We do not survey general-purpose global optimization techniques. Before delving into the detailed survey, we first list the reasons why there is a strong interest in computing the global solution of problems, even though the complexity may be high. First, the computed global solution is helpful in assessing the fundamental limits of the performance of the considered wireless communication system. Second, global optimization algorithms provide important benchmarks for performance evaluation of existing local and suboptimal algorithms for the same problem. The above two would be impossible without the global optimality guarantee. Third, fast global optimization algorithms can generate high-quality samples for end-to-end supervised learning, which is covered in Section \ref{sec:learning}.    

This section is organized as follows. We first introduce two most commonly used global optimization frameworks---namely, branch-and-bound (B\&B) \cite{lawler1966branch} and branch-and-cut (B\&C) \cite{padberg1991branch}---in Section \ref{subsection:framework}. These frameworks underlie all the problem-specific global techniques surveyed in this section.
Then, we review two vital components, bounds and cuts, within the above two frameworks. Specifically, we use a class of complex quadratic problems (CQPs) as an example to illustrate how to derive tight bounds in the B\&B framework by employing the SDR technique in Section \ref{subsection:cqp}, use mixed monotonic programming (MMP) as an example to illustrate how to get efficient bounds in the B\&B framework by exploiting the problem structure in Section \ref{subsection:mixed}, and use a class of mixed-integer problems as an example to illustrate how to generate valid cuts in the B\&C framework in Section \ref{subsection:cut}.

\subsection{Introduction to B\&B and B\&C Frameworks}\label{subsection:framework}
In this section, we briefly introduce the B\&B and B\&C frameworks, which are the two most popular frameworks for designing global optimization algorithms. All of the problem-specific global optimization algorithms and techniques to be surveyed in Sections \ref{subsection:cqp} to \ref{subsection:cut} lie within the above two frameworks. 

\subsubsection{B\&B Algorithmic Framework}
The B\&B algorithmic framework is an implicit enumeration procedure that employs a tree search strategy. During the enumeration procedure, the feasible region of unexplored nodes stored in a tree is partitioned into smaller subregions, and children subproblems over the partitioned subregions are explored recursively. Pruning rules are used to eliminate regions of the search space that cannot lead to a better solution. Once all nodes in the tree have been explored, the global solution is found and returned.   


Let us use the following example to illustrate how the B\&B algorithmic framework works. Consider an optimization problem of minimizing the objective function $f$ over the feasible set $\mathcal{Z}.$ The goal of B\&B is to find the global solution of \begin{equation}\label{problemfz}\bz^\ast\in\arg\min_{\bz\in \mathcal{Z}} f(\bz).\end{equation} To do so, B\&B builds a search tree of subproblems (i.e., the problem list $\mathcal{P}$) defined over subsets of the search space in an iterative fashion. More specifically, at each iteration, the algorithm selects a new subproblem defined over the set $\mathcal{Z}'\subset\mathcal{Z}$ to explore from the unexplored problem list $\mathcal{P}:$ 
\begin{itemize}
	\item If a solution $\bz'\in\mathcal{Z}'$ can be found (e.g., by some local optimization/heuristic algorithm) with a better objective value than the best known feasible solution, called the \emph{incumbent solution}, then the incumbent solution is updated to be $\bz'.$
	\item Otherwise, 
	\begin{itemize}
		\item if no solution better than the incumbent solution exists in $\mathcal{Z}'$, then the corresponding subproblem is pruned;
		\item else, children subproblems are generated by partitioning $\mathcal{Z}'$ into a set of subproblems defined over $\left\{\mathcal{Z}_t'\right\}_{t=1}^T$ and the newly obtained subproblems are added to the problem list $\mathcal{P}.$
	\end{itemize}  
\end{itemize}
The above procedure terminates until the problem list $\mathcal{P}$ becomes empty, and the incumbent solution is returned as the global solution.

The steps of a vanilla B\&B algorithm for solving problem \eqref{problemfz} with a continuous variable $\bz$ are summarized as follows:
\begin{itemize} 
				\item [(i)] {Initialize an outer box $\mathcal M_0 \supseteq \mathcal Z$ and a tolerance $\epsilon>0$, set the incumbent solution $\bar{\mat{z}}^0$ and value $\gamma_0 = f(\bar{\mat{z}}^0)$.}
				\item [(ii)] \label{alg:bb:selection} {Select box $\mathcal M_r$ that has the smallest bound $\beta(\mathcal M_r)$ (of the objective value of problem \eqref{problemfz}) for branching, i.e., $r=\arg \min_{j} \beta(\mathcal M_j).$}%
				\item [(iii)]  {Bisect $\mathcal M_r.$}
				\item [(iv)]  Reduce new boxes (optional).
				\item [(v)]  {Compute bound $\beta(\mathcal M) \le \min_{\mat{z}\in\mathcal M\cap\mathcal Z} f(\mat{z})$ for all new boxes $\mathcal M$.}
				\item [(vi)]  {Update the incumbent solution $\bar{\mat{z}}^r$ and value $\gamma_r = f(\bar{\mat{z}}^r)$.}
				\item [(vii)]  Delete infeasible ($\mathcal M\cap\mathcal Z = \emptyset$) and suboptimal (i.e., $\beta(\mathcal M) \ge \gamma_r+\epsilon$) new boxes.
				\item [(viii)] Terminate if no box is left or $\min_{\mathcal M} \beta(\mathcal M) \ge \gamma_r$. {Then, $\bar{\mat{z}}^r$ is a global $\epsilon$-optimal solution.}
			\end{itemize} 

Due to their modularity, B\&B algorithms are very flexible. The design choices comprise the subdivision procedure, the selection step, the bounding step, the reduction procedure, the feasibility check, and the finding of a feasible point. They need to be adapted to the properties and context of the considered global optimization problem. For a detailed overview of B\&B algorithms, please refer to the survey paper \cite{lawler1966branch} and the textbook \cite{wolsey1999integer}. 

Several remarks on the above B\&B algorithmic framework are in order. First of all, it is simple to see that the algorithm actually implicitly enumerates all the feasible solutions via a tree search strategy and hence the global optimality of the returned solution is guaranteed. However, the worst-case complexity of the B\&B algorithmic framework is generally exponential. In particular, the worst-case complexity of any B\&B algorithm in the above framework is ${\mathcal{O}}\left(CT^d\right)$ \cite{lawler1966branch}, where $T$ is the maximum number of generated children at any node, $d$ is the length of the longest path from the root of the tree to a leaf, and $C$ is the upper bound on the complexity of exploring/solving each subproblem. The length $d$ here usually depends on the given error tolerance $1/\epsilon.$ 

Second, there are usually two phases of any B\&B algorithm. The first is the search phase, where the algorithm seeks the (nearly) global solution. The second is the verification phase, in which the algorithm verifies that the found incumbent solution (in the first phase) is indeed (nearly globally) optimal. Note that an incumbent solution cannot be proven to be globally optimal if the unexplored problem list is nonempty. The verification of the global optimality of the incumbent solution is the price that needs to be paid in global optimization, which is unnecessary in local optimization. 


Last but not least, the pruning rule employed in the B\&B algorithm plays an essential role in its computational efficiency. In particular, if there is no solution better than the incumbent solution, then the corresponding subproblem can be safely pruned from the problem list and all of its children problems do not need to be explored. Therefore, an efficient pruning rule is helpful in reducing the total number of explored subproblems and accelerating the verification process.~
The most common way to prune is to compute a lower bound on the objective function value of each subproblem and use it to prune those subproblems whose lower bound is worse than the objective value at the incumbent solution. 

\subsubsection{B\&C Algorithmic Framework}
B\&C is another widely used global optimization algorithmic framework for solving linear integer programs.\footnote{For simplicity of presentation, we use the linear integer program as an example here. The B\&C algorithmic framework can be used to solve problems that involve mixed-integer variables and are not necessarily linear programs.} A key concept in B\&C is the cutting plane, also called valid cut or valid inequality. The cutting plane is defined as ``a linear constraint that can be added to an integer program to tighten the feasible region without removing any integer solutions'' \cite{lawler1966branch} (including the optimal solution of the original problem). The B\&C algorithmic framework often consists of two steps: using cutting planes to tighten the LP relaxations and running B\&B. More specifically, the B\&C algorithm first iteratively generates and adds cutting planes to the LP relaxation of the integer program and starts the B\&B process at some point (e.g., when the number of generated cuts is too large to be added or when it is computationally expensive to generate new cuts). Note that cutting planes are generated and added gradually based on the solution of the current LP relaxation problem. If the solution is already integer, then it must be optimal to the original problem; otherwise, new cutting planes are generated to exclude the current fractional solution to tighten the LP relaxation.~
It is evident that the efficiency of the B\&C algorithm considerably relies on the efficiency and quality of the generated cutting planes. 

In view of their central roles in global optimization algorithms, 
we survey some recent advances in problem-specific pruning rules and cutting planes~
for wireless communication system design. In particular, we review recent advances in deriving high-quality lower bounds for a class of CQPs by employing the SDR technique in Section \ref{subsection:cqp}, efficiently computing lower bounds for the MMP problems by using the problem's monotonicity structure in Section \ref{subsection:mixed}, and generating valid cutting planes for a class of mixed-integer problems in Section \ref{subsection:cut}.

\subsection{SDRs for a Class of CQPs}\label{subsection:cqp}
As discussed in Section \ref{subsection:framework}, an efficient pruning rule in the B\&B algorithm is of great importance to the algorithm's computational efficiency, and the most common way to prune is to estimate a lower bound on the (optimal) objective value of each subproblem. Since convex optimization problems possess favorable theoretical and computational properties and efficient and mature solvers, the lower bound on the objective value of each subproblem is often computed by solving a convex relaxation of the corresponding subproblem. The quality of the lower bound depends on the tightness of the convex relaxation. Designing convex relaxations that provide valid lower bounds with satisfactory tightness is an important research topic in global optimization. In this subsection, we review several SDRs for a class of nonconvex CQPs developed in \cite{9180096}.

We consider the following general CQP as in \cite{9180096}:
\begin{equation}\label{CQP}
\begin{aligned}
\min_{\bx\in \mathbb{C}^{n}}~&\bx^{\dag} \bQ \bx\\
\textrm{s.t.}~~&\ell_i \leq {|}x_i {|} \leq u_i,~ i=1,2, \ldots, n, \\
&\arg(x_i) \in \mathcal{A}_{i}, ~ i=1,2, \ldots, n,
\end{aligned}
\end{equation}
where $\bx=[x_1,x_2,\ldots,x_n]^\T \in \mathbb{C}^{n}$ is the $n$-dimensional complex (unknown) variable; 
$\ell_i$ and $u_i$~($i=1,2,\ldots,n$) satisfying $u_i\geq \ell_i\geq 0$ are $2n$ real numbers; $\mathcal{A}_{i}$~($i=1,2,\ldots,n$)~ is either an interval of the form $[\underline{\theta}_{i},\bar{\theta}_{i}]\subseteq [0, 2\pi)$ or a set of discrete points of the form $\{\theta_{i}^1,\theta_{i}^2,\ldots,\theta_{i}^M\}\subseteq [0, 2\pi)$; and $\arg(\cdot)$ denotes the argument of a complex number. Many problems arising from wireless communications and signal processing can be formulated as problem \eqref{CQP} with special choices of $\ell_i,~u_i,~\text{and}~\mathcal{A}_{i}$~($i=1,2,\ldots,n$); see \cite[Section II]{9180096} and the references therein. For example, the argument constraints are useful for specifying the phases of the symbols to be detected in the MIMO detection problem, or for specifying regions for branching in B\&B.

The difficulty of developing an SDR for CQP \eqref{CQP} that can provide a good lower bound lies in its last argument constraint. Indeed, an SDR for CQP \eqref{CQP} with the argument constraint dropped is also an SDR for the problem itself. However, the bound provided by the above naive SDR is generally not tight enough. The idea of developing an enhanced SDR for CQP \eqref{CQP} in \cite{9180096} is to represent the complex variable in polar coordinates and derive valid inequalities by exploiting the special structure of the argument constraint under the polar-coordinate representation.  
More specifically, we introduce the polar-coordinate representation of each variable $x_i=r_i \exp(\textrm{i}\theta_i)$ and a lifted matrix $\bX=\bx\bx^{\dag}\in \mathbb{C}^{n\times n}.$ 
Then, for each $i=1,2,\ldots,n,$ we get
\begin{equation}\label{CQeq5}
X_{ii}=r^2_{i}~\text{and}~\theta_i\in \mathcal{A}_{i}.
\end{equation} We now relax the two types of nonconvex constraints in \eqref{CQeq5} in order to obtain a convex relaxation of CQP \eqref{CQP}. 

First, for each $i=1,2,\ldots,n,$ consider the nonconvex set
{$$\mathcal{S}_{i}:=\left\{(X_{ii},r_i)\mid X_{ii}=r_i^2,~r_i\in [\ell_i,u_i]\right\}.$$}It has been shown in \cite{chen2017spatial} that the convex hull\footnote{The convex hull of a set is the smallest convex set that contains the given set.} of $\mathcal{S}_{i}$ can be represented as 
\begin{equation}\label{FBi}\textrm{Conv}(\mathcal{S}_i)=\left\{(X_{ii},r_i) \left| \begin{array}{@{}lll} X_{ii} \geq r_i^2,\\X_{ii} -(\ell_i+u_i)r_i +\ell_i u_i\leq 0\end{array}\right.\!\!\!\right\}.\end{equation}
%
%
%
Second, 
consider the nonconvex set
\begin{equation}\label{nonconvexsetAi}{\mathcal{T}_{i}:=\left\{(x_i ,r_i)\,|\,x_i =r_i e^{\textsf{i}\theta_i},~\theta_i\in \mathcal{A}_i,~r_i\geq0\right\}}.\end{equation} 
 We have the following results on the convex hull of $\mathcal{T}_{i}$ \cite{9180096}.~
In particular, for the continuous case where $\mathcal{A}_{i}=[\underline {\theta}_{i},\bar\theta_{i}]\subseteq [0, 2\pi),$ we have
\begin{equation}\label{eqgij}
\textrm{Conv} (\mathcal{T}_{i})=\left\{(x_{i},r_{i})\,\left|\,\begin{array}{@{}lll}a_{i} \mathrm{Re}\left(x_{i}\right)+ b_{i} \mathrm{Im}\left(X_{i}\right)\\[3pt]\geq c_{i}r_{i},\,|x_{i}|\leq r_{i}\end{array}\right.\!\!\!\right\},
\end{equation}
where
\begin{equation*}\label{thm2eq1}
\begin{aligned}&a_{i}=\cos\left(\frac{\underline {\theta}_{i}+\bar\theta_{i}}{2}\right),~b_{i}=\sin\left(\frac{\underline {\theta}_{i}+\bar\theta_{i}}{2}\right),\\ &c_{i}=\cos\left(\frac{\bar\theta_{i}-\underline {\theta}_{i}}{2}\right);\end{aligned}
\end{equation*}
for the discrete case where $\mathcal{A}_{i}=\{\theta_{i}^1,\theta_{i}^2,\ldots,\theta_{i}^M\}$ with $0\leq \theta_{i}^1 < \theta_{i}^2 < \cdots<\theta_{i}^M<2\pi,$
we have
\begin{equation}\label{thm2eq3}
\textrm{Conv}(\mathcal{T}_{i})=\left\{(x_{i},r_{i})\,\left|\,\begin{array}{@{}lll} a_{i}^m \mathrm{Re}\left(x_{i}\right)+ b_{i}^m \mathrm{Im}\left(x_{i}\right)\\[3pt]\leq c_{i}^m r_{i},\,m=1,2,\ldots,M \end{array}\right.\!\!\!\right\},
\end{equation}
where $\theta_{i}^{M+1}=\theta_{i}^{1}+2\pi$ and
\begin{equation*}\label{thm2eq2}
\begin{aligned}
&a_{i}^m=\cos\left(\frac{\theta_{i}^m+\theta_{i}^{m+1}}{2}\right),~b_{i}^m=\sin\left(\frac{\theta_{i}^{m}+\theta_{i}^{m+1}}{2}\right),\\
&c_{i}^m=\cos\left(\frac{\theta_{i}^{m+1}-\theta_{i}^{m}}{2}\right).
\end{aligned}
\end{equation*}
The valid cuts shown in \eqref{eqgij} and \eqref{thm2eq3} have been named argument cuts in \cite{lu2017efficient,lu2018argument} because they exploit the structure of the argument constraint. An illustration of the argument cuts in \eqref{eqgij} and \eqref{thm2eq3} can be found in \cite[Fig. 1]{9180096}.
Putting all of the above together, we obtain the enhanced SDR for CQP \eqref{CQP} as follows \cite{9180096}:
\begin{equation}\label{ECSDP}
\begin{aligned}
\min_{\bX,\,\br,\,\bx}~& \tr(\bQ\bX)\\
\textrm{s.t.}~~&\ell_i \leq  r_{i} \leq u_i,~ i=1,2,\ldots, n, \\
& (X_{ii},r_{i})\in \textrm{Conv}(\mathcal{S}_{i}),~i=1,2,\ldots,n,\\
& (x_{i},r_{i}) \in \textrm{Conv} (\mathcal{T}_{i}), ~i=1,2,\ldots,n,\\
&\bX\succeq \bx\bx^{\dagger}.
\end{aligned}
\end{equation} 

The SDR in \eqref{ECSDP} is closely related to other types of SDRs in the literature, e.g., \cite{chen2017spatial,lu2019tightness,xu2023new}. 
An SDR for an even more general CQP than \eqref{CQP} is provied in \cite[Section 3]{xu2023new}. 
As an extreme case where CQP \eqref{CQP} reduces to the MIMO detection problem \eqref{problem:mimodec} with $\mathcal{S}=\mathcal{S}_L,$ the corresponding SDR in \eqref{ECSDP} then reduces to (CSDP2) in \cite{lu2019tightness}. Thanks to the argument cuts in \eqref{thm2eq3}, (CSDP2) is shown to be tight\footnote{The tightness of the SDR here means that the gap between the SDR and problem \eqref{problem:mimodec} is zero and the SDR recovers the true vector of transmitted signals.} for the general case where $L\geq 3$ if the condition $\lambda_{\min}(\bH^\dagger\bH)\sin(\pi/L)\geq \|\bH^\dagger\bz\|_1$ holds true. The above sufficient condition for (CSDP2) in \cite{lu2019tightness} to be tight is intuitive. It basically says that if the channel matrix $\bH$ is well conditioned and if the constellation level $L$ and the level of the noise $\bz$ are not too large, then solving the corresponding SDR can find the global solution of problem \eqref{problem:mimodec}, which is also the true vector of transmitted signals $\bs$ in \eqref{rHv}.   

As a final remark on the use of the argument cuts in \eqref{eqgij} and \eqref{thm2eq3} and the SDR in \eqref{ECSDP} to deal with nonconvex CQPs, 
we note that significant efforts have been made in the literature 
to design efficient global optimization algorithms for this class of problems. 
Among them, \cite{lu2017efficient} and \cite{9180096} are most closely related to wireless communication applications. 
In particular, the argument cuts \eqref{eqgij} have been embedded in the B\&B framework in \cite{lu2017efficient} to globally solve the NP-hard single-group multicast problem; efficient B\&B algorithms based on the argument cuts \eqref{eqgij} and \eqref{thm2eq3} have been developed in \cite{9180096} to solve a class of nonconvex CQPs with signal processing and wireless communication applications. When applied to solve the MIMO detection problem \eqref{problem:mimodec} with $\mathcal{S}=\mathcal{S}_L,$ the proposed global optimization algorithm in \cite{9180096} significantly
outperforms the state-of-the-art tailored global optimization algorithm in the
hard cases (where the number of inputs and outputs is equal or the SNR is low).  

\subsection{Mixed Monotonic Programming}\label{subsection:mixed}

In this subsection, we continue with the discussion on computing a valid lower bound on the optimal objective value of each subproblem in order to do efficient pruning in the B\&B algorithmic framework. In Section \ref{subsection:cqp}, lower bounds are derived by developing convex relaxations of the corresponding subproblems that are as tight as possible. However, the computational cost of solving the convex relaxation problems (e.g., the SDP in \eqref{ECSDP}) might be high. Different from the previous subsection, the goal of this subsection is to derive the lower bound on the optimal objective value of each subproblem with a low computational cost to achieve high computational efficiency in computing the lower bound. This is possible when the problem at hand has certain special structure, e.g., monotonicity and mixed monotonicity. In particular, we use the MMP problem \cite{matthiesen2020mixed} as an example to illustrate how to exploit the monotonicity structure in MMP problems to obtain an easily computable lower bound. The results in this subsection are mainly from \cite{matthiesen2020mixed}.

We use problem \eqref{problemfz} as our example again, where the objective function $f:~\mathbb{R}^n \rightarrow \mathbb{R}$ is assumed to be continuous and the feasible set $\mathcal{Z}$ is assumed to be compact (i.e., closed and bounded). A given function $F:~\mathbb{R}^n\times \mathbb{R}^n \rightarrow \mathbb{R}$ is called a mixed monotonic function if it satisfies
\begin{equation}
\begin{aligned}
&F(\bz,\bw) \leq F(\bz',\bw),~\forall~\bz\leq \bz',\\
&F(\bz,\bw) \geq F(\bz,\bw'),~\forall~\bw\leq \bw'.
\end{aligned}
\end{equation} 
Moreover, problem \eqref{problemfz} is said to be an MMP problem if its objective function $f$ satisfies $f(\bz)=F(\bz,\bz)$ for all $\bz,$ where $F(\cdot,\cdot)$ is some mixed monotonic function defined in (a set containing) its feasible region. For the MMP problem, the lower bound can be easily obtained over rectangular sets. To be specific, let $\mathcal{B}=[\v{\ell},\v{u}].$ Then
\begin{equation*}
\min_{\bz\in \mathcal{B} \cap\mathcal{Z}} f(\bz) \geq \min_{\bz\in \mathcal{B}} F(\bz,\bz) \geq \min_{\bz,\bw\in \mathcal{B}} F(\bz,\bw)\geq F(\v{\ell},\v{u})
\end{equation*} gives a lower bound on the optimal objective value of the subproblem defined over $\mathcal{B} \cap\mathcal{Z},$ i.e., $\min_{\bz\in \mathcal{B} \cap\mathcal{Z}} f(\bz).$ 

Below we apply the MMP framework to globally solve the sum-rate maximization problem in the $K$-user interference channel. In fact, all we need to do is to find an MMP representation of the objective function of the interested problem and all the others are standard B\&B components.\footnote{There are two implementations of MMP including the complete B\&B algorithm available. The first is a C++ implementation available at \url{https://github.com/bmatthiesen/mixed-monotonic}. The second provides a Python framework for disciplined programming with MMP and B\&B, which can be easily applied and extended \url{https://github.com/Ciaoc/mmp_framework}.}
%
%
%
The sum-rate maximization problem takes a similar form as problem \eqref{problem:scheduling} but with all scheduling variables $\left\{\kappa_i\right\}$ being given and fixed. For ease of presentation, we explicitly write down the rate expression of user $k$ as follows:
\begin{equation}\label{ratek}
r_k(\bp)=\log_2\left(1+\frac{\alpha_kp_k}{\sigma_k^2+\sum_{j\in\cK} \beta_{kj}p_j}\right),
\end{equation} where $p_k$ is the transmit power of user $k,$ $\alpha_k\geq 0$ is the gain of the intended channel, and $\beta_{kj}\geq 0$ is the gain of the unintended channel for $j\neq k$. In the above, $\beta_{kk}\geq 0$ is included for modeling the self-interference or hardware impairment. It is simple to verify that 
\begin{equation}\label{ratekmmp}R_k(\bp,\bq)=\log_2\left(1+\frac{\alpha_kp_k}{\sigma_k^2+\beta_{kk}p_k+\sum_{j\neq k} \beta_{kj}q_j}\right)\end{equation} is an MMP representation of $r_k(\bp)$ in \eqref{ratek}. With the lower bounds provided by the above representation, the sum-rate maximization problem can now be globally solved by utilizing the MMP framework \cite[Algorithm 1]{matthiesen2020mixed}. 

It is worth mentioning the other existing global optimization algorithms for solving the same sum-rate maximization problem and comparing them with the MMP framework. One global algorithm is MAPEL \cite{qian2009mapel}, which approximates the original problem from outside by means of the polyblock algorithm (PA) \cite{tuy2000monotonic}. Another one is to use the monotonic optimization framework \cite{tuy2000monotonic}, which first rewrites $r_k(\bp)$ in \eqref{ratek} into difference-of-monotonic (DM) functions 
\begin{multline}
r_k^{\text{DM}}(\bp) =  \log_2\left({\alpha_kp_k+\sigma_k^2+\sum_{j\in\cK} \beta_{kj}p_j}\right) \\  -\log_2\left({\sigma_k^2+\sum_{j\in\cK} \beta_{kj}p_j}\right)\label{ratekdm}
\end{multline}
and applies the B\&B algorithm to solve the reformulated DM problem. It is interesting and somewhat surprising that the MMP bound is always better than the DM bound when they are applied to solve the sum-rate maximization problem. 
Again, the convergence speed of the B\&B algorithm depends strongly on the quality of the bounds, and tighter bounds generally lead to faster global optimization. This explains why the MMP algorithm \cite{matthiesen2020mixed} outperforms the B\&B algorithm equipped with the DM bound \cite{tuy2000monotonic} for solving the sum-rate maximization problem.

We conclude this subsection with further remarks on the MMP framework and representation. First, the MMP framework covers many existing problem formulations and frameworks as special cases, among which the most well-known one is the so-called DM programs, i.e., problem \eqref{problemfz} where the objective function $f$ can be written as $f=f^+-f^-$ with both of $f^+$ and $f^-$ being nondecreasing functions. An MMP representation of the objective function in DM programs is $F(\bz,\bw)= f^+(\bz)-f^-(\bw).$ Second, the MMP representation is not unique. In particular, if $F(\bz,\bw)$ is an MMP representation of $f(\bz),$ then 
$$\tilde F(\bz,\bw)=F(\bz,\bw)+\sum_{i}\left(z_i-w_i\right)$$
is also an MMP representation of $f(\bz).$ However, different MMP representations will lead to different bounds. A subtlety here is how to choose the MMP representation that leads to the tightest bound. Finally, we refer the reader to \cite{matthiesen2020mixed} for more detailed discussions on the MMP framework, including the functional operations that preserve the mixed monotonic properties and more application examples in wireless communication system design.

\subsection{Valid Cuts for Mixed-Integer Problems} \label{subsection:cut} 
Generating valid inequalities to strengthen the relaxation of a mixed-integer problem (MIP) is generally nontrivial, as it requires a judicious exploitation of the problem's special structure in order to tighten the corresponding relaxation yet without excluding the true solution. For mixed linear integer programs (MILPs), many different types of valid inequalities have been investigated in the literature. In particular, Gomory cuts \cite{gomory1958outline} have been extensively studied and included in all modern MIP solvers (e.g., Gurobi, CPLEX, and SCIP) due to its capability of significantly improving the solvers' practical numerical performance. Surveys on valid inequalities for general MIPs can be found in \cite{marchand2002cutting,cornuejols2008valid}. In this subsection, we use MIPs coming from wireless communication system design as examples to illustrate how to exploit structural information in the corresponding problems to generate valid inequalities. The results in this subsection are mainly from \cite{ni2018mixed}. 

Consider the JABF problem \eqref{problem:jabf}. Compared to relaxing both the binary variables and the nonconvex quadratic constraints in \eqref{problem:jabf} as in \cite{xu2014semidefinite}, an arguably better way is to keep the binary variables unchanged and apply the SDR to the nonconvex quadratic constraints, which leads to the following mixed-integer SDR:
\begin{equation}\label{problem:jabfsdr}\begin{aligned}
\displaystyle \min_{\tilde\bW,\,\bm{\beta}} &~ \tr(\tilde\bW) \\
\mbox{s.t.}~ &~ \displaystyle \tr(\tilde\bH_k\tilde\bW) \geq \beta_k,~k\in\cK,\\
&~\sum_{k\in\cK} \beta_k \geq \hat K,~\beta_k\in\left\{0,1\right\},~k\in\cK,\\
&~\tilde\bW\succeq \bm{0},
\end{aligned}\end{equation} 
where  $\tilde\bW=\begin{bmatrix}
\bW ~&\bw\\
\bw^{\dagger} ~&1\\
\end{bmatrix}\in \mathbb{S}_+^{M+1}$ and $\tilde\bH_k=\begin{bmatrix}
\bh_k\bh_k^\dagger ~&\bm{0}\\
\bm{0} ~&1\\
\end{bmatrix}\in \mathbb{S}_+^{M+1}$ for all $k\in\cK$. In the following, we focus on designing a B\&C algorithm for globally solving problem \eqref{problem:jabfsdr}. Then, based on the solution we can apply the Gaussian randomization procedure to obtain a feasible solution to the JABF problem \eqref{problem:jabf} with a provable guarantee \cite[Theorem 2]{ni2018mixed}.

The first step in the design of a B\&C algorithm for solving mixed-integer SDR \eqref{problem:jabfsdr} is to find a relaxation of the problem. This can be easily achieved due to the following fact: For any $\mathcal{T}\subset \mathbb{S}_+^{M},$ the constraint
\begin{equation}\label{inequalitysoc}\tr(\bT\bW)-\bw^\dagger \bT\bw\geq 0,~\bT\in\mathcal{T}\end{equation} is an outer approximation of the last constraint $\tilde\bW\succeq \bm{0}$ in \eqref{problem:jabfsdr}. As such, for any given $\mathcal{T}\subset \mathbb{S}_+^{M},$ the problem
\begin{equation}\label{problem:jabfsdr2}\begin{aligned}
\displaystyle \min_{\tilde\bW,\,\bm{\beta}} &~ \tr(\tilde\bW) \\
\mbox{s.t.}~ &~ \displaystyle \tr(\tilde\bH_k\tilde\bW) \geq \beta_k,~k\in\cK,\\
&~\sum_{k\in\cK} \beta_k \geq \hat K,~\beta_k\in\left\{0,1\right\},~k\in\cK,\\
&~\tr(\bT\bW)-\bw^\dagger \bT\bw\geq 0,~\bT\in\mathcal{T}
\end{aligned}\end{equation} is a relaxation of problem \eqref{problem:jabfsdr}. Moreover, with the decomposition $\bT=\bU\bU^\dagger$ at hand, each constraint in \eqref{inequalitysoc} 
can be expressed as the SOC constraint
$$\left\|\begin{bmatrix}
1-\tr(\bT\bW)\\
2\bU^\dagger \bw\\
\end{bmatrix}\right\|\leq 1+\tr(\bT\bW).$$ If the chosen set $\mathcal{T}$ in \eqref{problem:jabfsdr2} is a finite set of $\mathbb{S}_+^{M},$ then the problem is a mixed-integer SOCP, which can be efficiently solved (e.g., by Gurobi).

The second step in the design of a B\&C algorithm for globally solving mixed-integer SDR \eqref{problem:jabfsdr} is to iteratively generate valid inequalities and add them in \eqref{problem:jabfsdr2} to tighten the relaxation. More specifically, after obtaining an optimal solution $(\tilde \bW_{\mathcal{T}},\bm{\beta}_{\mathcal{T}})$ of problem \eqref{problem:jabfsdr2}, we solve problem \eqref{problem:jabfsdr} with $\bm{\beta}=\bm{\beta}_{\mathcal{T}},$ i.e., 
\begin{equation}\label{problem:jabfsdrinner}\begin{aligned}
\displaystyle \min_{\tilde\bW} &~ \tr(\tilde\bW) \\
\mbox{s.t.}~ &~ \displaystyle \tr(\tilde\bH_k\tilde\bW) \geq [\beta_{_{\mathcal{T}}}]_k,~k\in\cK,\\
&~\tilde\bW\succeq \bm{0},
\end{aligned}\end{equation}which is an inner approximation of problem \eqref{problem:jabfsdr}. The SDP \eqref{problem:jabfsdrinner} plays a central role in the design of the B\&C algorithm, as solving the SDP either verifies the global optimality of the incumbent solution $\bm{\beta}_{\mathcal{T}}$ or generates a valid inequality to eliminate $\bm{\beta}_{\mathcal{T}}$ and tighten the relaxation if $\bm{\beta}_{\mathcal{T}}$ is not optimal. In particular, we check the integrality gap (i.e., the difference between the optimal values) of problems \eqref{problem:jabfsdrinner} and \eqref{problem:jabfsdr2}. If the gap is zero, then $(\tilde \bW_{\mathcal{T}},\bm{\beta}_{\mathcal{T}})$ is the optimal solution of problem \eqref{problem:jabfsdr}; otherwise a valid inequality $\bT$ is generated based on the dual information of the SDP \eqref{problem:jabfsdrinner} and added to $\mathcal{T}$ to strengthen the relaxation problem \eqref{problem:jabfsdr2} \cite[Proposition 1]{ni2018mixed}.
Specifically, if the SDP \eqref{problem:jabfsdrinner} is feasible, then let $$\tilde \bT=\begin{bmatrix}
\bT ~&\bt\\
\bt^{\dagger} ~&t\\
\end{bmatrix}\in \mathbb{S}_+^{M+1}$$ be the optimal Lagrange multiplier corresponding to the constraint $\tilde\bW\succeq \bm{0};$ if not, then there exists a $\bm{\lambda}^*=[\lambda_1^*,\lambda_2^*,\ldots,\lambda_K^*]^\T$ such that $\tilde \bT=\sum_{k} \lambda_k^*\tilde\bH_k$ and $(\bm{\lambda}^*)^\T\bm{\beta}{_{\mathcal{T}}}<0.$ 


Since the total number of feasible binary solutions of problem \eqref{problem:jabfsdr} is finite and one binary solution is eliminated at each iteration, the above B\&C algorithm will return an optimal solution of problem \eqref{problem:jabfsdr} in a finite number of iterations. 
In the above algorithm, problems \eqref{problem:jabfsdrinner} and \eqref{problem:jabfsdr2} are two important subproblems, which need to be solved at each iteration and are closely related. In particular, solving problem \eqref{problem:jabfsdr2} can return a subset of users to serve based on which problem \eqref{problem:jabfsdrinner} is defined; solving problem \eqref{problem:jabfsdrinner} is to find the multicast beamforming vector to support the selected subset of users and solving it either verifies the optimality of the selected subset of users or returns a valid inequality for problem \eqref{problem:jabfsdr2} that cuts off the current suboptimal or infeasible solution. In the above algorithm, problem \eqref{problem:jabfsdr2} acts like a leader while problem \eqref{problem:jabfsdrinner} acts like a follower. Therefore, these two problems are named the master and follower problems in the literature. 
The above technique and idea can be extended to solve many other problems involving mixed-integer variables. For instance, an efficient global algorithm has been proposed for solving large-scale mixed-integer network slicing (NS) problems \cite{Zhang2017}. The algorithm proceeds by decomposing the original NS problem into the relatively easy function placement and traffic routing subproblems and iteratively solving these subproblems using the information obtained from each other.

%
%
%



\section{Distributed Optimization and Federated Learning}\label{sec:distributed}

In the past decade,  distributed optimization methods have garnered significant attention in wireless communications \cite{tychogiorgos2013non,lin2020distributed}.  
These methods provide the potential for scalability and efficiency by allowing multiple entities to solve global optimization problems using localized computations collectively. For example, in multi-cell coordinated systems or cell-free MIMO systems, the BSs collaborate to mitigate inter-cell interference, so as to improve the QoS of cell-edge users.  In contrast to centralized optimization methods, which require all users' CSI to be pooled at a central node,  distributed optimization methods can provide certain advantages such as reducing the backhaul information exchange \cite{boukhedimi2017coordinated,komulainen2013effective} and providing robustness against time-varying environments \cite{maros2017admm}.  Since distributed optimization methods can usually be implemented in parallel,  they are also low-complexity alternatives (in terms of computational time) for solving some large-scale wireless communication system design problems.  {Below, in Section \ref{sec: dist_opt part 1}, we present two distributed optimization methods---namely, the dual decomposition method \cite{falsone2017dual} and the alternating direction of method of multipliers (ADMM) \cite{boyd2011distributed}---and also their variants \cite{deng2017parallel,zhang2020proximal,chang2016asynchronous,hong2016convergence}, which are often adopted in distributed wireless designs. We then demonstrate their  applications in multi-cell coordinated beamforming.
	
	Distributed optimization methods will play an important role in future wireless networks.  For example,  edge intelligence,  which leverages the capabilities of AI at the network's edge,  is considered as a pivotal element of next-generation wireless networks \cite{shi2020communication}.  In intelligent edge networks,  AI services are not limited to centralized data centers but extend to edge nodes, enabling real-time decision-making and latency reduction. This is a vital technology for emerging applications like autonomous vehicles and augmented reality,  which demand ultra-low latency and high reliability.  Federated learning (FL) is a key enabler of edge intelligence.  FL is a distributed optimization methodology employed in wireless networks for collaborative AI model training across distributed edge devices.  Compared to the cloud-based centralized learning paradigm,  FL does not require users' data to be collected at the cloud center and therefore provides enhanced data privacy and security at the network's edge \cite{wang2023batch}.  However,  efficient implementation of FL is challenging because the learning process would involve iterative communications between the edge server and a massive number of user clients.  Besides, the local data owned by the clients may have different statistical distributions,  which can greatly degrade the learning performance \cite{wang2022quantized}.   In Section \ref{sec: dist_opt_part 2},  from the distributed optimization perspective,  we review the seminal FL algorithm FedAvg \cite{mcmahan2017communication} and present {its variants \cite{zhang2021fedpd,wang2023beyond}} that aim to improve the learning performance in heterogeneous edge networks.}

%
%

\subsection{Decomposition Methods}
\label{sec: dist_opt part 1}
\subsubsection{Dual Decomposition, ADMM, and Their Variants}
Dual decomposition \cite{falsone2017dual,boyd2011distributed} is a simple method to obtain a decentralized algorithm for convex optimization problems with separable structures.  Specifically,  
consider the problem
\begin{equation}\label{separable prob}\begin{aligned}
\min_{\bx} ~ &\sum_{i=1}^n f_i(\bx_i) \\
\text{s.t.~} ~ & \bx_i \in \mathcal{X}_i, \, i=1,2,\dots,n,  \\
&  \bA \bx = \bb, 
\end{aligned}\end{equation}
where $\{ f_i \}$ are convex functions,  $\{ \mathcal{X}_i \}$ are given convex sets, $\bA=[\bA_1,\bA_2,\ldots, \bA_n]$ are given matrices, and $\bx=[\bx_1^\T, \bx_2^\T, \dots, \bx_n^\T]^\T$ is the design variable.  
Both the objective function $\bx \mapsto \sum_{i=1}^n f_i(\bx_i)$ and the constraint $\bA\bx= \sum_{i=1}^n \bA_i \bx_i = \bb$ are separable with respect to  $ \{ \bx_i \} $.  The dual decomposition method aims to exploit the separable structure of problem \eqref{separable prob} via its Lagrangian dual.  Specifically,  the Lagrangian dual of problem  \eqref{separable prob}  decouples the problem into \( n \) individual subproblems,  each involves only a single variable \( \bx_i \) and its associated function \( f_i \) and constraint matrix $\bA_i$.  This enables parallel or distributed processing of each subproblem, followed by a coordination step to ensure that the global constraint is satisfied.  The obtained algorithm can be summarized as follows:
\begin{itemize}
	\item [(i)] \emph{Initialization:} Set ${\bm \lambda}^{0}={\bf 0}$ (initial Lagrange multiplier).
	\item [(ii)] \emph{Repeat until convergence:}
	\begin{itemize}
		\item For each \( i=1,2,\ldots, n \), solve the local problem
		\[
		\bx_i^{r+1} = \arg\min_{\bx_i \in \mathcal{X}_i} \left\{ f_i(\bx_i) + ({\bm \lambda}^{r})^\T \bA_i \bx_i \right\}.
		\]
		\item Update the Lagrange multiplier via
		\begin{align}\label{dd: dual ascent}
		{\bm \lambda}^{r+1} = {\bm \lambda}^{r} + \alpha_{r} (\bA \bx^{r+1} - \bb),
		\end{align} where \( \alpha_{r} \) is a step size.
	\end{itemize}	
	\item [(iii)] \emph{Output:} \( \bx_1^{r+1},\, \bx_2^{r+1}, \ldots, \bx_n^{r+1} \).
\end{itemize}
In general (e.g., when problem \eqref{separable prob} is not strictly convex), the update \eqref{dd: dual ascent} can lead to slow convergence.  Besides, the output $\{\bx_1^{r+1},\bx_2^{r+1}, \ldots, \bx_n^{r+1}\}$ is not guaranteed to be feasible (i.e., satisfying $\bA\bx=\bb$) \cite{boyd2011distributed}. Though being simple,  due to these issues, the dual decomposition method may become unfavorable in practice.  

The ADMM is an improved decomposition method that relaxes the strict convexity assumption and has a faster convergence rate.  The vanilla version of ADMM considers an optimization problem of the following form  
\begin{equation}\label{eq: ADMM problem}
\begin{aligned} 
\min_{\bx \in  \mathcal{X},\bz \in  \mathcal{Z}} ~ &  f(\bx) + g(\bz) \\
\text{s.t.~~~\,} ~ & \bA\bx + \bB\bz = \bc, \nonumber
\end{aligned}\end{equation}
where \( f \) and \( g \) are convex functions, and \( \bA \) and \( \bB \) are given matrices.  The ADMM is an iterative method that splits this problem into simpler subproblems, which can then be solved in a decoupled or even parallel fashion.   Unlike the dual decomposition method, the ADMM considers the augmented Lagrangian 
\begin{multline*}
\mathcal{L}(\bx,\bz, {\bm \lambda})  = ~ f(\bx) + g(\bz) + {\bm \lambda}^\T(\bA\bx + \bB\bz - \bc) \\
+ \frac{\rho}{2}\| \bA\bx + \bB\bz - \bc \|^2,
\end{multline*} where $\rho>0$ is the penalty parameter.
By updating the variables $\{\bx, \bz\}$ in an alternating manner and applying the method of multipliers to the constraints, the ADMM converges to a solution of the original problem under mild assumptions \cite{boyd2011distributed}. The ADMM algorithm is given below.
\begin{itemize}
	\item [(i)] \emph{Initialization:} Choose initial points \( \bx^{0} \) and \( \bz^{0} \), and set \( {\bm \lambda}^{0} = \bm{0} \).
	\item [(ii)] \emph{Repeat until convergence:}
	\begin{itemize}
		\item Update \( \bx \):
		\begin{align}\label{eqn: ADMM x update}
		\bx^{r+1} = \arg\min_{\bx \in \mathcal{X}} \mathcal{L}(\bx,\bz^{r} , {\bm \lambda}^{r} ).
		\end{align}
		\item Update \( \bz \):
		\[
		\bz^{r+1} = \arg\min_{\bz \in \mathcal{Z}} \mathcal{L}(\bx^{r+1},\bz , {\bm \lambda}^{r} ).
		\]
		\item Update the Lagrange multiplier:
		\begin{align}\label{dd: dual ascent ADMM}{\bm \lambda}^{r+1} = {\bm \lambda}^{r} + \alpha_{r} (\bA \bx^{r+1} + \bB\bz^{r+1} - \bc).
		\end{align}
	\end{itemize}
\item [(iii)]  \emph{Output:} \( \bx^{r+1} \) and \( \bz^{r+1} \).
\end{itemize}
Thanks to the augmented Lagrangian,  the update in \eqref{dd: dual ascent ADMM} is an inexact gradient ascent step,  enabling the ADMM to have a faster convergence rate than the dual decomposition method.

When the objective function and the constraint have separable structures,  e.g., $f(\bx) =  \sum_{i=1}^n f_i(\bx_i)$,  $\bA\bx= \sum_{i=1}^n \bA_i \bx_i,$ and $\mathcal{X}=\mathcal{X}_1 \times \mathcal{X}_2 \times \cdots \times \mathcal{X}_n$,   the update of $\bx$ in \eqref{eqn: ADMM x update} can be decomposed into $n$ Gauss-Seidel steps,  which are given by
\begin{align*} 
\bx^{r+1}_i = \arg\min_{\bx_i \in \mathcal{X}_i} \mathcal{L}(\bx_{< i}^{r+1}, \bx_i,\bx_{> i}^{r} , \bz^{r} , {\bm \lambda}^{r} )
\end{align*}
for $i=1,2,\ldots, n$.  Here,   $\bx_{< i}$ contains all $\bx_j$ with $j<i$ and 
$\bx_{> i}$ contains all $\bx_j$ with $j> i$.
A disadvantage of the Gauss-Seidel update is that the variables $\{\bx_i\}$ are updated one after another, which is not amenable for parallelization.  To have a parallel algorithm,  one can consider Jacobian-type updates.  However, a direct Jacobian ADMM is not guaranteed to converge in general.  To fix this,  the proximal ADMM method is proposed \cite{deng2017parallel,zhang2020proximal,hong2017prox}.  Specifically,  one can replace  \eqref{eqn: ADMM x update} by
\begin{align}\label{eqn: ADMM x update Proximal Jobian}
\bx^{r+1}= \arg\min_{\bx \in \mathcal{X}}\left\{ \mathcal{L}(\bx, \bz^{r} , {\bm \lambda}^{r} ) + \frac{1}{2}\|\bx - \bx^{r}\|_{\bf P}^2\right\},
\end{align}
where 
\[
\|\bx_i - \bx_i^{r}\|_{\bf P}^2=(\bx_i - \bx_i^{r})^\T {\bf P} (\bx_i - \bx_i^{r})
\]  and $ {\bf P}$ is a positive definite matrix.
In particular,  if one chooses $ {\bf P}$ to satisfy ${\bf P}= c {\bf I} - \rho \bA^\T\bA \succ {\bf 0}$ for some parameter $c>0$, then the update in \eqref{eqn: ADMM x update Proximal Jobian} can be decomposed into $n$ parallel subproblems.

\subsubsection{Application Example} Next, we present one application of the ADMM and its variants in distributed wireless system design, which is {multi-cell coordinated beamforming}.

 Consider the same cellular system as in Fig.~\ref{fig:commun_system}(b). The difference here is that data sharing is not allowed among different BSs, so as to reduce the signaling overhead. Assume that each user $k$ is assigned to a specific BS $b = b_k$ and let $\mathcal{K}_b$ denote the subset of users allocated to BS $b$. In this case, for each $k\in\cK,$ we have $\bv_{k,b}=\bm{0}$ for all $b\neq b_k.$ To simplify the notation in problem \eqref{problem:clustering}, we use $\bv_k$ to denote $\bv_{k,b_k}$. Then, the SINR of user $k$ is given by  \begin{multline} 
\operatorname{SINR}_k=~~~~~~\\ \frac{\left|\h_{k,b_k}^\dagger \bv_k\right|^2}{\sum_{j \in \mathcal{K}_{b_k} \backslash k}\left|\h_{k,b_k}^\dagger \bv_j\right|^2+\sum_{b \neq b_k} \sum_{i \in \mathcal{K}_b}\left|\h_{k,b}^\dagger \bv_i\right|^2+\sigma_k^2}. \label{SINR}
\end{multline} Let us introduce the inequality 
	\begin{align*}
	\tau_{k,b} \geq \sum\nolimits_{i \in \mathcal{K}_b}\left|\h_{k,b}^\dagger \bv_i\right|^2,
	\end{align*}
where the right-hand side denotes the inter-cell interference term from BS $b$ to user $k$ for $k\notin \cK_b$. Then, the SINR formula in \eqref{SINR} is modified as
	\begin{align*} 
	\operatorname{SINR}_k=  
	\frac{\left|\h_{ k, b_k}^\dagger \bv_k\right|^2}{\sum_{j \in \mathcal{K}_{b_k} \backslash k}\left|\h_{ k, b_k}^\dagger \bv_j\right|^2+\sum_{b \neq b_k} \tau_{k,b}+\sigma_k^2}.
	\end{align*}%
	Therefore, the minimum power beamforming design problem under the per-user SINR constraint can be reformulated as
\cite{tolli2010decentralized,maros2017admm}
	\begin{subequations}\label{Decouple Power Minimization}
		\begin{align} 
		\min_{\{\bv_k,\tau_{k,b}\}} ~& \sum_{b\in \mathcal{B}} \sum_{k \in \mathcal{K}_b}\left\|\bv_{k}\right\|_2^2 \\
		\text { s.t.~~~\,} ~& \operatorname{SINR}_k \geq \gamma_k,~k\in\cK, \label{QoS constraint}\\
		& \sum\nolimits_{i \in \mathcal{K}_b}\left|\h_{k,b}^\dagger \bv_i\right|^2 \leq \tau_{k,b},~k \notin \mathcal{K}_b,~b\in\cB, \label{decouple constraint}
		\end{align}
	\end{subequations}
	where constraint \eqref{decouple constraint} guarantees that the inter-cell
	interference generated from a given BS $b$ cannot exceed the user specific thresholds $\tau_{k,b}$ for all $k \notin \mathcal{K}_b$. The above reformulation can be handled by the ADMM, which will yield a distributed algorithm.

	Observe that the BSs are coupled in the SINR constraints \eqref{QoS constraint} by the interference terms $\left\{\tau_{k,b}\right\}$. By introducing local auxiliary variables and additional equality constraints, the coupling in the SINR constraints is transferred to the coupling in the equality constraints, which is easy to decouple by the dual decomposition or ADMM.
	Specifically,  note that each inter-cell interference term $\tau_{k,b}$ couples exactly two BSs, i.e., the serving BS $b_k$ and the interfering BS $b$. Therefore, it is enough to introduce local copies of $\tau_{k,b}$ for the two BSs,  i.e.,  $t_{k,b}^{(b)}$ and $t_{k,b}^{(b_k)}$,  and
	enforce the two local copies to be equal via 
	$t_{k,b}^{(b)}=\tau_{k,b}$ and $t_{k,b}^{(b_k)}=\tau_{k,b}$ \cite{shen2012distributed,maros2017admm}. 
	More compactly, define $\boldsymbol{\tau} \in \mathbb{R}^{K(B-1)}$ as an aggregate vector that contain all interference terms $\{\tau_{k,b}\}$,  and let $\mathbf{t}=\left[({\mathbf{t}^{(1)})}^\T, ({\mathbf{t}^{(2)})}^\T, \ldots, ({\mathbf{t}^{(B)})}^\T\right]^\T \in \mathbf{R}^{{2K(B-1)}}$, where
	$\mathbf{t}^{(b)}$ contains $\{t_{k,b}^{(b)}\}_{k\notin \cK_b}$ and
	$\{t_{k,b'}^{(b)}\}_{k \in \cK_b,  b'\neq b}$.
	{Then,  the consistency between $\mathbf{t}$ and $\boldsymbol{\tau}$ can be compactly expressed using the equality $\mathbf{E} \boldsymbol{\tau}=\mathbf{t}$}, where $\mathbf{E} \in \mathbb{R}^{2K(B-1) \times K(B-1)   }$ is a matrix whose elements are $\{0,1\}$ that maps the elements of $\boldsymbol{\tau}$ in the positions corresponding to the copies in $\bt$. 
	Consequently, problem \eqref{Decouple Power Minimization} can be reformulated as
	\begin{equation}\label{Fully Couple Power minimization}
		\begin{aligned}
		\min _{\left\{\bv_k\right\}, \mathbf{t}, \boldsymbol{\tau}} ~& \sum_{b \in \mathcal{B}} \sum_{k \in \mathcal{K}_b}\left\|\bv_k\right\|^2 \\
		\text { s.t.~~~} ~& \operatorname{SINR}_k^{(b)} \geq \gamma_k,~k \in \mathcal{K}_b,~b\in\cB,\\
		& \sum\nolimits_{i \in \mathcal{K}_b}\left|\h_{k, b}^\dagger \bv_{i}\right|^2 \leq t_{k, b}^{(b)},~k \notin \mathcal{K}_b,~b\in\cB, \\
		& \mathbf{E} \boldsymbol{\tau}=\mathbf{t},
		\end{aligned}
	\end{equation}
	where the variables $\{t_{b, k}^{(b)}\}_{k\notin \cK_b}$ and the terms
	\begin{align*}
	&\operatorname{SINR}_k^{(b)} =  \frac{\left|\h_{ k, b}^\dagger \bv_k\right|^2}{\sum_{j \in \mathcal{K}_{b} \backslash k}\left|\h_{ k, b}^\dagger \bv_j\right|^2+\sum_{b' \neq b} t_{k, b'}^{(b)}+\sigma_k^2}
	\end{align*}
	are local for each BS $b$.
	
	Notice that the objective and the constraints in \eqref{Fully Couple Power minimization} are now separable with respect to $\left\{\bv_i\right\}_{i\in  \mathcal{K}_b}$ and $\bt^{(b)}$ across the BSs.
	Thus,  problem \eqref{Fully Couple Power minimization} can be solved distributedly at each BS $b$ using the ADMM: 
	\begin{subequations}
		\begin{align*}
		\min _{\bt^{(b)},\left\{\bv_k\right\}_{k\in  \mathcal{K}_b}}~& \sum_{k \in \mathcal{K}_b}\left\|\bv_k\right\|^2+\left(\boldsymbol{\nu}_b\right)^\T\left(\bt^{(b)}-\mathbf{E}_b \boldsymbol{\tau}\right) \notag\\
	&	\quad\quad\quad\quad\quad~~~~~~~~+\frac{\rho}{2}\left\|\bt^{(b)}-\mathbf{E}_b \boldsymbol{\tau}\right\|^2 \\
		\text { s.t.~~~~~}~~& \operatorname{SINR}_k^{(b)} \geq \gamma_k,~k \in \mathcal{K}_b,\\
		& \sum\nolimits_{i \in \mathcal{K}_b}\left|\h_{k, b}^\dagger \bv_{i}\right|^2 \leq t_{k, b}^{(b)},~k \notin \mathcal{K}_b.
		\end{align*}
	\end{subequations}
	Here, $\boldsymbol{\nu}=[\boldsymbol{\nu}_1^\T,\boldsymbol{\nu}_2^\T,\ldots, \boldsymbol{\nu}_B^\T]^\T\in \mathbb{R}^{2K(B-1)}$ is the Lagrange multiplier associated with the equality constraint $\boldsymbol{\tau}=\mathbf{t}$ in \eqref{Fully Couple Power minimization}. 
	Once each BS $b$ obtains $\left\{\bv_i\right\}_{i\in  \mathcal{K}_b}$ and $\bt^{(b)}$, they will share the relevant elements within $\bt^{(b)}$ with other BSs, which are further used to compute {$\boldsymbol{\tau}=\mathbf{E}^{+} \mathbf{t}$}, where $\mathbf{E}^{+}$ denotes the pseudo-inverse of $\mathbf{E}$.
	Then, we perform the update $\boldsymbol{\nu}_b \leftarrow \boldsymbol{\nu}_b+\mu\left(\mathbf{t}^{(b)}-\mathbf{E}_b \boldsymbol{\tau}\right)$ until convergence, where $\mu>0$ is the step size to update the multiplier.
	
	While problem \eqref{Decouple Power Minimization} can also be handled by the dual decomposition method as shown in \cite{tolli2010decentralized},  it is demonstrated in \cite{maros2017admm} that the ADMM can track the solution variation in a dynamic environment with time-varying CSI. The ADMM can also be applied,  together with the SDR technique,  to handle the multi-cell coordinated robust beamforming problem under imperfect CSI; see \cite{shen2012distributed} for details. It is noteworthy that distributed optimization methods can also be employed to develop algorithms that leverage parallel computing resources to tackle large-scale optimization problems, such as the multi-UAV power and trajectory control problem discussed in \cite{shen2020multi}.
{\subsection{Federated Learning in Wireless Edge Networks}\label{sec: dist_opt_part 2}
	
	%

	
	Consider a wireless edge network, as illustrated in Fig.~\ref{FL_fig1}, where an edge server orchestrates $N$ edge clients to collaboratively address a distributed learning problem via FL.
	The problem of interest is given by
	\begin{align}\label{FL objective function}
	\min \limits_{{\mathbf{w}} \in \mathbb{R}^{m} }F({\mathbf{w}}) = \sum\limits_{i = 1}^N p_i F_i({\mathbf{w}}).
	\end{align}
	Here, $p_i$ represents the weight assigned to the $i$-th client, which satisfies $p_i \geq 0$ and $\sum_{i=1}^N p_i = 1$.
	The parameter $\mathbf{w} \in \mathbb{R}^{m}$ signifies the ${m}$-dimensional model parameter targeted for learning.
	The local cost function $F_i(\cdot) = \mathbb{E}_{\mathcal{D}_i}[\mathcal{L}(\cdot;\mathcal{D}_i)]$ is the expectation of a (possibly nonconvex) loss function $\Lc$ and operates on the local dataset $\mathcal{D}_{i}$. The global cost function $F(\cdot) = \mathbb{E}_{\mathcal{D}}[\mathcal{L}(\cdot;\mathcal{D})]$ considers the global dataset $\mathcal{D} \triangleq \bigcup_{i = 1}^N \mathcal{D}_i$.
	When utilizing mini-batch samples ${\bm \xi}_{i}$ with size $b$, the local loss function is defined as $F_i(\cdot;{\bm \xi}_{i}) = \frac{1}{b} \sum_{j = 1}^{b} \mathcal{L}(\cdot;\xi_{ij})$, where $\xi_{ij}$ represents the $j$-th randomly selected sample from the dataset of client $i$, and $\mathcal{L}(\cdot;\xi_{ij})$ is the model loss function with respect to $\xi_{ij}$.
	
	In this subsection, we first present the seminal FL algorithm \textsc{FedAvg}\cite{mcmahan2017communication} for solving problem \eqref{FL objective function} and then analyze the factors that influence its convergence performance. 
	Finally, we present several improved FL algorithms. 
	\begin{figure}[t]
		\centerline{\includegraphics[width = 0.3\textwidth]{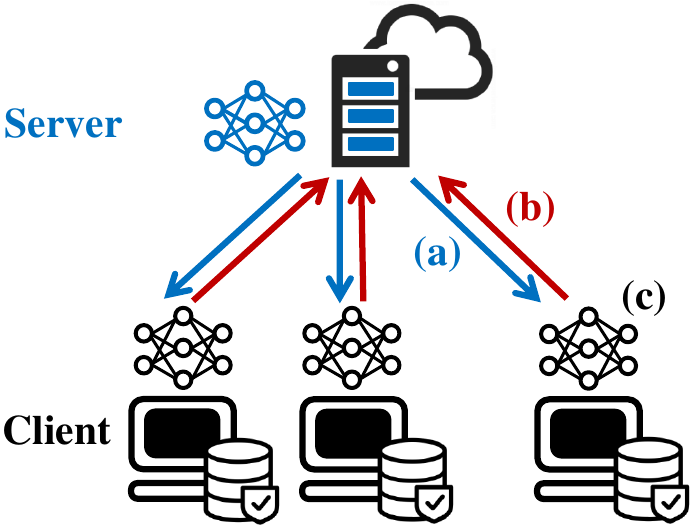}}
		\caption{{An illustration of an FL wireless edge network, where an edge server orchestrates multiple edge clients to solve a distributed learning problem collaboratively via FL.}}
		\label{FL_fig1}
	\end{figure}
	
	%
	\subsubsection{\textsc{FedAvg} Algorithm}
	
	\textsc{FedAvg} is an extension of the consensus-based distributed stochastic gradient descent (SGD) method \cite{9085431} to the star network as depicted in Fig.~\ref{FL_fig1}.  It involves three essential steps in each communication round:
	\begin{enumerate}
		\item [(i)] \emph{Broadcasting}:
		In the $r$-th communication round, the server randomly chooses $K$ clients, represented by the set ${\mathcal{K}}_{r}$, where $|{\mathcal{K}}_{r}|=K$.
		It then broadcasts the global model ${\bar {\mathbf{w}}}^{r-1}$ from the previous iteration to each client in ${\mathcal{K}}_{r}$.

		\item [(ii)] \emph{Local model updating}: Each client $i \in {\mathcal{K}}_{r}$ updates its local model using local SGD.
		This involves the $E$ consecutive SGD updates
		\begin{align*} 
		\begin{aligned}
		& \mathbf{w}^{r,0}_{i}  =  {\bar {\mathbf{w}}}^{r-1}, \\
		&  \mathbf{w}^{r,t}_{i}   =  {\mathbf{w}}^{r,t-1}_{i}  -  \alpha \nabla F_{i}({\mathbf{w}}^{r,t-1}_{i};{\bm \xi}^{r,t}_{i}) ,~t  = 1, 2, \ldots , E,
		\end{aligned}
		\end{align*}
   where $\alpha>0$ represents the learning rate.
		
		\item [(iii)] \emph{Aggregation}:
		The selected clients upload their locally updated model $\mathbf{w}_i^{r,E}$ to the server, which then aggregates these models to produce a new global model based on a specific aggregation principle.
	\end{enumerate}
	
	Notably, \textsc{FedAvg} employs two aggregation schemes, depending on whether all clients participate or not:
	\begin{itemize}
		\item \emph{Full participation}:
		All clients actively participate in the aggregation process, i.e., ${\mathcal{K}}_{r} = {\cal N} \triangleq \{ 1,2,\ldots,N \}$ for all $r$.
		The global model is updated by
		\begin{align}\label{full_participation}
		{\bar {\mathbf{w}}}^{r} = \sum\limits_{i = 1}^N p_i \mathbf{w}^{r,E}_{i}.
		\end{align}
		However, this scheme may pose feasibility challenges due to a limited communication bandwidth for uplink channels, given the large number of participants.
		
		\item \emph{Partial participation}:
		With $|{\mathcal{K}}_{r}| \ll N$, the global model is updated by
		\begin{align}\label{partial_participation}
		{\bar {\mathbf{w}}}^{r} = \frac{1}{K} \sum\limits_{i \in {\mathcal{K}}_{r}} {\mathbf{w}}^{r,E}_{i} \text{.}
		\end{align}
		Here, all $K$ clients in ${\mathcal{K}}_{r}$ are selected with replacement based on the probability distribution $\{ p_i\}_{i=1}^N$.
		It is important to note that the averaging scheme in \eqref{partial_participation} provides an unbiased estimate of ${\bar {\mathbf{w}}}^{r}$ in \eqref{full_participation} \cite{li2019convergence}.
	\end{itemize}
	
	\begin{figure*}
		\begin{multline}\label{FL_theorem_1_bound}
		\frac{1}{R} \sum_{r = 1}^{R}
		\big\| \nabla_{\mathbf{w}} F({\bar {\mathbf{w}}}^{r-1}) \big\|^2 
		\leq ~\frac{496 L  \left(F({\bar {\mathbf{w}}}^{0}) - \underline{F} \right)}{11  \left( TK \right)^{\frac{1}{2}} }
		+  \underbrace{ \left(  \frac{ 39 }{ 88  \left( T K \right)^{\frac{1}{2}} }
			+  \frac{1}{ 88  \left( T  K \right)^{\frac{3}{4}} }  \right) \frac{\sigma^2}{b}  }_{ \text{\footnotesize (a) (caused by mini-batch SGD)} }\\   
		+ \underbrace{  \left(  \frac{4 }{ 11 \left( TK \right)^{\frac{1}{2}} } +  \frac{1}{ 22 \left( TK \right)^{\frac{3}{4}} }   \right) \sum \limits_{i = 1}^{N}  p_i D_i^2 }_{ \text{\footnotesize (b) (caused by data heterogeneity)} } 
		+ \underbrace{ \frac{31 }{ 22 T^{\frac{1}{4}} K^{\frac{5}{4}}} \sum_{i=1}^N p_i D_i^2 }_{ \text{\footnotesize (c) (caused by partial participation)} }.
		\end{multline}
		\hrulefill\end{figure*} 
	
	\subsubsection{Performance Analysis} Several factors influence the performance of \textsc{FedAvg}, including the number of clients $K$, the number of local updating steps $E$, and data heterogeneity \cite{li2019convergence}.
	Moreover, there are interactive relationships between data heterogeneity and other training factors.
	For instance, a larger $E$ exacerbates the negative impact of data heterogeneity, while a smaller $E$ increases the communication cost of transmitting model parameters. To conduct a thorough analysis of the influence of data heterogeneity on FL's convergence, we can utilize the difference between local and global function gradients, i.e., ${\mathbb E}[\| \nabla F_i({\mathbf{w}}) - \nabla F ({{\mathbf{w}}})\|^2]$,  as a metric to quantify data heterogeneity \cite{wang2022quantized}. Up to now, extensive research has been conducted on FL's convergence; see, e.g., \cite{wang2022quantized, chen2020joint, salehi2021federated, zhu2020one, reisizadeh2020fedpaq, zheng2020design, liu2020distributed}.
	Here, we present a key result that unveils the fundamental properties of FL, whose validity has been examined in \cite{wang2022quantized}.
	
	To begin, let us state the assumptions:
	\begin{itemize}
		\item Each local function $F_i$ is lower bounded by $\underline{F}$ and the local gradient $\nabla F_i$ is Lipschitz continuous with a constant $L$.
		\item The local SGD is unbiased, i.e., ${\mathbb E}[ \nabla  F_i({\mathbf{w}},{\xi}_{ij}) ] = \nabla  F_i({\mathbf{w}})$, and has a bounded variance, i.e., ${\mathbb E}[\| \nabla  F_i({\mathbf{w}},{\xi}_{ij}) - \nabla  F_i({\mathbf{w}})\|^2]  \leq \sigma^2$.
		\item The data heterogeneity metric is upper bounded, i.e., ${\mathbb E}[\| \nabla F_i({\mathbf{w}}) - \nabla F ({{\mathbf{w}}})\|^2] \leq D_i^2$ for all $i \in {\cal N}$.
	\end{itemize}
Let $R$ denote the number of iterations and $T = RE$ denote the total number of SGD updates per client. Suppose that
	$\alpha = K^{\frac{1}{2}} / (8L{T}^{\frac{1}{2}}) $ and $E \leq T^{\frac{1}{4}}/K^{\frac{3}{4}}$. Then, the inequality \eqref{FL_theorem_1_bound} (at the top of the page) holds \cite{wang2022quantized}. The terms (a) and (b) in \eqref{FL_theorem_1_bound} reveal that the influence of mini-batch SGD variance ${\sigma^2}/{b}$ and data heterogeneity $\{D_i\}$ can be alleviated by increasing the number of selected clients $K$.
	This allows \textsc{FedAvg} to achieve a \emph{linear speed-up} with respect to $K$.
	Furthermore, due to the presence of the partial participation term (c) in \eqref{FL_theorem_1_bound}, the convergence rate is $\mathcal{O}(1/T^{\frac{1}{4}})$.
	In the scenario where full participation (i.e., \eqref{full_participation}) is adopted, the term (c) would disappear, leading to an improved convergence rate of $\mathcal{O}(1/T^{\frac{1}{2}})$.

	\subsubsection{Improved FL Algorithms}
	In recent research,  wireless resource allocation in non-ideal wireless environments has garnered attention within the context of FL, where diverse perspectives have been explored.
	For instance, the work \cite{chen2020joint} delves into the impact of packet error rates on the convergence of \textsc{FedAvg} and proposed a novel approach that integrates joint resource allocation and client selection to enhance the convergence speed of \textsc{FedAvg}. 
	On another front, research efforts have been directed toward exploring compressed transmission through quantization and evaluating its influence on FL's performance.
	For instance, the work \cite{reisizadeh2020fedpaq} proposes \textsc{FedPAQ}, a communication-efficient FL method that transmits the quantized global model in the downlink, with a subsequent analysis of the quantization error's effect on FL's convergence. The work
	\cite{zheng2020design} explores layered quantized transmissions for communication-efficient FL, where distinct quantization levels are assigned to various layers of the trained neural network.
	Different from these, the work \cite{wang2022quantized} considers both transmission outage and quantization error concurrently, undertaking joint allocation of wireless resources and quantization bits to achieve robust FL performance. In the quest to enhance FL's performance within heterogeneous data networks, researchers have explored more advanced algorithms that aim to surpass the capabilities of the conventional \textsc{FedAvg}. See \cite{wang2020tackling,gong2022fedadmm,zhang2021fedpd} and the references therein for improved FL algorithms which can better tackle data and system heterogeneity in problem \eqref{FL objective function}.

}

\section{Optimization via Learning}\label{sec:learning}

{In the last few years}, new machine learning (ML) and AI techniques have powered nothing short of a technological revolution in a number of application areas including speech recognition~\cite{hinton2012deep}, image classification~\cite{krizhevsky2012imagenet}, and natural language processing~\cite{liu2012sentiment}. In particular, well-trained deep neural networks (DNNs) are capable of utilizing limited knowledge about the underlying model to effectively transform a large amount of data to latent {informative feature} spaces. {Remarkably}, deep learning-based AI has exceeded human-level performance in many nontrivial tasks.

Unlike classic communication modeling and computational tools that are mainly {model-driven}, ML-based methods such as DNNs and deep reinforcement learning (DRL) are largely {data-driven} \cite{hinton2012deep, krizhevsky2012imagenet}.
A natural question then arises: {\it Can data-driven ML/AI-based methods significantly enhance the capacity and performance of communication networks?}
Recent surge of research suggests that these methods can achieve significant gains for tasks such as encoding/decoding, equalization,  power control, and beamforming.

One of the most important tasks in wireless networking is to determine, in each time period, which subset of links to activate and what subcarriers and transmit powers those links should use.
Evidently, deactivating a link or a subcarrier is equivalent to setting its transmit power to zero.
While there are 
closed-form solutions under 
a few special settings, %
optimal spectrum and power allocations 
are NP-hard to compute in general interference-limited networks~\cite{luo_complexity}.
{Most of the best-known power allocation algorithms such as WMMSE~\cite{WMMSE2}, SCALE~\cite{papand09}, 
	FlashLinQ \cite{FlashLinQ}, ITLinQ
	\cite{ITLinQ}, majorization-minimization \cite{FP:MM},} 
and those that have been surveyed in this work in the previous sections,  
require complete model knowledge and are computationally very {challenging}. 
If the model parameters are 
only partially known, e.g., when the channel coefficients are time-varying, 
a principled, efficient design is yet to be found in the literature~\cite{
	hong12survey,
	bjornson13,
	Razaviyayn13Stochastic}.
Today's cellular networks dynamically allocate subcarriers to each link and step up/down the transmit power primarily based on {this link's own} receiver feedback, which is far from being globally optimal. 

To address both computational complexity and model uncertainty issues,  data-driven approaches {such as DNN}  can provide a much-needed solution for next-generation wireless networks. To be concrete, consider the multi-user multi-carrier interference network with $K$ transmitters, and each using power ${p}^{\ell}_k\ge 0$ to transmit to its associated receiver on the $\ell$-th subcarrier. Let ${h}^{\ell}_{kj}\in\mathbb{C}$ denote the channel between transmitter $j$ and receiver $k$ on subcarrier $\ell.$
{
	Then, the sum-rate maximization problem is given by
	\begin{equation}\label{eq:wsr}
		\begin{aligned}
		\max_{\left\{p_k^{\ell}\right\}}~ 
		& {\rm WSR}({\bm p}, {\bm h})\triangleq\sum_{k} w_k \sum_{\ell}
		\log\left(1+\mbox{SINR}_k^{\ell}\right)  \\
		{\rm s.t.~}~
		& \sum_{\ell} p_k^{\ell} \leq P_{k},~k\in\cK, 
		\end{aligned}
	\end{equation} 
where \begin{align}\label{eq:snr}
	\mbox{SINR}_k^{\ell}
	=
	\frac{|{h}_{kk}^{\ell}|^2p^{\ell}_k}{\sum_{j\neq k}|h_{kj}^{\ell}|^2 p_j^{\ell}+\sigma_k^2}
	\end{align} is the SINR of receiver $k$ over subcarrier $\ell,$ and $w_k$ and $P_k$ are the weight and power budget of user $k,$ respectively.
	This problem and its various generalizations such as  the BF problem \eqref{problem:sumrate} {are} NP-hard~\cite{
		luo_complexity,hong12survey,liu2014complexity}.
	A closer look at the popular and computationally more affordable WMMSE algorithm~\cite{WMMSE2} reveals a number of relatively expensive operations including taking the magnitude, thresholding, re-weighting, and matrix inversion.  More importantly, implementing 
	{the WMMSE algorithm} requires precise knowledge of the parameters (e.g., channel coefficients) and may use an unknown number of iterations---which varies {from instance to instance}. 
	A  question that many researchers have started asking	around  2017 is:	{\it Can neural networks help, and {if so, to what} extent?}  Since then, extensive literature has been developed to address both the computational and model uncertainty issues.

	In the rest of this section, we focus on learning-based power control and beamforming methods that leverage the instantaneous CSI to enhance the performance of wireless systems. In Section \ref{sec:no:csi}, we switch to more recent developments in learning methods that do not require the CSI. 

	\subsection{Black-Box Based Approaches}
	
	
	{The first line of work started with the \emph{learning to optimize} approach proposed in \cite{sun2018learning}.}  In this approach, an (potentially computationally expensive) optimization algorithm is treated as a {nonlinear mapping}---which takes problem specification as the input, and outputs the (hopefully optimal) decision variables.  Formally, let ${\cal T}(\bm h)=\bm p^*$ denote a nonlinear relationship of the input (i.e., the channel coefficients) and output (i.e., the optimized powers) for an algorithm solving \eqref{eq:wsr}.
	Due to DNNs' superior ability to learn compact representation of nonlinear relations~\cite{hornik1989multilayer}, in principle it is possible to use it as a ``black-box''  to learn the relation ${\cal T}(\cdot)$ using a DNN without going into iterations to mimic the ``lower level operations''.  If we could use a very simple network, say a few layers and neurons to well-approximate a power control algorithm 
	which normally runs for more than 100 iterations,  then substantial saving in real-time computation can be achieved.
	In \cite{sun2018learning}, a DNN-based approach is developed to approximate WMMSE {in the special case of a single carrier.
		Specifically,}  a supervised learning approach is used, where the training pairs are generated using {simulated channel and the WMMSE algorithm} (where the $i$-th snapshot of the simulated channel is denoted as $\bm{h}^{(i)}$, and the resulting WMMSE solution is denoted as $\bm{p}^{(i)}$), and then they are used to train a DNN that mimics the behavior of WMMSE. Let $\tau(\cdot,\bm \theta)$ denote the DNN, where $\bm \theta$ collects all the parameters of the DNN.  Then, the training problem can be expressed as
	\begin{equation}\label{eq:supervise}
		\begin{aligned}
		\min_{\bm \theta}&~ \sum_{i=1}^{I}\|\tau({\bm h}^{(i)}, {\bm \theta}) - {\bm p}^{(i)}\|^2\\
		\mbox{\rm s.t.~} &~ \tau({\bm h}^{(i)}, {\bm \theta})  \in \mathcal{P},
		\end{aligned}
	\end{equation}
	where $\cal{P}$ denotes the feasible set of the transmit power vectors and $I$ is the total number of training samples.

	
	{The approach is tested on a variety of scenarios in \cite{sun2018learning}, including real-data experiments, and the results are very encouraging. The {key findings} are as follows: (i) It is indeed possible to {closely} approximate a highly complex iterative power control algorithm by using a relatively simple DNN (in this case, a network with only three hidden layers). (ii) 
		{The DNN-based implementation is typically 25 to 250 times faster than 
			the best C language implementation of the WMMSE.}}
	
	Subsequently, a number of works have been developed to improve the \emph{learning to optimize} techniques discussed above. For example,  in \cite{liang2019towards}  an {unsupervised learning} approach is developed, which directly optimizes some system utilities such as the WSR over the training set. More specifically,  the training problem is given by
	\begin{equation}\label{eq:unsup}
		\begin{aligned}
		\min_{\bm \theta}~& -  \sum_{i=1}^{I} {\rm WSR}(\tau({\bm h}^{(i)}, {\bm \theta}), {\bm h}^{(i)})\\
		\mbox{\rm s.t.~}~ & \tau({\bm h}^{(i)}, {\bm \theta})  \in \mathcal{P},
		\end{aligned}
	\end{equation}
	where ${\rm WSR(\cdot)}$ is defined in \eqref{eq:wsr}. It has been shown that by using the negative WSR as the loss function, it is possible to find power allocation strategies whose performance goes beyond that of WMMSE. It is worth highlighting that the above unsupervised approach has been designed specifically for the interference management problem because the task of wireless system utility optimization (which includes the WSR maximization as a special case) offers a natural training objective to work with. Since it does not require any existing algorithms to help generate high-quality labels, it is much preferred when training samples are difficult to generate. {On the other hand, the {associated} training objective appears to be difficult to optimize, since the WSR is a highly nonlinear function with respect to the transmit power or the beamformer, which in turn is a highly nonlinear function of the DNN parameters. Therefore, in the future, it is worth understanding the tradeoffs between the two formulations \eqref{eq:supervise} and \eqref{eq:unsup}.

		In \cite{Eisen20}, the fully connected neural networks used in the previous works are replaced by certain random edge graph neural network (REGNN), which performs convolutions on random graphs created by the network's fading interference patterns. REGNN-based policies maintain an important permutation equivariance property, facilitating their transference to different networks. The key benefit of the proposed architecture is that only a small neural network is needed, and the dimensionality of the network does not scale with the network size. It is worth mentioning that there are other recent works that apply graph neural networks to learn algorithms that are capable of learning globally optimal beamformers; see, e.g., \cite{10043790,wu2024efficient}. In \cite{xia2019deep}, the supervised deep learning approach is extended from the power control problem to a multi-user beamforming problem by utilizing convolutional neural networks and expert knowledge such as uplink-downlink duality (which has been reviewed in Section \ref{subsection:duality}). In particular, three beamforming neural networks are developed for optimizing the SINR, power minimization, and sum rate. Similarly as in \eqref{eq:supervise}, the beamforming neural networks employ supervised learning for SINR and power minimization and a hybrid approach for sum-rate maximization. In \cite{Sun22}, in view of the fact that the previous learning-based algorithms have only been developed in the static environment, where parameters like the CSI  are assumed to be constant, a methodology for continuous learning and optimization in certain ``episodically dynamic'' settings is introduced, where the environment changes in ``episodes'', and in each episode the environment is stationary. The work proposed to incorporate the technique of  {continual learning}  into the model to enable incremental adaptation to new episodes without forgetting previous knowledge. 
		{By further utilizing certain specific structures of the optimal beamforming solution (e.g., the low-dimensional structure and/or the invariance property under the permutation of users' indices) and embedding these structures into the network, the constructed neural network can have better scalability to different numbers of transmit antennas and BSs and can tackle more difficult QoS constraints. Some recent progress in this direction has been made in \cite{zhang2023learning,10494519}.} 

		\subsection{Unfolding-Based Approaches}\label{subsection:unfolding}
		
		Different from the above {black-box} DNN approach,  where DNN is used as a black box to approximate the input-output relationship of certain algorithms or systems, another line of work leverages the deep unfolding technique, 
		which builds DNNs based on finer-grained approximation of a known iterative algorithm with a finite number of iterations.  Specifically,  the neural network to be built will have multiple stages, where each stage consists of function blocks that imitate a given step of the target optimization algorithm.
		For example, the work \cite{samuel2019learning} unfolds the GP algorithm to build learning networks for the MIMO detection problem in \eqref{problem:mimodec}.
		%
		For a single-cell multi-user beamforming problem, the work \cite{hu2020iterative} proposes a learning network by unfolding the WMMSE algorithm.
		To overcome the difficulty of matrix inversion involved in the WMMSE algorithm, they approximate the matrix inversion by its first-order Taylor's expansion. Another recent work \cite{unfolding_2} proposes to unfold the WMMSE algorithm to solve the coordinated beamforming problem in MISO interference channels. In \cite{Zhu23}, certain GP algorithm is unfolded for the multi-user beamforming problem. {Again, by utilizing certain low-dimensional structure of the optimal beamforming solution, the constructed neural network can be made independent of the numbers of transmit antennas and BSs.} 

		Overall, the advantage of these deep unfolding methods is that they can leverage existing algorithms to guide the design of neural networks. In this way,  the number of parameters to be learned can be much smaller as compared to black-box-based DNN methods.

		\section{Learning-Based Optimization Without Explicit Channel Estimation}\label{sec:no:csi}
		
			While the focus of the previous section is on using the neural network to {mimic} a sophisticated optimization solver, the true benefit of the ML approach for optimizing communication system design goes much further. In this section, we point out that the practical advantage of the
		ML-based solver lies not necessarily in that a data-driven approach may provide
		a more efficient way to solve complex optimization problems, but more importantly,
		a learning-based approach allows communication channels to be modeled and to
		be parameterized differently (and potentially more effectively), so that
		relevant channel characteristics that are otherwise difficult to build into an
		analytic model can now be taken into account in the optimization process. In
		fact, the learning-based approach can allow the optimization of wireless communication
		systems to be performed {without} explicit channel estimation. This
		ability to bypass the CSI estimation process is where the true promise of
		ML lies. 
		
		In many optimization problems for wireless communication system design, the estimation
		of CSI is a highly nontrivial process for the following reasons. As modern systems
		move toward massive MIMO with many antenna elements, while also incorporating novel
		devices such as RIS with many tunable reflectors,
		the number of parameters in the overall channel has exploded. Yet, as
		next-generation wireless services increasingly demand agility to support
		ultra-reliable low-latency communications and cater toward high-mobility
		applications where the channel coherence time is severely limited, the amount of
		time available for CSI acquisition has effectively been shortened, making the
		estimation task ever more challenging. Furthermore, modern wireless communication
		networks often involve a large number of independent transmitter-receiver
		links. To facilitate interference management, the CSI between each transmit
		and each receive device would need to be estimated and collected at a
		centralized controller.  The coordination required for channel estimation and
		feedback will become increasingly complex as the network size grows. 
		
		Therefore, the bottlenecks in the optimization of wireless communication networks are 
		often not only the efficiency of the optimization algorithms for achieving either
		global or local optimal solutions of a particular system-level optimization
		problem---they could well also be the availability of an accurate CSI across the entire network. In this section, we first discuss the issue of channel modeling, then highlight 
		several approaches of using learning-based methods for the optimization of wireless communication networks that are {model-free}.  
		
		\subsection{Channel Modeling and CSI Estimation} 
		
		Communication engineers have invested heavily in the study of channel models. Cellular
		and WiFi standards include sophisticated electromagnetic propagation models
		under which the transceiver designs must perform well. At a system level, radio
		propagation maps for outdoor and indoor environments have been carefully
		developed and used for deployment planning purposes. However, these established
		channel models are typically statistical in nature.  At a link level, when
		optimizing the transceivers for a specific channel realization, we must rely on
		pilots to estimate the channel within the coherence time. 
		
		One of the key questions for channel estimation is how to parameterize a wireless
		channel.  For a MIMO channel with $M$ antennas at the transmitter and $M$
		antennas at the receiver, the conventional method is to capture the complex
		channel coefficients from each transmit antenna to each receive antenna. Thus, an $M \times
		M$ channel has ${\cal O}(M^2)$ complex parameters. While such a parameterization may be
		suitable for a rich-scattering environment when antenna spacing is at least
		half wavelength apart so that the channels across the antennas are
		uncorrelated, at higher frequencies such as the mmWave band, the propagation environment becomes increasingly sparse.  This means that the channels across
		the antennas would exhibit strong correlations, and the overall MIMO channel can
		be parameterized by a much smaller number of parameters.  Toward this end,
		sparse channel models and sparse optimization techniques (which have been reviewed in Section \ref{subsec:sparse}) have proved to be 
		useful for CSI estimation in such channels.
		
		A convenient approach to the modeling of sparse channels is to use a
		ray-tracing model, in which the wireless propagation environment is
		characterized by a limited number of rays from the transmitter to the receiver
		via the reflective paths. However, these model assumptions are susceptible to
		variations in the deployment scenario, e.g., it is difficult to determine the
		number of paths in advance. Also, as the Bayesian parameter estimation process
		would require a prior distribution on the model parameters, it is not
		obvious how these prior distributions should be chosen. 
		
		In general, choosing the most suitable channel model is an art
		rather than science.  There is a delicate balance between choosing a model with
		many parameters, which may be more accurate but also makes channel estimation
		harder, versus choosing a model with fewer parameters, which may be less
		accurate but makes parameter estimation easier. 
		Moreover, as a mobile transceiver can easily move from a limited scattering
		location to a rich-scattering location, identifying the suitable channel model
		for each specific situation is a highly nontrivial task.
		
		\subsection{Model-Free Optimization} 
		
		\begin{figure*}[t]
			\centering
			\includegraphics[width=11cm]{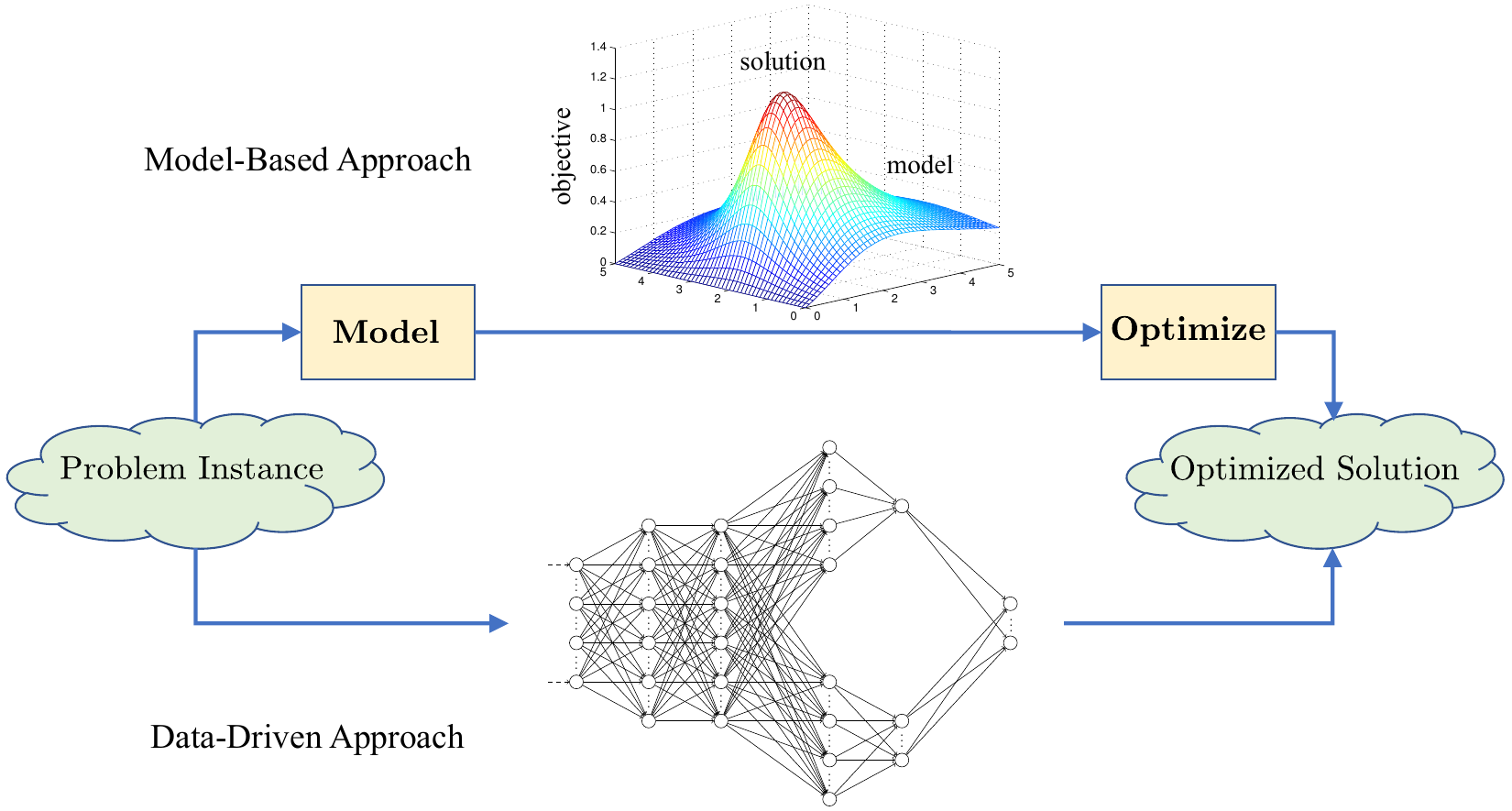}
			\caption{Traditional wireless system design follows the paradigm of
				model-then-optimize. The ML-based approach is capable of
				directly learning the optimal solution based on a representation of 
				the problem instance. The neural network is 
				trained over many problem instances, by adjusting its weights according to 
				the overall system objective as a function of the representation of 
				the problem instances \cite{wei_role}.}
			\label{fig:ML_model}
		\end{figure*}
		
		A learning-based approach can potentially circumvent the difficulty posed by
		model-based optimization. Instead of optimizing the transceiver parameters such
		as power and beamforming based on the CSI acquired in the channel estimation phase, 
		modern neural networks can be efficiently trained to allow
		the possibilities of taking a variety of relevant information about the
		channel as the input, while producing an optimized solution based on these inputs.
		
		This new data-driven paradigm is illustrated in Fig.~\ref{fig:ML_model} \cite{wei_role}. 
		While traditional optimization methods must rely on a specific parameterization of
		the channel, the data-driven approach can take any representation of the problem
		instance as the input, then map the problem instance to an optimization
		solution. This opens up the possibility of using not only the CSI but also relevant
		information such as the locations of mobile devices, visual images of the
		surrounding environment, or sensing data from radar/lidar in autonomous
		vehicles, to aid the specification of the propagation channel. 
		
		This ability for the neural network to merge the multitude of different kinds of
		information is a key advantage of the proposed data-driven paradigm.  In
		effect, once properly trained over many problem instances, the first layers of
		the neural network can act as {feature-extraction layers} to find the most
		prominent features of the optimization problem, while the later layers would
		act as {optimization layers} to find an optimized solution.  Such an
		approach allows the potential of reducing the reliance or completely eliminating
		the need for explicit channel estimation. This is where ML-based
		optimization would have the potential to have the largest impact.  
		
		In the remainder of this section, we survey several examples of how a variety
		of different information can be taken as the problem specification to allow for
		an effective solution of the optimization problem. 
		
		\subsubsection{Scheduling and Power Control Using Geographic Information}
		
		As already discussed in detail in the previous section, power control for the interference channel is one of the long-standing optimization problems in the wireless domain. In fact, when formulated as an integer
		programming problem of deciding whether a device should be on or off, it can be
		readily seen that the sum-rate maximization problem (and many of its
		variations) is NP-hard \cite{luo_complexity,NET-036}.
		
		However, the difficulty of the power control problem goes beyond algorithmic
		complexity.  In an interfering environment with $K$ transmitter-receiver pairs,
		the acquisition of the CSI would require transmitting pilots from each of
		the $K$ transmitters to each of the $K$ receivers, in order
		to estimate the ${\cal O}(K^2)$ channel parameters. Not only would such a channel
		acquisition phase consumes valuable coherence time, it also requires careful
		coordination, which is often costly in itself. Further, these estimated channel	parameters need to be collected at one central location so that a centralized
		optimization problem may be formulated and solved. Finally, the solution needs
		to be communicated back to the devices.  These tasks are often cost-prohibitive
		in a distributed networking environment.
		
		The core of the power control problem is {scheduling}, i.e., to decide
		which set of transmitter-receiver pairs should be active at any given time, so
		as to balance the need for throughput provisioning and interference mitigation.
		To this end, the work \cite{spatial_wei} makes a crucial observation that the
		transmission decision of each transmitter-receiver pair is essentially a function
		of the locations of nearby transmitters relative to the receiver and the locations of
		nearby receivers relative to the transmitter. Such {geographic location
			information} already tells us a great deal about the interference level each receiver would experience, and likewise the interference pattern these
		transmitters would emit depending on which ones are
		turned on. Thus, instead of using exact CSI to formulate an optimization
		problem of maximizing the network utility, it ought to be possible to provide
		the location information as the input to a neural network, and to train the
		neural network to arrive at an approximately optimal solution.  This is an example where precise channel information, which is difficult to obtain, may be replaced by geographic spatial maps of the potentially interfering transmitters
		and the potentially interfered receivers as a representation of the
		optimization problem. These maps already contain sufficient amount of
		information to derive a reasonable schedule, as shown in \cite{spatial_wei}. 
		The benefit of not having to
		estimate CSI would outweigh the cost in terms of loss in optimality.
		
		The idea of modeling the spatial relationship between transceiver pairs as a graph in order to aid a network-wide optimization has found relevance in many
		related works, e.g., in using graph embedding based on the distances between
	    nodes as features to perform link scheduling \cite{graph_geoffrey}, and in using 
		GNNs to account for the interference landscape in a network 
		\cite{Eisen20}, \cite{gnn_jun}. Neural networks have also been found useful for
		estimating the radio map of a complex environment \cite{radio_net}.
		
		\subsubsection{Beamforming and RIS Reconfiguration with Implicit Channel Estimation}
		
		The traditional optimization paradigm always assumes a channel estimation phase
		based on the pilots, followed by an optimization phase based on the estimated
		channel.  The CSI serves as the intermediary interface between the two phases.
		However, as already mentioned, choosing the most appropriate channel model and the
		channel estimation method involves many tradeoffs, so it is not a trivial task. 
		
		The capability of neural networks for taking diverse types of information as
		the inputs to the optimization process gives rise to a new possibility.  Instead
		of explicitly estimating the channel based on the received pilots and then performing
		optimization, a better
		idea is to feed the received pilots directly into the neural network and to train
		the neural network to produce an optimized solution based on the information about
		the channel implicitly contained in the received pilots. Such a model-free
		optimization paradigm would bypass explicit channel estimation and let the neural network perform both feature extraction (i.e., finding the most relevant
		information about the channel) and optimization together at the same time.
		
		The optimization of RIS is an example in
		which this new approach can be much more effective than the traditional channel
		estimation-based approach. The deployment of RIS in a communication setting 
		allows real-time re-focusing of electromagnetic beam from a transmitter through
		the reflecting surface to an intended receiver, thereby enhancing the overall SINR.
		In a traditional optimization paradigm, the channel between the transmitter and
		the receiver would need to be parameterized by all the reflective paths.
		To explicitly estimate these channel parameters would have cost a large number
		of pilots \cite{ris_csi_jie}. If instead, the received pilots are used as a
		representation of the channel as the input to the neural network, then it would 
		result in a much more efficient optimization process.
		
		Experimentally, this approach has been shown in \cite{ris_tao} to be able to
		produce optimized configurations for the RIS using a much smaller number of pilots
		than traditional channel estimation-based approaches. The model-free approach
		produces a higher overall rate than sophisticated manifold optimization and
		block coordinate descent techniques \cite{ris_manifold_2,
			ris_opt_yafeng} for optimizing RIS---when the channel estimation error is taken
		into account. Interestingly, the neural network produces highly interpretable
		results; it can effectively track the users and further cancel the
		interference between the users. It is also possible to use this framework to
		include scheduling \cite{scheduling_david}, which is a difficult discrete
		optimization problem.  Moreover, the locations of the users can be used as
		an additional input to the neural network to further reduce the length of pilots.

		\subsubsection{Sensing, Localization, and Beam Alignment with Massive MIMO}
		
		Model-free deep learning-based optimization has an important role to play not
		only for wireless communication applications but also for sensing and localization
		tasks, which are crucial application areas for future wireless networks. The
		framework discussed earlier in this section applies naturally to sensing
		applications, because model-free deep learning is capable of implicitly
		estimating the channel, and channel estimation is an example of sensing.  In
		sensing applications, typically a known signal is transmitted, then possibly
		reflected off a target object, and finally received and processed.  The goal 
		is to identify or to characterize some properties of the transmitter, or 
		the reflecting object, or the environment, based on the received signal.
		Traditional optimization techniques can be used if a model of the target or 
		the environment is available, and if the sensing objective can be stated in a
		mathematical form (e.g., in terms of the MSE). However, because the
		validity of these models is subject to assumptions, it would have
		been much more preferable to train a neural network to accomplish the same
		task. The premise here is of course that the training data for realistic
		sensing scenarios are available or can be easily generated. Historically,
		image/speech recognition is among the first successful applications of deep
		learning \cite{hinton2012deep,krizhevsky2012imagenet}---broadly speaking, these are all sensing tasks. 
		
		Sensing applications for which deep learning has been shown to provide
		substantial benefit, as compared to handcrafted model-based optimization,
		include localization and mmWave massive MIMO 
		initial beam alignment \cite{alignment_foad, active_foad}, which is  
		a problem of designing a beamformer to align with the incoming ray during
		the pilot stage. 
		
		A further consideration in many sensing tasks is that sensing operations
		are often sequential in nature. The sensing strategies can be adaptively
		designed depending on the observations made so far. The sequential optimization
		of sensing strategies, if formulated as an analytic optimization problem, would
		have been a high-dimensional problem that is impossible to solve
		analytically. Given the right neural network architecture, however, they can be
		readily tackled using deep learning methods. For example, the work \cite{twosided_tao}
		demonstrates that in a massive MIMO channel where both the transmitter and
		receiver are equipped only with a single radio-frequency chain (so they
		can only perform analog beamforming), it is possible to design a sequence of
		analog sensing beamformers so that the transmitter and receiver can jointly
		discover the strongest direction in a high-dimensional channel. In effect, deep
		learning is capable of performing singular value decomposition over the
		air---without explicitly estimating the channel matrix.
		
		The dynamic nature of the sensing task is especially important in applications
		involving object tracking. In this realm, 
		the work \cite{vision_alkhateeb} demonstrates that a deep learning approach can
		incorporate visual imaging data for beam tracking and beam alignment. This
		speaks to the utility of learning-based optimization---the ability to incorporate imaging data for RF beamforming would have been very difficult to achieve using traditional model-based
		approaches.

		\subsection{Neural Network Architecture Considerations}
		
		A crucial consideration in the design of deep learning methods for solving
		optimization problems is the choice of the neural network architecture. The general
		principle is that the neural network architecture should match the structure of
		the optimization task at hand. As already being alluded to, the GNN
		\cite{gnn_jun,Eisen20} that captures the spatial relationship
		between the interfering links is a well-suited architecture for scheduling
		and power control in device-to-device ad-hoc networks and for beamforming
		pattern design in RIS-assisted communication scenarios. By tying the weights of different
		branches of the GNN together, it would facilitate faster
		training and a more generalizable solution. For sequential optimization in sensing applications, neural network
		architectures such as long short-term
		memory \cite{lstm} have shown to be effective for capturing the correlation
		over time \cite{localization_david, alignment_foad, active_foad,twosided_tao}.
		Modern attention-based neural network architectures can
		potentially offer even further improvements. 
		
The field of modern ML is evolving rapidly. Unquestionably, future wireless communication system design will incorporate elements of learning-based approaches soon and will likely go beyond the present model-based methodology.

\section{Open Problems and Research Directions}\label{sec:open}


In this section, we present some open problems and future research directions for mathematical optimization theory and algorithms for wireless communication system design.

\subsection{Open Problems}
While many advanced mathematical optimization theory and various algorithms have been developed in the past decades, there are still many open problems. 

\begin{itemize}
	\item With the help of quadratic and Lagrangian dual transforms reviewed in Section \ref{subsection:fractional}, we can transform a complicated low-dimensional (e.g., sum-rate maximization) problem into an equivalent, relatively easy high-dimensional problem and further apply the AO algorithm to efficiently solve the latter. It is pointed out in \cite{FP1} that, when applied to solve a univariate toy example, the AO algorithm (based on the two FP transforms) converges sublinearly and thus slower than the conventional Dinkelbach's algorithm. A rigorous proof of the convergence rate of the AO algorithm when applied to more general (sum-rate maximization) problems remains open. Another question is how to accelerate the AO algorithm to achieve a faster convergence rate.
	
	\item As mentioned at the end of Section \ref{subsection:duality}, reformulating the problem under consideration into a convex form (if necessary) and exploring its solution structure in algorithm design are two major technical obstacles for duality-based algorithms to work \cite{liu2021UplinkdownlinkDualityMultipleaccess}. However, the constantly evolving structures and nature of optimization problems due to architecture and networking innovations in wireless communication systems lead to significant challenges in the application of duality-based algorithms. More efforts are still needed to push the boundary of uplink-downlink duality towards more general scenarios (e.g., ISAC and RIS-assisted systems) and develop duality-based algorithms for solving (possibly nonconvex) optimization problems in these scenarios. 
	

\item While we have explored various distributed optimization methods for wireless communication system design in Section \ref{sec: dist_opt part 1}, it is worth noting that only a few of them have found practical implementation. This limited deployment can be attributed to the challenges posed by the interconnection between BSs in multi-cell systems, where backhaul bandwidth constraints often lead to significant communication delays.
In real-world scenarios, BSs cannot engage in frequent message exchanges due to these constraints, so iterative algorithms in Section \ref{sec: dist_opt part 1} are not favored.  Consequently, a critical question arises: How can we effectively optimize the transmission strategies of multiple BSs with minimal or even no message exchanges \cite{han2020distributed}? Similar challenges also arise in the context of user scheduling within multi-cell systems, particularly for cell-edge users who require coordinated optimization across multiple BSs to be simultaneously allocated to the same resource block. These issues underscore the complexity of achieving efficient wireless system design and user scheduling in practical resource-constrained environments.

	\item 
 Most evidence on the effectiveness of neural networks is empirical.
	There are still many open questions, such as whether there can be any
	theoretical guarantee in the performance of learning-based approaches, how to
	choose the best neural network architecture that would require the fewest training
	samples, how to account for constraints in a data-driven approach
	\cite{graph_constraint_mark}, how to combine data-driven and model-driven methodologies (an example of which is by unfolding existing
	algorithmic structures \cite{unfolding_1, unfolding_2}; see Section \ref{subsection:unfolding} for more details), the possible role
	of reinforcement learning in solving optimization problems, etc.  
\end{itemize}

\subsection{Research Directions}
In this subsection, we point out some potential directions for future work on next-generation wireless communication system design.
\subsubsection{{Distributed Signal Processing and Optimization for Extremely Large-Scale Antenna Array (ELAA) Systems}}

To support multiple services with diverse and customized QoS requirements in next-generation wireless communication systems, there is a growing trend in increasing the number of antennas at BSs, which has led to the emergence of ELAA systems. However, as mentioned in Section \ref{sec:distributed},
as the number of antennas increases, traditional centralized baseband processing (CBP) architectures encounter bottlenecks in terms of high fronthaul costs and computational complexities. To address these challenges, decentralized baseband processing (DBP) architectures have emerged as a promising approach \cite{li2017decentralized, jeon2019decentralized, sanchez2020decentralized, sarajlic2019fully, zhang2021decentralized,  alegria2021trade}. In the DBP architecture, the antennas at the BS are divided into several antenna clusters, each equipped with an independent and more affordable baseband processing unit (BBU) and connected with other BBUs as a star network or as a daisy-chain network.

Compared to the CBP architecture, the DBP approach has several advantages. First, the DBP architecture only requires distributed units (DUs) to exchange some locally processed (low-dimensional) intermediate results, thereby reducing the interconnection cost. Second, since each DU only needs to process a low-dimensional received signal, the computational complexity in each DU can be significantly reduced. Last but not least, the DBP architecture improves the scalability and robustness of ELAA systems, as adding or removing antenna elements simply amounts to adding or removing computing units.

Despite the promising initial advancements in ELAA systems with DBP architectures, interconnection costs, which increases rapidly with the expansion of the array size, is a key issue.  More specifically, most of developed distributed algorithms are based on iterative implementations that suffer from frequent message exchanges and high computational complexities \cite{li2017decentralized, jeon2019decentralized, sanchez2020decentralized, sarajlic2019fully, zhang2021decentralized,  alegria2021trade,zaib2016distributed},  although some attempts have been made recently to overcome these bottlenecks \cite{10281371,10196411}. In addition, tight synchronization is required among distributed nodes and corresponding synchronization signals must be implemented across the DUs. For example, to perform coherent beamforming, high-accuracy time synchronization and phase calibration are crucial \cite{larsson2023phase,rashid2022frequency}.  
Furthermore, when additional components such as network-controlled repeaters, RISs, and backscatter communication are introduced to distributed MIMO systems \cite{chen2023static}, the integration of these techniques with the distributed architecture will present new challenges. All of these are fresh opportunities for optimization algorithm development.

{The ELAA system is used in the above as an example to illustrate that new wireless communication applications and scenarios will lead to new mathematical optimization problems and drive the development of distributed signal processing and optimization theory and algorithms.  
Indeed, there are many interesting applications as well as signal processing and optimization problems in all of the six major usage scenarios of 6G (see Fig.~\ref{fig:6Gusecases}), which call for new and novel signal processing and optimization theory and algorithms. As Alan V. Oppenheim reminds us\cite{Oppenheim2012}, ``\emph{There will always be signals, they will always need processing, and there will always be new applications, new mathematics, and new implementation technologies.}''
}
}
\subsubsection{Quantum Optimization and ML \cite{Kim21,Botsinis19,Nawaz19,Narottama23}}
{ML and AI techniques have significantly changed and will continue to change the way that mathematical optimization problems are formulated and solved. As reviewed in Sections \ref{sec:learning} and \ref{sec:no:csi}, data-driven approaches have provided an efficient way of solving complex optimization problems from wireless communication system design that cannot be accurately modeled and/or efficiently solved by traditional optimization approaches. There are new ideas on the horizon that can change the research landscape for mathematical optimization. An example of this is quantum computing. A grand research challenge, as well as opportunity, is the co-design of quantum computer architectures and quantum optimization and ML algorithms such that optimization problems can be efficiently solved by quantum optimization and ML algorithms on quantum computers.   
}

Emerging paradigms of ML, quantum computing, and quantum ML, and their synergies with wireless communication systems might become enablers for future networks. 
This speculative vision of a quantum internet is outlined in \cite{Gyongyosi22}. On one hand, quantum information theory will give rise to new optimization problem formulations  \cite{Holevo2013}. For the optimization community, it often leads to novel and exciting mathematical optimization problems involving matrix-valued functions (e.g., functions involving input density matrices or operators). Quantum-assisted optimization for wireless communications and networking has already been investigated in \cite{Kim21,Botsinis19}. On the other hand, quantum-assisted (e.g., annealing-based) computational models can lead to more efficient solutions to problems in wireless communications and networking. Typical optimization problems include quantum-assisted multi-user detection, quantum-aided multi-user transmission in combination with multiple-access technologies including channel estimation, quantum-assisted indoor localization for mmWave and visible light communications, and quantum-assisted joint routing, load balancing, and scheduling \cite{Botsinis19}. 

Finally, quantum ML \cite{Nawaz19} defines complex artificial neural network structures to perform quantum supervised, unsupervised, reinforcement, federated, and deep learning. The work \cite{Narottama23} presents a perspective on quantum ML methodologies and their applications for wireless communications.

\section{Conclusion}\label{sec:conclusion}
Mathematical optimization is a powerful modeling and solution tool for the design of wireless communication systems. Mathematical optimization theory, algorithms, and techniques play central roles in formulating the right optimization problems behind wireless communication system design, obtaining structural insights into their solutions,  developing efficient, provable, yet interpretable algorithms for solving them, as well as understanding analytic properties of optimization problems and convergence behaviors of optimization algorithms. This paper provides a survey of recent advances in mathematical optimization theory and algorithms for wireless communication system design. More specifically, we review recent advances in nonconvex nonsmooth optimization (including fractional programming, sparse optimization, proximal gradient algorithms, penalty methods, and duality-based algorithms), global optimization (including branch-and-bound and branch-and-cut algorithms), distributed optimization (and federated learning), learning-based optimization (with and without CSI), and their successful application examples in wireless communication system design. More importantly, a goal of this paper is to give guidance on how to choose and/or develop suitable algorithms (and neural network architectures) for solving structured optimization problems from wireless communications and to promote the cross-fertilization of ideas in mathematical optimization and wireless communications.

\ifCLASSOPTIONcaptionsoff
\newpage
\fi



%
\bibliographystyle{IEEEtran}
\bibliography{proximal_gradient_2, sparse_2, reference_onebit_2, liuieeebibfiles20210702_2, chenbib20210424_2, FP20210702_2, networkslicing20210702_2, tsp_duality_2, reference_dce_2, research_2,ref_wei_3, journalshort, qaplp_2, glopt_2,ref_2,THC20231007_2,introduction202310_2,reference_FL_2} %

\end{document}